\documentclass[%
superscriptaddress,
preprint,
 amsmath,amssymb,
 aps,
pre
]{revtex4-1}

\usepackage{graphicx}
\usepackage{dcolumn}
\usepackage{bm}
\usepackage{hyperref}
\usepackage{chngcntr}

\allowdisplaybreaks 
\begin{document}


\title{Chaotic semi-express buses in a loop}
\author{Vee-Liem Saw}
\email{Vee-Liem@ntu.edu.sg}
\affiliation{Division of Physics and Applied Physics, School of Physical and Mathematical Sciences, 21 Nanyang Link, Nanyang Technological University, Singapore 637371}
\affiliation{Data Science and Artificial Intelligence Research Centre, Block N4 \#02a-32, Nanyang Avenue, Nanyang Technological University, Singapore 639798}
\author{Luca Vismara}
\email{vism0001@e.ntu.edu.sg}
\affiliation{Complexity Institute, Interdisciplinary Graduate Programme, 61 Nanyang Drive, Nanyang Technological University, Singapore 637335}
\affiliation{Division of Physics and Applied Physics, School of Physical and Mathematical Sciences, 21 Nanyang Link, Nanyang Technological University, Singapore 637371}
\author{Lock Yue Chew}
\email{lockyue@ntu.edu.sg}
\affiliation{Division of Physics and Applied Physics, School of Physical and Mathematical Sciences, 21 Nanyang Link, Nanyang Technological University, Singapore 637371}
\affiliation{Data Science and Artificial Intelligence Research Centre, Block N4 \#02a-32, Nanyang Avenue, Nanyang Technological University, Singapore 639798}
\affiliation{Complexity Institute, 61 Nanyang Drive, Nanyang Technological University, Singapore 637335}
%

\date{\today}

\begin{abstract}
Urban mobility involves many interacting components: buses, cars, commuters, pedestrians, trains etc., making it a very complex system to study. Even a bus system responsible for delivering commuters from their origins to their destinations in a loop service already exhibits very complicated dynamics. Here, we investigate the dynamics of a simplified version of such a bus loop system consisting of two buses serving three bus stops. Specifically, we consider a configuration of one bus operating as a normal bus which picks up passengers from bus stops $A$ and $B$, and then delivers them to bus stop $C$, whilst the second bus acts as an express bus which picks up passengers only from bus stop $B$ and then delivers them to bus stop $C$. The two buses are like asymmetric agents coupled to bus stop $B$ as they interact via picking up passengers from this common bus stop. Intriguingly, this \emph{semi-express} bus configuration is more efficient and has a lower average waiting time for buses, compared to a configuration of two normal buses or a configuration of two express buses. We reckon the efficiency arises from the chaotic dynamics exhibited in the semi-express system, where the tendency towards anti-bunching is greater than that towards bunching, in contradistinction to the regular bunching behavior of two normal buses or the independent periodic behaviour of two non-interacting express buses.


\end{abstract}

\maketitle


\textbf{Bus systems play an important role in moving people efficiently within cities. A recent discovery by means of multi-agent reinforcement learning revealed that a \emph{semi-express} bus configuration in a loop service would reduce commuters' average waiting time for a bus to arrive, compared to normal or fully express buses. Here, we study its dynamics to see how this works, illuminate the intricate mechanisms involved, and show that the transition to significant improvement in the average waiting time occurs at the edge of chaos as the demands for services at bus stops are varied. A semi-express bus configuration is useful as it does not confuse passengers and drivers, compared to other active and adaptive intervention strategies commonly implemented by bus operators.}

\section{Introduction}

The rise of artificial intelligence in the last decade has spurred serendipitous advances in a raft of disparate research areas. Such rapid development works mutually where it benefits other traditional areas as well as the latter providing insights towards the inner working of the former. For instance, the use of machine learning has aided physicists in discovering (or rediscovering) new physics concepts \cite{Iten20}. Conversely, applications of physics in machine learning have been implemented in recent studies as regression problems, for example in predicting the temperature of a lake \cite{Karp18} as well as the inverted pendulum (a cart-pole setup) and tumor growth dynamics \cite{Singh19}. The incorporation of physical laws into the otherwise arguably black-box machine learning algorithms has been demonstrated to vastly improve its performance and produce physically meaningful results. Perhaps most remarkably, significant breakthrough in image recognition of rotated or transformed images being equivalent to the original image has finally been achieved \cite{Cohen19,Cheng19}, thanks to an abstract mathematical proof of an invariant quantity. That framework on gauge-invariance is essentially based on the mathematics of \emph{Albert Einstein's Theory of General Relativity} on Gravitation \cite{Ein15,Vee16}. All these thus illustrate the intimate symbiosis between computer scientists and physicists leading to collective and emergent cutting-edge developments in both fields.

One urban complexity problem with enormous implications is on public transportation systems which move numerous people in cities worldwide. A perennial problem is that buses tend to bunch together which reduce its efficiency, in contrast to them being spread out evenly along their service route. Extensive research carried out over the past several decades contributed towards understanding why buses frequently end up bunching \cite{Newell64,Chapman78,Powell83,Gers09,Bell10,Vee2019,Chew2020}, including a physical theory based on coupled oscillators that describes bus bunching and its stability \cite{Vee2019,Chew2020}. Various strategies have also been proposed to rectify this problem \cite{Abk84,Ros98,Eber01,Hick01,Fu02,Bin06,Mukai08,Daganzo09,Cor10,Cats11,Gers11,Bart12,Chen15,Ibarra15, Chen16,Moreira16,Wang18,Ale18,Menda19,Del09,Del12,Zhao16,Sun18,Vee2019b,Vee2019c,Vee2019d,Li91,Eber95,Fu03,Sun05,Cor10,Liu13,Furth85,Furth85b,Eber95,Eber98,Liu13,Quek2020,Aramsiv20}.

In particular, our work in Ref.\ \cite{Aramsiv20} presents a theory of express buses where a bus or group of buses serve a fixed subset of bus stops, with these subsets being disjoint. We then find by reinforcement learning \cite{Sutton} that under some conditions, express buses perform better than normal buses where the latter would end up bunching into a single platoon. Beyond our original expectations, however, a simple example shows that a \emph{semi-express} configuration where one bus serves all bus stops but another bus only serves \emph{some} bus stops turns out to be the best, in terms of minimising the average time a commuter waits at a bus stop for a bus to arrive. That surprising and counter-intuitive result motivates a formal study of semi-express buses to understand their complex behaviour. This paper is in essence inspired by that novel reinforcement-learning-discovered achievement of semi-express buses, which we will show here that it is in fact a \emph{chaotic system}. Hence, we experience the aforementioned symbiotic relationship between physics and artificial intelligence.

We present the formulation of a simple semi-express bus system in the next section. There, we study the system with $M_O=2$ origin bus stops $A$ and $B$, followed by $M_O=2$ origin bus stops $A$ and $B$ plus $M_D=1$ destination bus stop $C$ to show its chaotic dynamics when served by $N=2$ semi-express buses. The former simplified version where alighting is not required has less states to enumerate compared to the latter. It also turns out that it admits periodic orbits in the regime $k_A<k_B$ and windows of periodic orbits embedded within chaos in the regime $k_A>k_B$. The two parameters $k_A$ and $k_B$ are, respectively, the ratios of the people arrival rates at bus stops $A$ and $B$ to the loading rate. These periodic orbits are destroyed when alighting is included at bus stop $C$. Subsequently in Section \ref{analytic}, we derive an approximate analytical map for the semi-express system which allows us to calculate the Liapunov exponents to show that it behaves chaotically. The approximate analytical map also allows for the calculation of the average waiting time of commuters for a bus to arrive at a bus stop. We can compare this with the cases of normal and express buses (Appendix \ref{expresstheory}) to show that semi-express buses are superior for $k_A\gtrsim k_B$. Then, Section \ref{semiexpressdiscussion} discusses these chaotic results, before concluding the paper.

We provide in Appendix \ref{expresstheory} of this paper, a construction of the theory of fully express buses but with a different paradigm from the view adopted in Ref.\ \cite{Aramsiv20}. Here, a bus stop is exclusively treated as an origin bus stop (where people only want to board a bus) or a destination bus stop (where people want to alight at). Under this framework, we can calculate the times spent by buses at various such origin or destination bus stops as well as the average time a commuter has to wait at a bus stop for a bus to arrive, in the case of express buses where they do not interact with each other. It turns out that the locations of these bus stops as well as where people want to go would become irrelevant, with regards to the analytical results for express buses. Consequently, one can place an origin bus stop arbitrarily close to a destination bus stop such that they are effectively merged into a typical bus stop where there are both people who want to board and alight. This view of separating the origin and destination natures of a bus stop is useful, as we study the simplest non-trivial $A+B\rightarrow C$ system where there are two origin bus stops $A$ and $B$, with one destination bus stop $C$ (see Section \ref{semiexpress}). This is one of the systems where reinforcement learning in Ref.\ \cite{Aramsiv20} discovers a semi-express configuration of two buses that minimises the average waiting time better than normal or fully express configurations. Appendix \ref{expressdiscussion} discusses the implications of our theory of these express buses, where we highlight special symmetric cases showing how express buses are superior to normal buses, as well as pointing out how our framework also applies to normal bus stops which are both an origin and a destination. Several subsequent appendices \ref{appenB}-\ref{appenF} are included to deal with greater technical details as well as summarising the exact state transition rules for the semi-express bus system.

\section{Interacting semi-express buses}\label{semiexpress}

Consider perhaps the simplest non-trivial setup of a loop comprising $M_O=2$ origin bus stops $A$ and $B$ with $M_D=1$ destination bus stop $C$, served by $N=2$ buses $X$ and $Y$. We shall refer to this as an ``$A+B\rightarrow C$ system''. Normal buses would end up bunching and this pair of buses simultaneously pick up people from $A$ and $B$, subsequently allowing all of them to alight at $C$. Express buses is the case where say, $X$ picks up people from $A$ and sends them to $C$ whilst $Y$ picks up people from $B$ and sends them to $C$. Here, $X$ and $Y$ are non-interacting as they mind their own businesses independently picking people up from $A$ and $B$, respectively.

Suppose now that $X$ picks up people from \emph{both} $A$ and $B$, but $Y$ remains picking up people only from $B$. Then $Y$ is still an express bus, but $X$ is a normal bus. We refer to such a system as \emph{semi-express}, since it comprises buses which are express and not express. In more general setups with more origin bus stops and buses, each bus itself could be \emph{express} in the sense that they only serve their respective subsets of origin bus stops where no bus serves all bus stops. However, unlike the non-interacting express buses as we present in Appendix \ref{expresstheory}, they may here interact if their subsets have a non-null intersection. Then, one can refer to this as \emph{interacting express buses}.

In this semi-express system, $X$ and $Y$ interact via $B$ because $X$ shares the load with $Y$ at $B$. The more people $X$ picks up from $B$, then the less people $Y$ picks up from $B$ which allows it to complete the loop faster. Consequently, $Y$ may be the one picking up more people next time (or maybe not), which may (or may not) reduce the time $X$ would have to spend stopping at $B$ next time, creating complex behaviour. In contrast to normal buses where they eventually bunch into a single platoon and move together, here $X$ has to also serve $A$ by itself which breaks it away from $Y$. Thus, whilst the interaction at $B$ induces a bunching proclivity between $X$ and $Y$, the additional stopping at $A$ only for $X$ induces an anti-bunching effect.

A closed loop of bus stops can always be isometrically mapped to a unit circle, such that the evolving positions of the buses are effectively represented by their phases $\theta_X(t),\theta_Y(t)\in[0,2\pi)$ on the unit circle. Let each bus move with constant angular velocity $\omega=2\pi/T$ along the loop, unless of course when they are stationary at a bus stop to pick up people. We can determine the exact evolution of the bus system by brute-force enumeration of what the next state of the system is, given its current state. We define a \emph{state} as the moment a bus just leaves a bus stop, with its phase difference defined as $\Delta:=(\theta_Y-\theta_X)\textrm{ mod }2\pi$ at that moment, measured in radians. For example if bus $X$ just leaves bus stop $A$, then this state is $XA$ with phase difference $\Delta_{XA}=(\theta_Y-\theta_X)\textrm{ mod }2\pi$ measured at the moment when $X$ just leaves $A$. In the event where both $X$ and $Y$ happen to be at some bus stops (could be bunched at the same bus stop), we only consider a state to be when the second bus leaves the bus stop, in other words the moment when no bus is at any bus stop any more. For example if $X$ leaves $A$ before $Y$ leaves $B$, then the state is $YB$ with the phase difference $\Delta_{YB}=(\theta_Y-\theta_X)\textrm{ mod }2\pi$ measured at the moment when $Y$ leaves $B$. In the event where both buses leave simultaneously, one of the appropriate states is chosen as the next state.

\begin{figure}
\centering
\includegraphics[width=13cm]{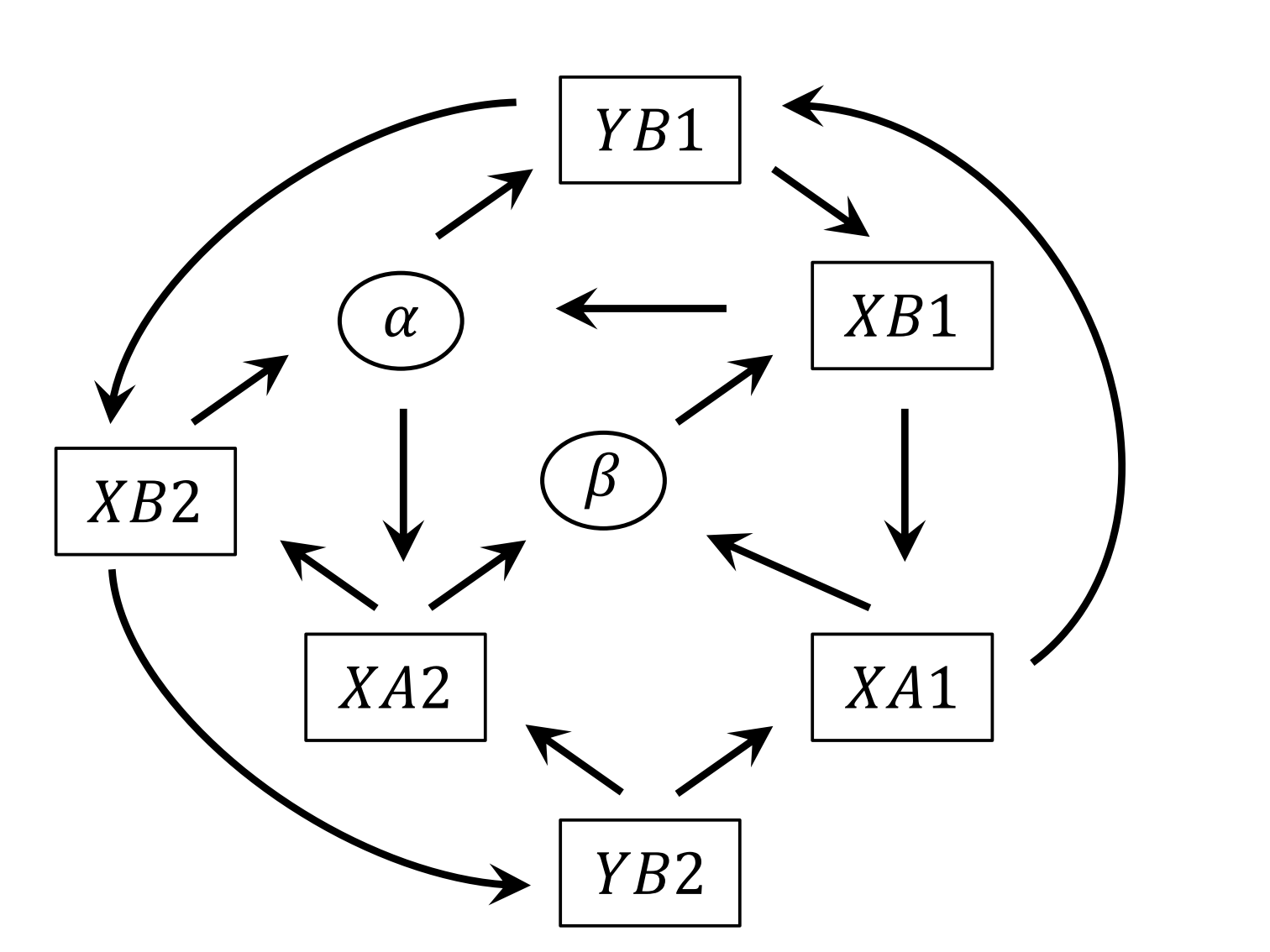}
\caption{The transition graph for the $AB$ system. There are six distinct states, together with $\alpha$ which denotes the situation when the two buses are currently at the two \emph{antipodally} located bus stops, and $\beta$ which denotes the situation when the two buses are \emph{bunched} at bus stop $B$. These $\alpha$ and $\beta$ are referred to as \emph{configurations}, as they conveniently denote unique situations in this $AB$ system. They are not considered as \emph{states} because they are not situations where one bus just leaves a bus stop with the other bus not being at some bus stop. Later in the $A+B\rightarrow C$ system, we do not correspondingly define such configurations $\alpha,\beta$ explicitly because the two buses can bunch at $B$ or $C$, and there are several ways the two buses can be both stopping at two out of three bus stops. In other words, it is only in the $AB$ system that $\alpha$ uniquely refers to $X$ being at $A$ and $Y$ being at $B$; and $\beta$ uniquely refers to $X$ and $Y$ bunching at $B$.}
\label{fig1}
\end{figure}

Before dealing with the $A+B\rightarrow C$ system, it is instructive to consider a simpler system without alighting, i.e. a loop comprising only two bus stops $A$ and $B$. This smaller system, referred to as the ``$AB$ system'', has only two bus stops. This leads to six distinct states, where knowing its current state with its $\Delta$ would uniquely determine its next state and its next phase difference $\Delta'$. For definiteness, we place $A$ and $B$ antipodally, so the time taken to traverse between them is $T/2$. In this simplified system, $X$ picks up people from both $A$ and $B$ whilst $Y$ only picks up people from $B$. After a bus finishes with picking up people, they leave the bus stop and resume motion on the loop to the next bus stop. For $Y$, it stops at $B$ to pick up people until there is nobody left and then takes time $T$ going around the loop to return to $B$ and pick up new people. These six distinct states are $XA1$, $XA2$, $XB1$, $XB2$, $YB1$ and $YB2$. Here, ``$XA$'' refers to the situation where $X$ just leaves $A$, and generally the suffix ``1'' refers to the situation where $\Delta\leq\pi$ whilst the suffix "2" refers to the situation where $\Delta\geq\pi$. The transition graph, obtained by brute-force enumeration, is shown in Fig.\ \ref{fig1}.

The bus stops $A$ and $B$ have people arrival rates of $s_A$ and $s_B$ people per second, respectively. Loading up people occurs at a rate of $l$ people per second, and $k_A:=s_A/l,k_B:=s_B/l$. These rates are constant, so there is no stochasticity involved. The number of people at a bus stop is zero when a bus just leaves after picking up everybody, and then begins to accumulate until a next bus comes around to pick up everybody again. In contrast to non-interacting express buses (see Appendix \ref{expresstheory}), this points to the \emph{phase difference} between the two buses being an important quantity as it plays a role in determining the number of people accumulated at a bus stop since a previous bus had left. Given the moment a bus just leaves a bus stop, we can determine the next state, i.e. which bus will arrive at some bus stop, and calculate how many people it has to pick up. This determines the time $\tau_i$ it spends stopping. After spending time $\tau_i$, it then leaves this bus stop and defines a new state with a new phase difference. The calculations are simple and straightforward algebraic manipulations, albeit tedious. These detailed transition rules are explained in Appendix \ref{appenB} and summarised in Figs.\ \ref{fig15}-\ref{fig20}. The corresponding transition graph for the realistic $A+B\rightarrow C$ system where everyone who boarded a bus would alight at a third bus stop $C$ ($A,B,C$ are separated by $2\pi/3$ on the circle) is vastly more complicated to be drawn in a two-dimensional plane, so we present it as a table in Table \ref{table1}. Here, the suffixes ``$1,2,3$'' generally refer to $\Delta\leq2\pi/3,2\pi/3\leq\Delta\leq4\pi/3,4\pi/3\leq\Delta$, respectively. The detailed transition rules for this are summarised in Figs.\ \ref{fig21}-\ref{fig35} in Appendix \ref{appenC}.

\begin{table}
\centering
\begin{tabular}{|c|c|c|c|c|c|c|c|c|c|c|c|c|c|c|c|}
\hline
State & $XA1$ & $XA2$  & $XA3$  & $XB1$  & $XB2$  & $XB3$  & $XC1$  & $XC2$  & $XC3$  & $YB1$  & $YB2$  & $YB3$  & $YC1$  & $YC2$  & $YC3$\\
\hline
& $XB1$ & $XB2$ & $XB1$ & $XC1$ & $XC2$ & $XC3$ & $XA1$ & $XA2$ & $XA3$ & $XB1$ & $XA2$ & $$XC1 & $XC1$ & $XB2$ & $XA1$ \\
To & $YB1$ & $YC1$ & $XB3$ & $YC1$ & $XC3$ & $YB2$ & $XA2$ & $YB1$ & $YC2$ & $XB2$ & $XA3$ & $XC3$ & $XC2$ & $XB3$ & $XA3$ \\
& & $YC2$ & & $YC3$ & $YB2$ & $YB3$ & $YB1$ & $YB2$ & $YC3$ & $YC1$ & $YC2$ & $YC3$ & & & \\
\hline
& $XC1$ & $XC1$ & $XC3$ & $XA1$ & $XA2$ & $XA3$ & $XB1$ & $XB2$ & $XB2$ & $XA1$ & $XB2$ & $XB3$ & $XA2$ & $XA2$ & $XB1$ \\
From & $YC3$ & $XC2$ & $YB2$ & $XA3$ & $YB1$ & $YC2$ & $YB3$ & $YC1$ & $XB3$ & $XC1$ & $XB3$ & & $XB1$ & $XC3$ & $XC3$ \\
& & $YB2$ & $YC3$ & $YB1$ & $YC2$ & & $YC1$ & & $YB3$ & $XC2$ & $XC2$ & & $YB1$ & $YB2$ & $YB3$ \\
\hline
\end{tabular}
\caption{Table of state transitions for the $A+B\rightarrow C$ system. There are fifteen distinct states when a bus leaves a bus stop (with the other bus not being at a bus stop). From each state, we list its next possible states as well as the possible states preceding it.}\label{table1}
\end{table}

We present some interesting results first for the $AB$ system where windows of periodic orbits exist, and then for the $A+B\rightarrow C$ system where they do not exist. Subsequently, we derive an approximate analytical description for both the $AB$ and $A+B\rightarrow C$ systems that would enable the calculation of the Liapunov exponents. This would turn out to imply sensitivity to initial conditions and chaotic behaviour of the semi-express bus system.

\subsection{The \texorpdfstring{$AB$}{AB} system}

The $AB$ system allows us to focus on studying the effect of the interaction of bus $X$ and bus $Y$ via bus stop $B$. We begin the dynamics with the following initial condition: $X$ just leaves $A$ and $Y$ just leaves $B$ at time $t=0$ with $\tau_{XA_0}=\tau_{YB_0}=0$ (the durations that $X$ stopped at $A$ and $Y$ stopped at $B$, respectively), so there are zero people at both bus stops and $\Delta_0=\pi$. We then evaluate the system according to the transition graph given in Figs.\ \ref{fig1}, \ref{fig15}-\ref{fig20}, over $10,000$ iterations and only plot the last $500$ iterates, i.e. we assume that by $9,500$ iterations any transient has been excluded. In fact, we tested this with different initial conditions and they give essentially identical results. All values are expressed in units of $T$. The interaction of the two buses at one common bus stop in this semi-express setup leads to complex chaotic dynamics with aperiodic evolution.

\begin{figure}
\centering
\includegraphics[width=16cm]{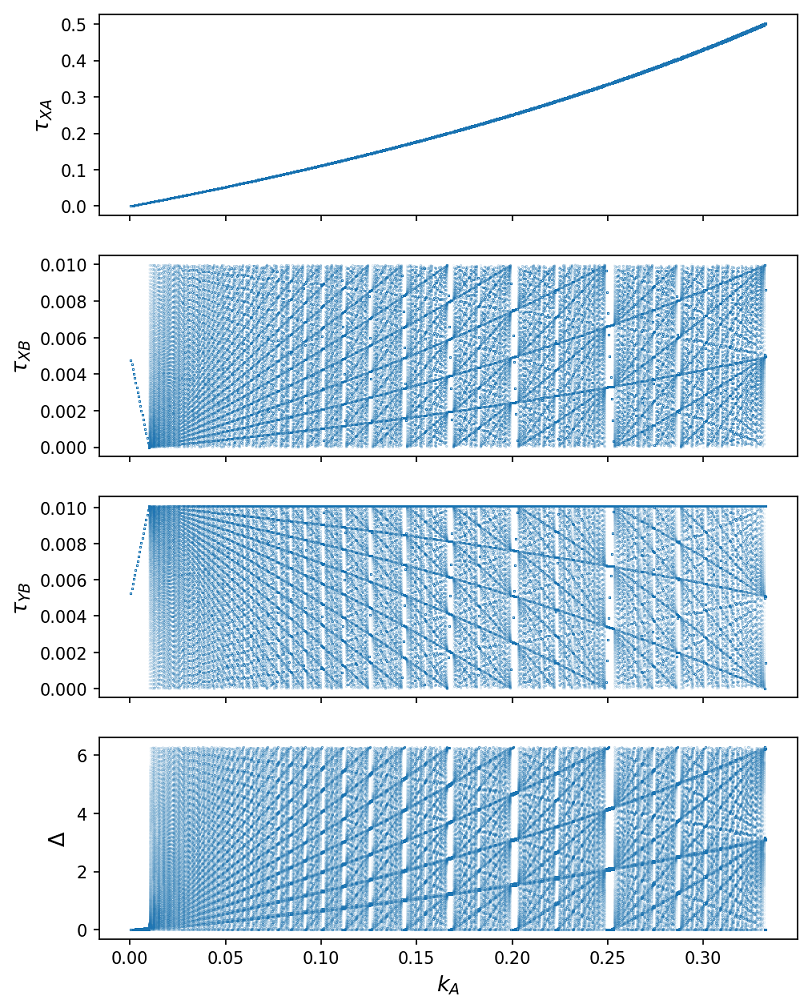}
\caption{Values taken by $\tau_{XA}$, $\tau_{XB}$, $\tau_{YB}$ and $\Delta$ for various values of $k_A$ from $0$ to $0.3325$, in an increment of $0.0005$. The value for $k_B$ is kept at $0.01$. This is the exact $AB$ system.}
\label{fig2}
\end{figure}

Fig.\ \ref{fig2} shows the values taken by $\tau_{XA}$, $\tau_{XB}$, $\tau_{YB}$ which respectively denote the time that bus $X$ stops at $A$, at $B$, and the time that bus $Y$ stops at $B$. It also shows the values taken by the phase difference $\Delta$ between the buses when a bus leaves a bus stop (and the other bus is not at a bus stop). Here, $k_B$ is kept fixed at $0.01$, with $k_A$ given a value that starts from $0$ and increased by $0.0005$ for each complete run, up till $0.3325$. For higher values of $k_A$, the state transition rules given by Fig.\ \ref{fig1} break down due to the assumption that when a bus is at a bus stop, the other bus at most traverses only one bus stop (see Appendix \ref{appenB}). Notice also that $\tau_{XA}\sim0.5$ near this upper limit, which is about $1/2$ of a revolution where a bus may traverse two bus stops if $k_A$ is even stronger than $0.3325$. Whilst the plot for $\tau_{XA}$ appears to be a smooth curve for all $k_A$, this is not really true for $k_A>k_B$ other than the windows of periodic orbits. Typically, it comprises a smear of points at a much smaller scale (of the order of $0.001$) which appear like a single point because the value of $\tau_{XA}$ itself is of the order of $0.1$. This similarly occurs in Figs.\ \ref{fig3}-\ref{fig5}.

As all these quantities are bounded, we find that the bus system is essentially always in chaos, except when $k_A<k_B$ and for some windows of values for $k_A>k_B$ where the system cycles in \emph{periodic orbits}. Details on such periodic orbits are presented in Appendix \ref{appenD}. Later, we provide an analytical approximation to calculate the Liapunov exponents to show that the system is in fact essentially chaotic (that analytical approximation turns out to destroy the existence of any periodic orbits).

\subsection{The \texorpdfstring{$A+B\rightarrow C$}{A + B -> C} system}

Similar to the previous boarding-only case, we adopt the initial condition that $X$ just leaves $A$ and $Y$ just leaves $B$ at time $t=0$ with $\tau_{XA_0}=\tau_{XC_0}=\tau_{YB_0}=0$, so there are zero people at both origin bus stops and $\Delta_0=2\pi/3$. We then evaluate the system according to the transition rules given in Table \ref{table1} and Figs.\ \ref{fig21}-\ref{fig35} over $10,000$ iterations and only plot the last $500$ iterates. All values are expressed in units of $T$.

\begin{figure}
\centering
\includegraphics[width=16cm]{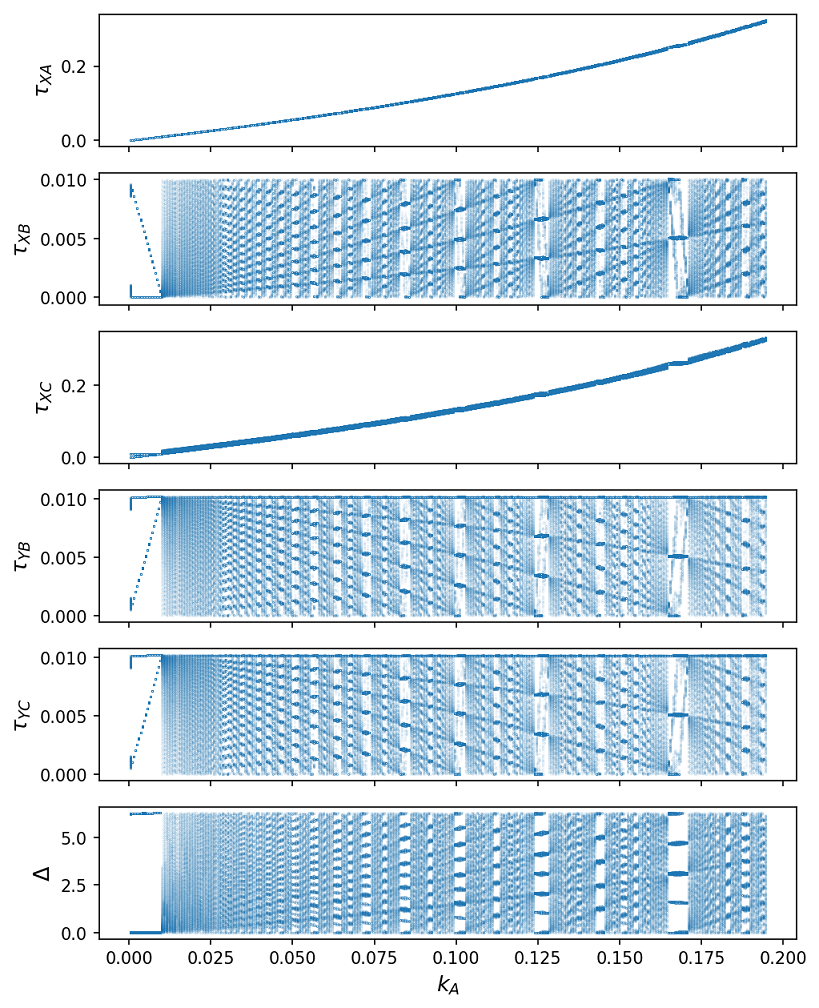}
\caption{Values taken by $\tau_{XA}$, $\tau_{XB}$, $\tau_{XC}$, $\tau_{YB}$, $\tau_{YC}$ and $\Delta$ for various values of $k_A$ from $0$ to $0.1945$, in an increment of $0.0005$. The value for $k_B$ is kept at $0.01$. This is the exact $A+B\rightarrow C$ system.}
\label{fig3}
\end{figure}

Fig.\ \ref{fig3} shows the values taken by $\tau_{XA}$, $\tau_{XB}$, $\tau_{XC}$, $\tau_{YB}$, $\tau_{YC}$ which respectively denote the time that bus $X$ stops at $A$, at $B$, at $C$, and the time that bus $Y$ stops at $B$ and at $C$. Here, $\tau_{XC}=\tau_{XA}+\tau_{XB}$ since the number of people alighting at $C$ is the sum of the numbers boarded from $A$ and $B$ for bus $X$, and $\tau_{YC}=\tau_{YB}$ since the number of people alighting at $C$ is equal to the number of people who boarded from $B$ for bus $Y$. It also shows the values taken by the phase difference $\Delta$ between the buses when a bus leaves a bus stop (and the other bus is not at a bus stop). Here, $k_B$ is again kept fixed at $0.01$, with $k_A$ given a value that starts from $0$ and increased by $0.0005$ for each complete run, up till $0.1945$. For higher values of $k_A$, the state transition rules break down due to the assumption that when a bus is at a bus stop, the other bus at most traverses only one bus stop (see Appendix \ref{appenC}). Notice also that $\tau_{XA}\sim0.3333$ near this upper limit, which is about $1/3$ of a revolution where a bus may traverse two bus stops if $k_A$ is even stronger than $0.1945$.

As all these quantities are bounded, we find that the bus system is essentially always in chaos. Unlike the $AB$ system however, there is no periodic orbit for any value of $k_A$! Even for $k_A<k_B$, close examination reveals that $\Delta$ and the various $\tau_i$ do not cycle in a finite set of fixed points, but take values that appear to form a fractal set. When a point is zoomed in, it actually turns out to comprise two or more points. When one of these points is zoomed in further, it turns out to again comprise two or more points, and so on. There is also no window of periodic orbits for $k_A>k_B$. What seemingly look like periodic orbits are actually a smear of points.

There is a crucial difference to why there is no periodic orbit when there is a bus stop $C$ for people to alight, as compared to the $AB$ system where the buses only pick up people. Here, the time spent at $C$ has a ``memory'' based on the number of people already on board. So if $X$ and $Y$ happen to bunch at $C$, they do not necessarily leave together. They only leave together from $B$ if they bunch there because they share loading. Nevertheless, the smear of points do look to be bounded within localised pockets, despite not being periodic orbits.

Thus, we find the benefits of first studying a simpler albeit arguably unrealistic system in understanding the dynamics of the semi-express system. Apart from being easier to deal with with less number of states to enumerate, the $AB$ system admits nice and analytically calculable periodic orbits in the regime where $k_A<k_B$, as well as windows of periodic orbits in the regime where $k_A>k_B$. Such periodic orbits are destroyed when a destination bus stop $C$ is included due to the memory of the number of people on board from the origin bus stops. Without first concretely understanding the $AB$ system, we might not have appreciated the fact that these localised chaotic points are actually periodic orbits in the absence of bus stop $C$.

\section{An analytical approximation to the semi-express bus system}\label{analytic}

The exact evolution of the bus system cannot be written in terms of analytical equations. This is due to the next state being conditional upon the phase difference. More specifically, if a bus has to traverse some other bus stops before arriving at the intended bus stop to pick up people, then there would be slightly more people to pick up arising from having to stop at those intermediate bus stops. In the $A+B\rightarrow C$ system, if $\Delta<2\pi/3$, then there is no intermediate bus stop to traverse. But if $2\pi/3<\Delta<4\pi/3$, then there is one intermediate bus stop to traverse. If $\Delta>4\pi/3$, then there are two intermediate bus stops to traverse. Writing a computer programme with conditional statements is fine. However, this cannot be written as a single analytical equation.

To have any hope of analytically studying such an interacting system of buses to glean insightful understanding on its complex dynamics, we would construct an approximate analytical map with the aim of calculating the eventual quantities of interest, viz. the average waiting time $W$ and times spent by buses at bus stops $\tau_i$ of this system. The map allows us to show that the system is in fact \emph{chaotic}, with the quantities $W$, $\tau_i$ evolving aperiodically and being sensitive to initial conditions.

Before constructing the map for this $A+B\rightarrow C$ system, it is again instructive to consider the simpler $AB$ system where $X$ only picks up people from $A$ and $B$, and $Y$ only picks up people from $B$, i.e. there is no alighting required. The reason for first working with this is its simplicity in illustrating the key ideas to derive a 6-d map that describes the time evolution of the relative positions between $X$ and $Y$ on the loop (i.e. their phase difference) as well as how long they spend stopping at $A$ and $B$, respectively. With this understanding, generalisation to the $A+B\rightarrow C$ system is straightforward, producing a 10-d map that describes the time evolution of the phase difference between $X$ and $Y$ and how long they spend stopping at $A$, $B$ and $C$, respectively. This method can be systematically extended to bus systems with more buses and bus stops.

\begin{figure}
\centering
\includegraphics[width=16cm]{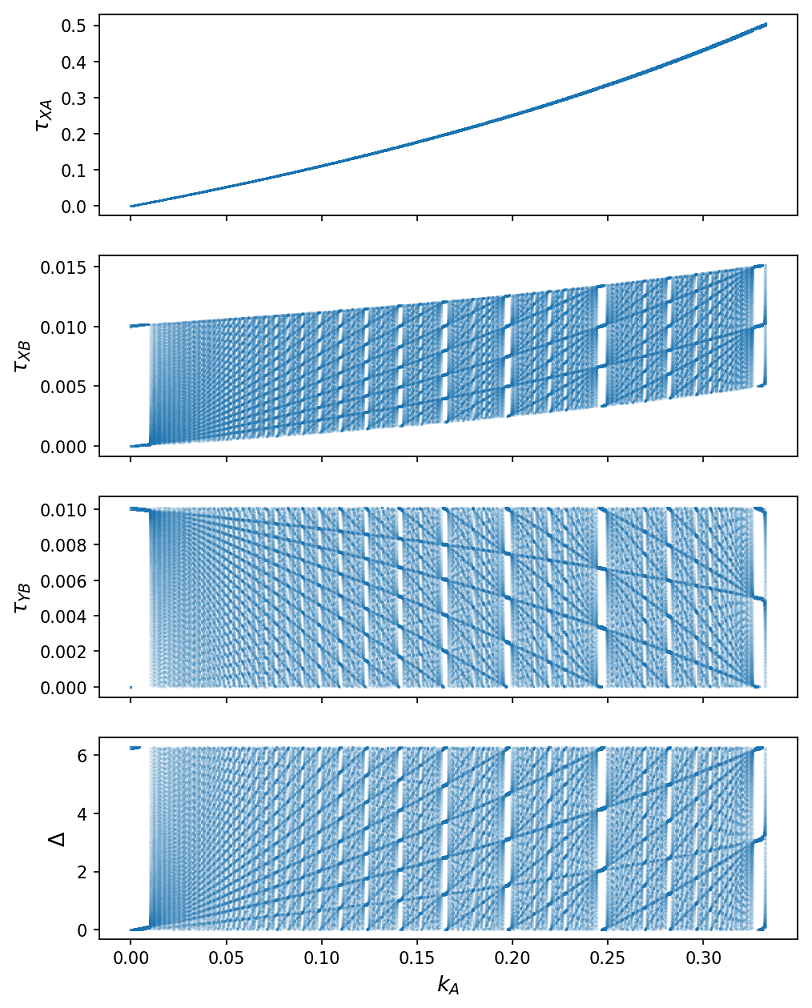}
\caption{Values taken by $\tau_{XA}$, $\tau_{XB}$, $\tau_{YB}$ and $\Delta$ for various values of $k_A$ from $0$ to $0.3325$, in an increment of $0.0005$. The value for $k_B$ is kept at $0.01$. This is an analytical approximation of the $AB$ system.}
\label{fig4}
\end{figure}

\begin{figure}
\centering
\includegraphics[width=16cm]{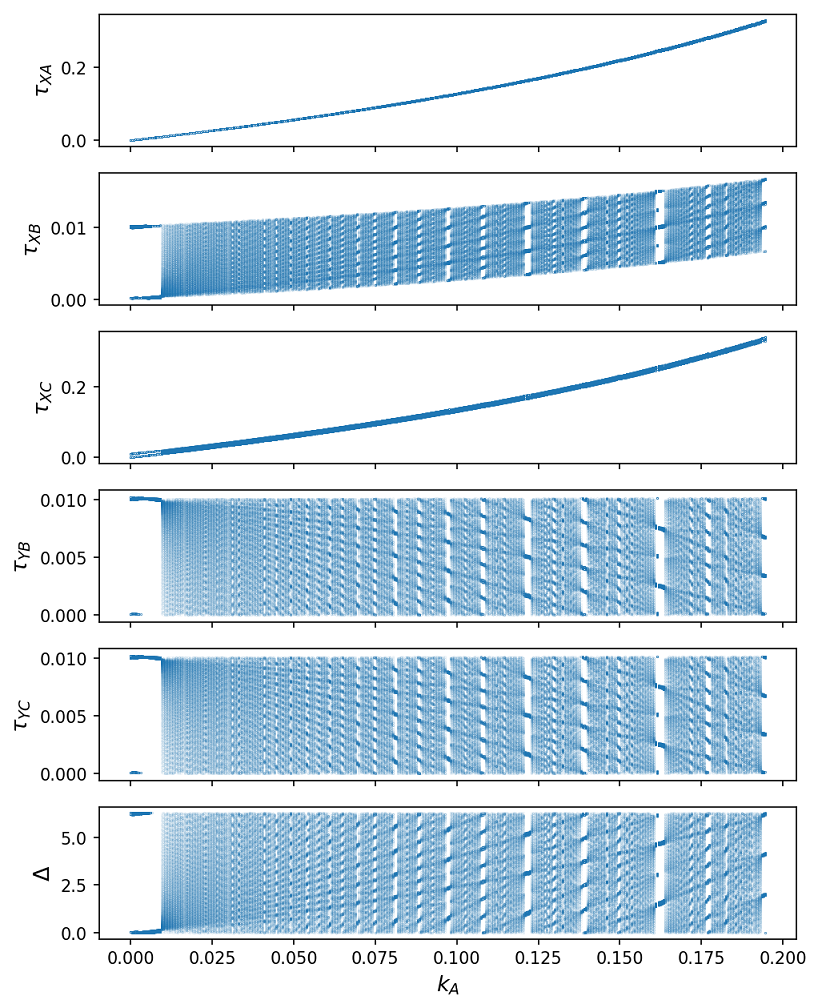}
\caption{Values taken by $\tau_{XA}$, $\tau_{XB}$, $\tau_{XC}$, $\tau_{YB}$, $\tau_{YC}$ and $\Delta$ for various values of $k_A$ from $0$ to $0.1945$, in an increment of $0.0005$. The value for $k_B$ is kept at $0.01$. This is an analytical approximation of the $A+B\rightarrow C$ system.}
\label{fig5}
\end{figure}

\begin{figure}
\centering
\includegraphics[width=13cm]{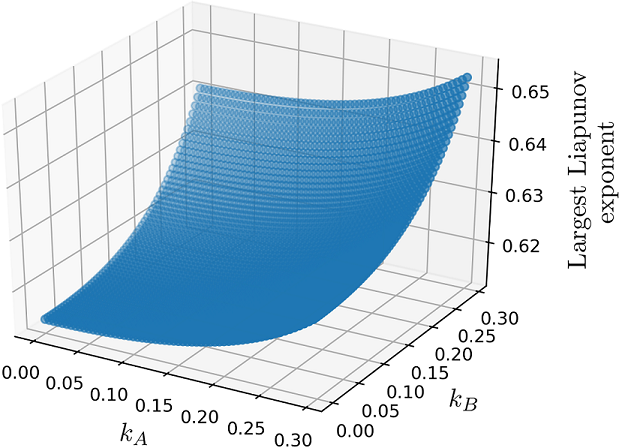}
\caption{The largest Liapunov exponent of the 10-d map is always positive over various values of $k_A,k_B\in(0,0.3)$. Incidentally, none of the 10 Liapunov exponents ever has value $0$ for all parameters $k_A,k_B$.}
\label{fig6}
\end{figure}

The analytical approximation to the $AB$ system is derived in Appendix \ref{appenE}, and that for the $A+B\rightarrow C$ system is presented in Appendix \ref{appenF}. Here, we show the corresponding results which capture the essential dynamics of the exact system, as displayed in Fig.\ \ref{fig4} for the $AB$ system and Fig.\ \ref{fig5} for the $A+B\rightarrow C$ system. In fact with the analytical equations, we can calculate the largest Liapunov exponent for the $A+B\rightarrow C$ system and find that it is \emph{always positive for all values of $k_A$ and $k_B$} (see Fig.\ \ref{fig6}), with none of the Liapunov exponents being zero. This implies sensitivity on initial conditions, i.e. \emph{the system is chaotic}. Similar results are true for the $AB$ system as well, implying no periodic orbit under the analytical approximation to the exact $AB$ system.

\subsection{Average waiting time \texorpdfstring{$W$}{W} for the semi-express \texorpdfstring{$A+B\rightarrow C$}{A+B to C} bus system}

\begin{figure}
\centering
\includegraphics[width=16cm]{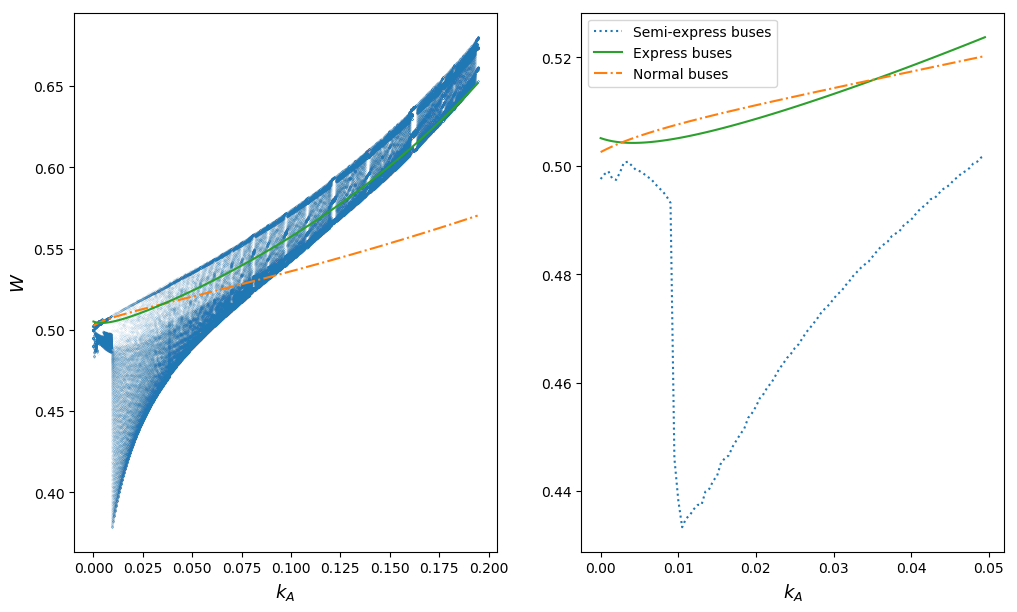}
\caption{The average waiting time $W$ for a bus to arrive at a bus stop, for various values of $k_A$, with $k_B=0.01$. Left: $k_A\in[0,0.1945]$, and each blue point corresponds to the average waiting time for that loop of the semi-express buses. Different loops would lead to different average waiting times, which fluctuate chaotically. Right: Each blue point is the average across different loops. This plot also zooms into $k_A\sim k_B$. Similar results are found in Ref.\ \cite{Aramsiv20} using a time-step-based simulation.}
\label{fig7}
\end{figure}

\begin{figure}
\centering
\includegraphics[width=16cm]{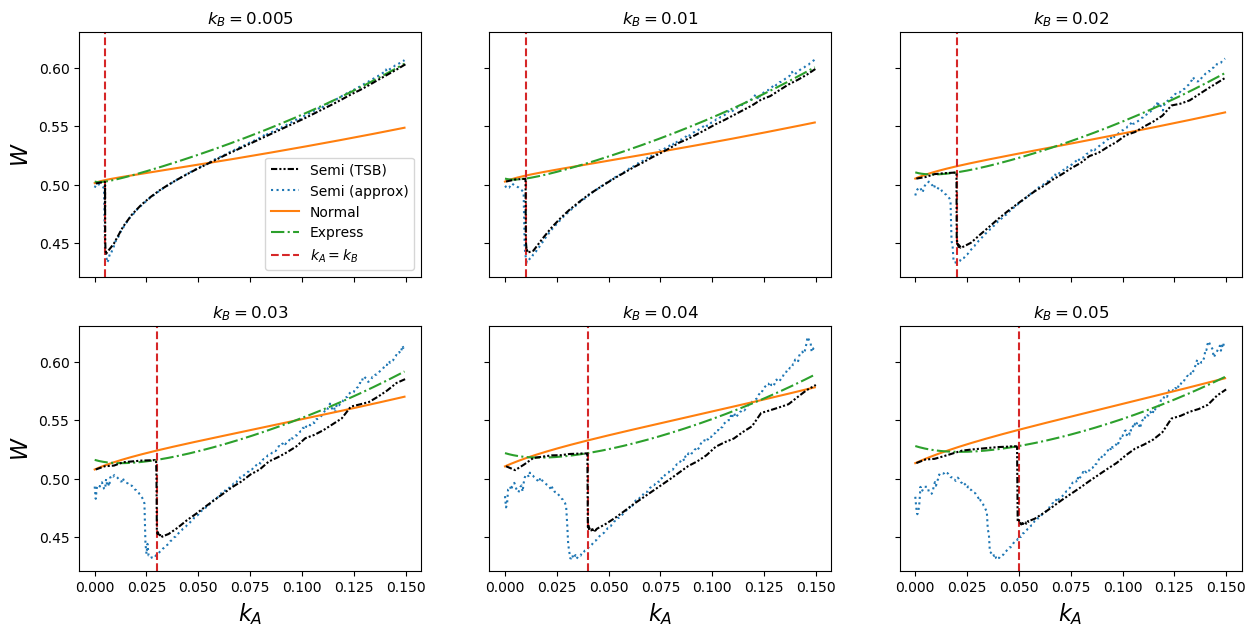}
\caption{Graphs of the average waiting time $W$ for a bus to arrive at a bus stop, for various values of $k_A$, with $k_B=0.005,0.01,0.02,0.03,0.04,0.05$. For each graph, the significant improvement to $W$ occurs at $k_A=k_B$, as seen from TSB.}
\label{fig8}
\end{figure}


The analytical approximation allows for an analytical calculation of the average waiting time of commuters for a bus to arrive at the bus stop. This quantity can be calculated from the following:
\begin{gather}\label{semiexpressW}
W=\frac{k_AW_A+k_BW_B}{k_A+k_B},
\end{gather}
where $W_A$ and $W_B$ are the average waiting times at $A$ and $B$ respectively, so $W$ is just the average waiting time over both bus stops, weighted by how many people there are --- which are proportional to $k_A$ and $k_B$, respectively. Let us now evaluate $W_A$ and $W_B$ so that Eq.\ (\ref{semiexpressW}) can be calculated.

Since the people arrival rates at $A$ and $B$ are assumed to be constant, then $W_A$ and $W_B$ are just half times the longest waiting time at $A$ and $B$, respectively. For $A$, this is
\begin{align}
W_A&=\frac{1}{2}(T+\tau_{XB}+\tau_{XC})\label{howtogetWa}\\
&=\left(\frac{1-k_A}{2k_A}\right)\tau_{XA},\label{WA}
\end{align}
where we have used Eq.\ (\ref{auxA}) to simplify. In Eq.\ (\ref{howtogetWa}), bus $X$ would take a time of $T+\tau_{XA}+\tau_{XB}+\tau_{XC}$ to complete a loop. So this total time minus $\tau_{XA}$ is how long the unluckiest person has to wait, since after bus $X$ just leaves $A$, then it takes $T+\tau_{XB}+\tau_{XC}$ to arrive at $A$ again.

For $B$ on the other hand, this is the weighted average of the average waiting time for $X$ ($W_{XB}$) and that for $Y$ ($W_{YB}$), where the weights are proportional to how long $X$ stops at $B$ and how long $Y$ stops at $B$, i.e. $\tau_{XB}$ and $\tau_{YB}$, respectively. So
\begin{align}
W_{XB}&=\frac{1}{2}\left(\frac{\Delta_{YB}}{\omega}+\tau_{XC}+\tau_{XA}\right)\\
&=\left(\frac{1-k_B}{2k_B}\right)\tau_{XB},
\end{align}
where we have used Eq.\ (\ref{aux1}) to simplify. Similarly,
\begin{align}
W_{YB}&=\frac{1}{2}\left(T-\frac{\Delta_{XB}}{\omega}+\tau_{YC}\right)\\
&=\left(\frac{1-k_B}{2k_B}\right)\tau_{YB},
\end{align}
where we have used Eq.\ (\ref{aux2}) to simplify. Hence, $W_B$ is
\begin{align}
W_B&=\frac{\tau_{XB}W_{XB}+\tau_{YB}W_{YB}}{\tau_{XB}+\tau_{YB}}\\
&=\left(\frac{1-k_B}{2k_B}\right)\left(\frac{\tau_{XB}^2+\tau_{YB}^2}{\tau_{XB}+\tau_{YB}}\right).\label{WB}
\end{align}

Plugging in Eqs.\ (\ref{WA}) and (\ref{WB}) into Eq.\ (\ref{semiexpressW}) gives an expression for $W$ for this $A+B\rightarrow C$ semi-express bus system, which can be numerically calculated by iterating the map Eqs.\ (\ref{aux1})-(\ref{endmap}) to obtain the relevant $\tau_i$'s evolution. Therefore, Fig.\ \ref{fig7} is plotted using Eq.\ (\ref{semiexpressW}). The simplifications of $W_A,W_{XB},W_{YB}$ using the defining map's Eqs.\ (\ref{auxA}), (\ref{aux1}), (\ref{aux2}) imply that given $k_A$ and $k_B$, then $W$ in Eq.\ (\ref{semiexpressW}) only depends on three independent quantities $\tau_{XA},\tau_{XB},\tau_{YB}$, viz. the durations buses spend stopping at origin bus stops to pick up people. In this figure, we also plot the analytical results for the average waiting times from Appendix \ref{expresstheory} for normal buses and express buses serving these $M_O=2$ origin bus stops with $M_D=1$ destination bus stop. The semi-express system clearly leads to the lowest average waiting time in the regime where $k_A\gtrsim k_B$, though the average waiting time for every loop may fluctuate chaotically and occasionally exceed the average waiting times of normal and express buses.

Fig.\ \ref{fig8} shows the graphs corresponding to the right plot of Fig.\ \ref{fig7} for various values of $k_B$. Here, these plots include the average waiting times $W$ obtained by a time-step-based (TSB) simulation that directly measures the waiting times of commuters at bus stops for a bus to arrive \cite{Aramsiv20}. This provides the actual measurements for $W$ as the basis for comparison with the analytical approximation given by Eqs.\ (\ref{semiexpressW}), (\ref{WA}) and (\ref{WB}). Generally, semi-express buses are better than express buses or normal buses when $k_A\gtrsim k_B$ for various $k_B$ and the significant improvement in the semi-express buses happens at $k_A=k_B$ as seen by the actual measurements based on TSB. For $k_B$ less than $\sim0.2$, the analytical approximation is in good agreement with TSB. For larger $k_B$, it predicts the transition to significant improvement in $W$ happening at some $k_A$ less than $k_B$.

For $k_A\gg k_B$, the asymmetry between demands from the two origin bus stops is too large. Since $A$ has much stronger demand than $B$ but the former is only being served by $X$, then $Y$ is relatively underutilised as it serves a small demand from $B$. In this regime, normal buses are the best with the lowest average waiting time. In other words, the chaotic semi-express system is only superior if $k_A\gtrsim k_B$ such that the ``unbunching force'' due to $A$ holding back $X$ is sufficiently strong, as well as not being excessively large to render $Y$ irrelevant. For $k_A<k_B$, the unbunching force is insufficient and the two buses are almost always bunched. So semi-express buses are only marginally better than normal or fully express buses if $k_A<k_B$.


\section{Discussion on chaotic semi-express buses}\label{semiexpressdiscussion}

\subsection{Transition into \texorpdfstring{$k_A>k_B$}{kA>kB}}

Consider the $AB$ system. If $X$ and $Y$ bunch at $B$ and then leave together, after $T/2$ $X$ would stop at $A$ for duration $\tau_{XA}$ with $Y$ proceeding on and eventually returning to $B$. Let $k_A$ be sufficiently small such that $X$ would leave $A$ and then bunch with $Y$ at $B$ before $Y$ gets to leave $B$. This semi-express system is in a periodic orbit, as discussed earlier. We would like to determine the critical value of $k_A$ such that if $k_A$ exceeds this value, then this periodic orbit ceases to exist.

This critical $k_A$ is defined by the situation where $X$ just reaches $B$ the moment $Y$ leaves $B$. Hence, $X$ spends zero stoppage time at $B$, i.e. $\tau_{XB}=0$. From the expression that we have found for $\tau_{XB}$ in Eq.\ (\ref{XBperiod2}), this gives $k_A=k_B$. This is why we see the transition in Fig.\ \ref{fig2} when $k_A$ exactly matches $k_B$. If $k_A$ exceeds this critical value, then $X$ is held at $A$ for too long such that when it eventually reaches $B$, $Y$ would have already left and the two buses successfully unbunch.

Let us now deal with the $A+B\rightarrow C$ system. The corresponding critical $k_A$ is defined by the situation where $X$ just reaches $B$ the moment $Y$ leaves $B$. They then move together and arrive at $C$, spending the same amount of time stopping there, and eventually leave $C$ together. Hence, $X$ spends zero stoppage time at $B$, i.e. $\tau_{XB}=0$. Then, $\tau_{XC}=\tau_{XA}$ since the number of people alighting at $C$ from $X$ is the same as the number of people who boarded $X$ from $A$. Similarly, $\tau_{YC}=\tau_{YB}$. Note that this critical situation decouples into a fully express system where $X$ effectively serves only $A$ to $C$ with $Y$ serving $B$ to $C$. Such a fully symmetric situation with $X$ bunching with $Y$ at $B$, then moving together to $C$ and subsequently leaving together from $C$ implies that $\tau_{XC}=\tau_{YC}$. This necessitates $\tau_{XA}=\tau_{YB}$, i.e. $k_A=k_B$. Thus, we have the critical $k_A$ being equal to $k_B$ where $X$ is held briefly enough at $A$ and cannot unbunch from $Y$ if $k_A$ is weaker than $k_B$. If $k_A$ exceeds $k_B$, then $X$ cannot reach $B$ before $Y$ leaves $B$, allowing the two buses to unbunch. This is why a transition happens as seen in Fig.\ \ref{fig3} when $k_A$ exactly matches $k_B$.

This critical $k_A=k_B$ represents a transition into chaos or the \emph{edge of chaos}. From Figs.\ \ref{fig7}-\ref{fig8}, it appears that it is precisely at the edge of chaos that the average waiting time of the semi-express system is minimal. Note that the $AB$ version comprises periodic orbits for $k_A<k_B$ and begins to behave chaotically from $k_A>k_B$. Hence, the edge of chaos at $k_A=k_B$ is viewed from this, as the memory effect makes the $A+B\rightarrow C$ system admit no periodic orbits.

\subsection{Windows of periodic orbits in the \texorpdfstring{$AB$}{AB} system, the corresponding \texorpdfstring{$A+B\rightarrow C$}{A+B->C} system, and their analytical approximations: chaotic attractors}

Windows of periodic orbits exist in the $AB$ system for $k_A>k_B$, and the orbits are periodic for any $k_A<k_B$. The addition of a destination bus stop $C$ destroys this due to the memory of the number of people to alight based on the number of people picked up previously. Apart from that, the analytical approximations whereby the order of events are forced to be in a fixed ordering and one single $\tau_i$ being specified regardless of the historical evolution of states also turn out to destroy the existence of any periodic orbit. Let us look more closely at these trajectories and compare between these cases.

\begin{figure}
\centering
\includegraphics[width=16cm]{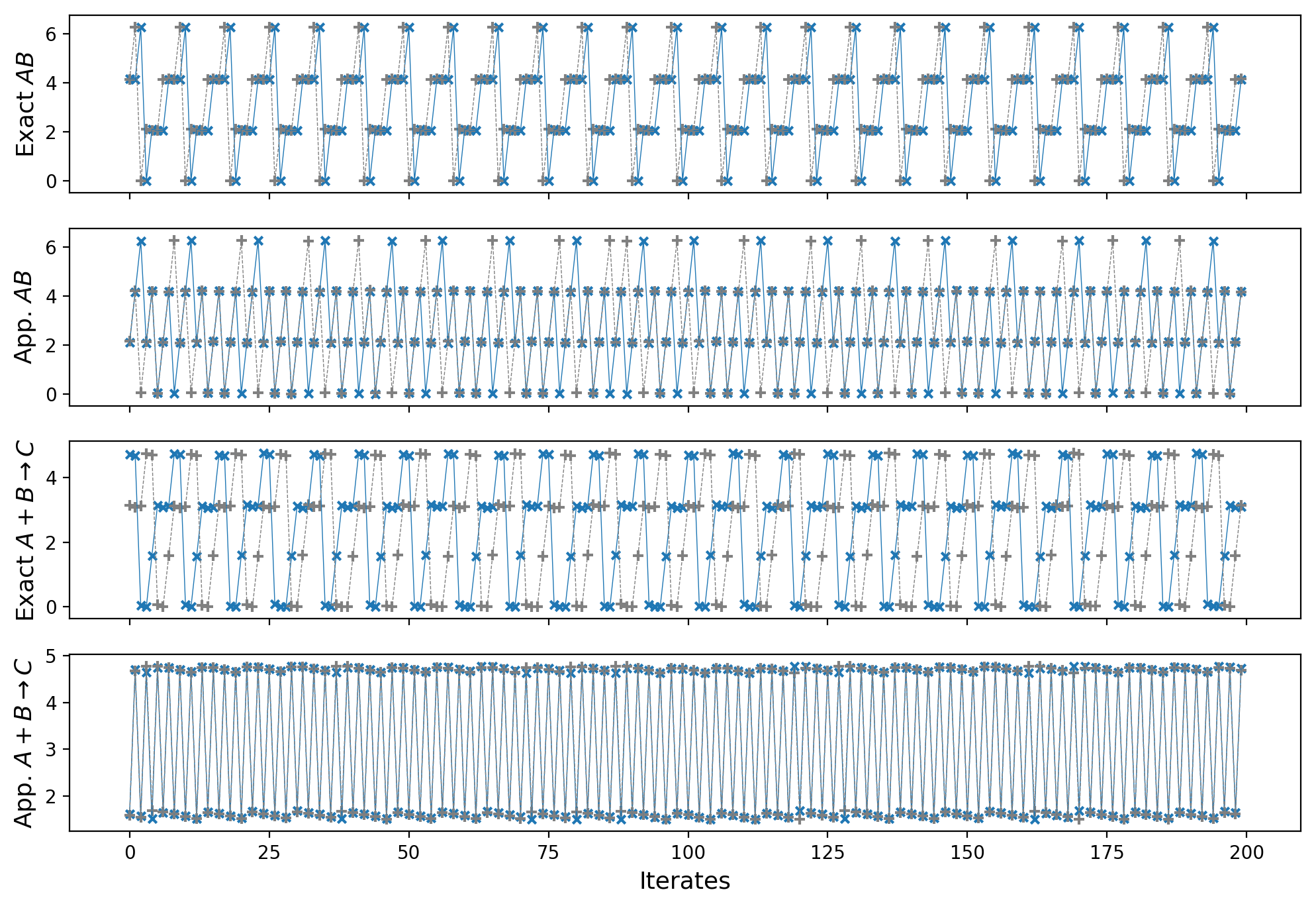}
\caption{Sequences of the last $200$ iterates of $\Delta$ for (from top to bottom): Exact $AB$ system ($k_A=0.25$), analytical approximation of the $AB$ system ($k_A=0.248$), exact $A+B\rightarrow C$ system ($k_A=0.17$), analytical approximation of the $A+B\rightarrow C$ system ($k_A=0.163$). Each point denotes the value of $\Delta$, with lines connecting them indicating the evolution through time. These $k_A$ are all within their respective largest window of periodic or \emph{almost} periodic orbits. Each graph shows the trajectories arising from two different nearby initial conditions. Note that for the exact $AB$ system, both trajectories cycle through the same set of period-$8$ points. For the rest, the trajectories are all irregular.}
\label{fig9}
\end{figure}

\begin{figure}
\centering
\includegraphics[width=16cm]{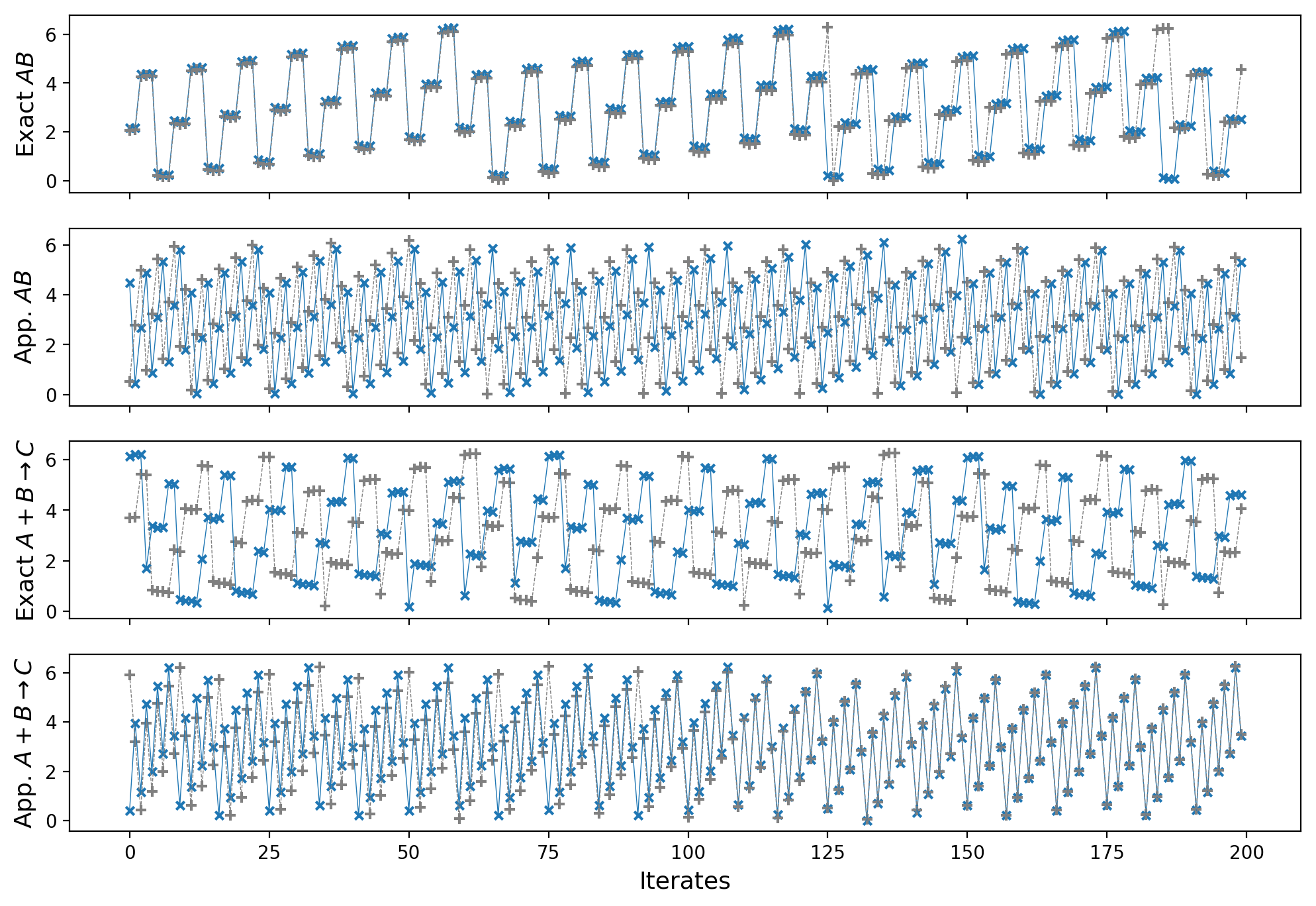}
\caption{The plots corresponding to Fig.\ \ref{fig9} where these do not lie within a window (from top to bottom): Exact $AB$ system ($k_A=0.26$), analytical approximation of the $AB$ system ($k_A=0.26$), exact $A+B\rightarrow C$ system ($k_A=0.175$), analytical approximation of the $A+B\rightarrow C$ system ($k_A=0.175$).}
\label{fig10}
\end{figure}

Fig.\ \ref{fig9} shows the sequences of the last $200$ iterates of $\Delta$ starting from two different nearby initial conditions, taken from the largest window of periodic orbits for the exact $AB$ system (top plot). Here, $k_A=0.25$. The values of $\Delta$ cycle through $8$ fixed points. These correspond to those worked out analytically in Appendix \ref{appenD} with the system cycling through the states given by Eq.\ (\ref{period8-1}). With the inclusion of bus stop $C$ in the $A+B\rightarrow C$ system, the third plot ($k_A=0.17$) shows how the system now cycles aperiodically. Sometimes $\Delta$ goes through $8$ values but it occasionally takes $17$ values, before returning to the value $0$ where they bunch at $B$ and leave together. An inspection on the sequence of states (Table \ref{table1}) also reveals that it sometimes cycles through one set of states, but occasionally goes through a different set of states. These two sets of states are:
\begin{gather}
(1)\ XB1\rightarrow XC1\rightarrow XA2\rightarrow YC2\rightarrow XB2\rightarrow XC3\rightarrow YC3\rightarrow XA1,\textrm{and back to }XB1\\
(2)\ XB1\rightarrow XC1\rightarrow XA2\rightarrow YC2\rightarrow XB2\rightarrow XC3\rightarrow YC3\rightarrow XA1\rightarrow YB1\rightarrow XB1\rightarrow\nonumber\\XC1\rightarrow XA2\rightarrow YC2\rightarrow XB2\rightarrow XC3\rightarrow YC3\rightarrow XA1,\textrm{and back to }XB1.
\end{gather}
The switches between $(1)$ and $(2)$ are irregular.

For the analytical approximations to the $AB$ (second plot in Fig.\ \ref{fig9}, $k_A=0.248$) and $A+B\rightarrow C$ (bottom plot in Fig.\ \ref{fig9}, $k_A=0.163$) systems, $\Delta$ also takes varying values in an aperiodic manner. This leads to a smear of localised points that seem to surround some special points serving as attractors. In other words, there is a set of points which serve as chaotic attractors upon which the trajectories approach and stay close to, even though the actual values are non-repeating and aperiodic. Moreover, different initial conditions eventually lead to trajectories ending up near these chaotic attractors. The corresponding figure for trajectories not within a window is shown in Fig.\ \ref{fig10}. They all gradually fill up space where $\Delta\in[0,2\pi)$.

\begin{figure}
\centering
\includegraphics[width=16cm]{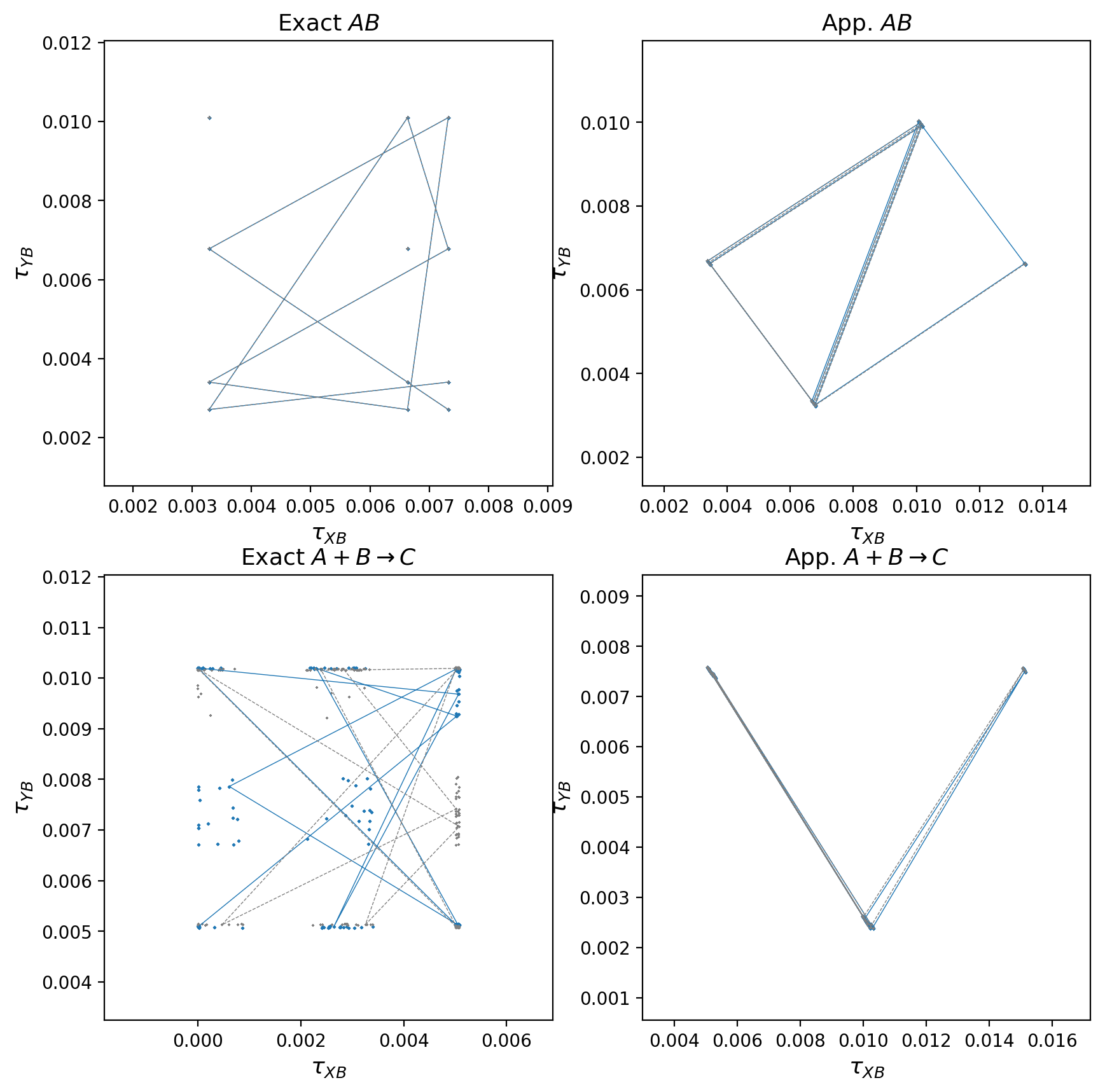}
\caption{The plots of $\tau_{YB}$ versus $\tau_{XB}$ corresponding to Fig.\ \ref{fig9} where these lie within a window.}
\label{fig11}
\end{figure}

\begin{figure}
\centering
\includegraphics[width=16cm]{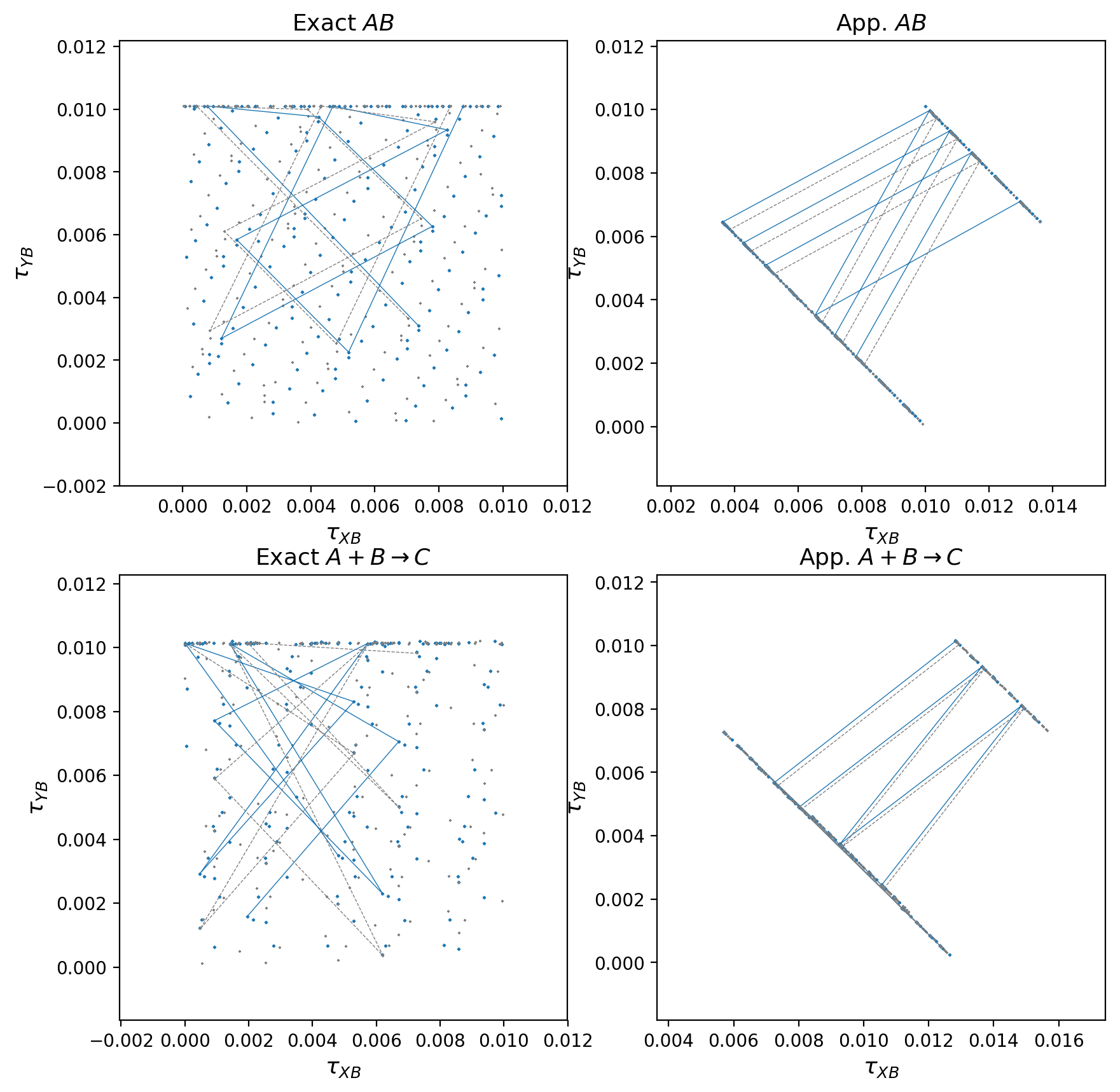}
\caption{The plots of $\tau_{YB}$ versus $\tau_{XB}$ corresponding to Fig.\ \ref{fig10} where these do not lie within a window.}
\label{fig12}
\end{figure}

Figs.\ \ref{fig11}-\ref{fig12} show plots of $\tau_{YB}$ versus $\tau_{XB}$ for the various systems. The former figure corresponds to the parameters in Fig.\ \ref{fig9}, i.e. within a window; whilst the latter figure corresponds to parameters in Fig.\ \ref{fig10}, i.e. not within a window. Lines are drawn to show how $10$ iterates evolve through time. (We do not draw the lines connecting all 200 points, as they would clutter the plots.) Generally, these $\tau_i$ jump around. With the exception of the exact $AB$ system in a window which cycles periodically (top left in Fig.\ \ref{fig11}), these two trajectories arising from different nearby initial conditions end up with vastly different outcomes.

Interestingly, for the analytical approximations, these plots lie on diagonal lines (with some small thickness) with gradient $\approx-1$. This is because the order of events are assumed to be fixed with no overtaking allowed. Consequently, the load at $B$ is shared by both $X$ and $Y$ when they stop at $B$ separately, such that we have $\tau_{XB}+\tau_{YB}\approx$ some constant. There are two diagonal lines because when bunching occurs at $B$, they do not share loading, and that second line is $\tau_{XB}+\tau_{YB}\approx2\ \times$ that constant. In this case, both $X$ and $Y$ would stop at $B$ for the usual duration picking up people at a rate of $l$ as if only one bus is there. We do not implement shared loading when they bunch in the analytical approximation as a conditional clause is required to trigger shared loading at a rate of $2l$, which would break the analyticity of the map. On the other hand for the exact systems, $Y$ tends to overtake $X$ when $k_A>k_B$. Therefore, there are more $\tau_{YB}$ compared to $\tau_{XB}$. In fact, recall that the exact $AB$ system cycles in periodic orbits when $k_A=0.25$. Here, $\tau_{XB}$ has period $3$ whilst $\tau_{YB}$ has period $4$ (see Appendix \ref{appenD}). This is why we see $3\times4=12$ points in the plot of $\tau_{YB}$ versus $\tau_{XB}$ at the top left of Fig.\ \ref{fig11}, as these are all the possible combinations that arise when plotted.

\subsection{The proliferation of chaos in a bus loop system}

In a bus loop system, buses continually go round and round the loop whilst serving the bus stops. When mapped onto a unit circle such that positions of the buses are identified by their phases $\theta_i$, the phase difference between a pair of buses take bounded values $\Delta\in[0,2\pi)$, where $\Delta$ gets a modulo by $2\pi$.

A dynamical system is chaotic if it has the following three properties \cite{Yorke96}:
\begin{enumerate}
\item The trajectories remain bounded, i.e. they do not go off to infinity at all times.

\item The trajectories are not asymptotically periodic, i.e. they do not end up cycling through a finite set of points.

\item The trajectories are sensitive to initial conditions (i.e the largest Liapunov exponent is positive), with none of the Liapunov exponents being zero (ruling out quasi-periodicity).

\end{enumerate}

The phase difference $\Delta$ satisfies property 1. Whilst this does not necessarily imply that the time $\tau_i$ a bus spends at a bus stop also must remain finite, a realistic bus system must not have buses staying put at a bus stop indefinitely. Therefore, property 1 holds for $\Delta$ and $\tau_i$ for reasonable values of $k_A,k_B$.

For property 2, a formal proof is generally hard to obtain \cite{Yorke96}. Nevertheless, Figs. \ref{fig2}-\ref{fig5} provide strong numerical evidence that these quantities are generally aperiodic as they fill up space (and remain bounded). Curiously, there are some parameters where those quantities are bounded into discrete pockets, though they still appear to fill up those pockets. By looking at the state transitions, they are generally non-repeating for the $A+B\rightarrow C$ system. Nevertheless in the $AB$ system, for $k_A<k_B$ and in the windows of periodic orbits for $k_A>k_B$, the state transitions do cycle around like those given by Eqs.\ (\ref{period2}), (\ref{period4}), (\ref{period8-1}), (\ref{period8-2}).

Finally, property 3 is satisfied by the analytical calculation of the Liapunov exponents where none of them are zero and the largest one is positive. The largest Liapunov exponent is in fact positive over the entire 2-d parameter space of $(k_A,k_B)$. The analytical map has a constant Jacobian, similar to the skinny Baker map \cite{Yorke96}. The modulo $2\pi$ on this analytical map is comparable to the ``$2x\textrm{ mod }1$'' map given in example $3.6$ in Ref.\ \cite{Yorke96} which is not continuous at $x=1/2$. These maps all contain chaotic orbits.

The complex chaotic behavior of bus loop systems should perhaps not be surprising, given that the loopy nature and the naturally finite dwell time of buses at bus stops would guarantee the values do not shoot off to infinity. Some form of \emph{interaction} between \emph{asymmetric} agents is a means of leading to aperiodicity. In the absence of interaction, asymmetric agents like express buses which serve different disjoint subsets of bus stops would go about with their respective periodic evolutions. On the other hand, symmetric agents like normal buses which interact would end up all bunching into a single platoon. In the semi-express example that we presented, $X$ serves both origin bus stops but $Y$ only serves one of them so they are asymmetric agents that interact at one of the bus stops --- exhibiting chaotic dynamics. These buses' evolutions turn out to be sensitive to initial conditions for all values of the parameters such that it appears to be \emph{always} chaotic.

\subsection{Time-step-based simulation versus event-based simulation}

When carrying out simulations of a bus loop system, a time-step-based algorithm would discretise the loop into some finite number of cells for the bus to land on at every time step. To study chaotic behaviour, however, precision is crucial since trajectories fill up the space of possible values. The finite discretisation of the loop necessarily rounds off the quantities such that minute differences would just be recorded as being the same. Failure to track such discrepancies to the required precision would falsely lead to the quantities cycling through a finite set of rounded off values, giving the ostensible impression of the absence of chaos. On the other hand, boosting precision by increasing the number of cells would proportionately lengthen the simulation time to the point where it may become painfully slow whilst still not meeting the required precision \cite{Sand98,Pincus09}.

In this paper, we did not show aperiodic trajectories by means of simulating a bus loop system. Instead, we enumerated the exact transition rules for the states as well as derived an approximate map which captures the essential dynamics and then calculated the Liapunov exponents analytically. This is complemented by \emph{computations carried out to iterate the state transition rules as well as the approximate analytical map} as opposed to simulating the bus loop system directly.

If one wishes to carry out simulations on the bus loop system, perhaps a viable approach would be to implement an event-based algorithm that tracks the events directly (viz. when a bus is at a bus stop). Evolution of such an algorithm is in terms of events instead of a fixed time step, which limits the resolution due to restricted computing speed and finite precision. Thus for more complicated setups where the analytical approach is impossible, event-based simulation may be employed to study its complex and possibly chaotic dynamics.

Such an event-based algorithm does not require enumeration of what the next state is, given the current state. All it needs to track is the time a previous bus had left a bus stop, which is the information required to calculate the number of people a next bus has to pick up. Although such an algorithm being programmed to run is scalable to systems comprising many bus stops served by many buses, it does not provide information on the evolution of the states which would not offer insights via a systematic logical analysis that we have presented in this paper. Nevertheless, these two approaches are complementary. Once an event-based simulation explores larger systems and points towards interesting properties, one may subsequently study its properties with greater depth via enumeration of the state transition rules if desired.

\section{Concluding remarks}\label{conclusion}

This paper presents a real-world problem of bus loop systems, where we considered a semi-express configuration. The simplest semi-express system comprises one normal bus serving two origin bus stops with the other bus only picking up from the second origin bus stop. Such a semi-express setup was discovered by a reinforcement learning algorithm (beyond what the authors originally expected \cite{Aramsiv20}) to produce the lowest average waiting time of commuters at a bus stop for a bus to arrive, in a system with two origin bus stops and one destination bus stop. By considering some simplifying assumptions but still capturing its essential dynamics, we derived a 10-d map in this paper to describe this semi-express system and showed that it behaves chaotically.

Although bus systems are known to exhibit chaotic dynamics due to other kinds of setups \cite{Nagatani02,Nagatani03,Nagatani03b,Nagatani03c,Nagatani06}, this is perhaps the first demonstration of chaos for a semi-express system, viz. a mechanism of interacting asymmetric agents. This system is important since it is the most efficient configuration as found by reinforcement learning even beyond just two origin bus stops to one destination bus stop served by two buses \cite{Aramsiv20}. We have also unraveled the understanding on how chaotic motion arises through interaction of asymmetric agents where the ``unbunching force'' $k_A$ must be stronger than the ``bunching force'' $k_B$ for it to be chaotic and improve the average waiting time. Furthermore, we argued that the conditions for chaos of bus loop systems are fairly easy to achieve such that we should perhaps be surprised by the absence of chaos rather than its presence.

In this semi-express setup, chaos appears to be salutary in terms of lowering the average waiting time. On one extreme, normal buses which end up bunching into one single platoon can be regarded as an ``ordered'' situation. On the other extreme, non-interacting express buses resemble a ``random'' situation whereby each express bus does not care about the others and the whole system comprises independently moving units. It turns out that a chaotic ``in between'' situation beats either of these extremes. The most optimal situation occurs at the critical condition $k_A=k_B$, i.e. at the edge of chaos.

Chaos has profound implications in the real world, especially for a bus loop system \cite{Ball03}. For instance, a primary objective of bus operators is to maintain regular scheduling of their fleet of buses such that they are able to report consistent arrival times at bus stops to facilitate commuters' travel plans. A ramification of chaotic motion is the aperiodic fluctuations of time taken for a bus to complete a loop, so there is no way to reliably predict when a bus will arrive. This erratic behaviour emerges completely in the absence of any noise. To enforce regular bus arrival times at bus stops, active intervention strategies like holding \cite{Abk84,Ros98,Eber01,Hick01,Fu02,Bin06,Mukai08,Daganzo09,Cor10,Cats11,Gers11,Bart12,Chen15,Ibarra15, Chen16,Moreira16,Wang18,Ale18,Menda19,Vee2019d}, no-boarding \cite{Del09,Del12,Zhao16,Sun18,Vee2019b,Vee2019c,Vee2019d}, stop-skipping \cite{Li91,Eber95,Fu03,Sun05,Cor10,Liu13}, deadheading \cite{Furth85,Furth85b,Eber95,Eber98,Liu13} which are adaptive real-time or when the phase difference goes beyond some prescribed bound, would seem necessary to maintain stable anti-bunched configurations of buses in a loop \cite{Chew2020}. Nevertheless, semi-express buses \emph{do not actively interfere} with prescribed bus services to the various origin bus stops. In other words, unlike active interventions like no-boarding and holding, semi-express buses do not confuse the passengers as it is clear that this bus or that bus serves or does not serve this bus stop. Besides that, it also does not confuse the bus drivers and they can carry out their duty without being bothered repeatedly on implementing various actions.

Real bus systems are subjected to noise. Buses go through traffic and people arrival rates at various bus stops are non-uniform but perhaps follow a Poisson distribution, for example \cite{Quek2020}. On top of that, it is known that human-driven buses tend to cruise at different natural speeds due to differing driving styles \cite{Vee2019}. It is certainly interesting and important to investigate these effects on realistic bus systems, especially whether the chaotic behaviour is negated such that semi-express buses always end up bunching due to the presence of noise and/or different natural speeds. Whilst the introduction of stochasticity may render an analytical treatment as presented here to be a formidable task, an event-based simulation approach should provide numerical results to reveal what complex behaviour may arise so that we can better understand the dynamics of real bus systems. This will be a direction for future research on such bus systems.

\appendix


\section{A bus loop system with \texorpdfstring{$M_O$}{MO} origin bus stops and \texorpdfstring{$M_D$}{MD} destination bus stops}\label{expresstheory}

Assumptions:
\begin{enumerate}
\item There is a loop with $M_O+M_D$ bus stops, where these bus stops are arbitrarily located along the loop.

\item Each of the $M_O$ bus stops, denoted by $\alpha_i$ where $i=1,\cdots,M_O$, has people arriving at a fixed rate of $s_i$.

\item Each person from $\alpha_i$ has a probability $\zeta_{ij}$ of heading to one of the $M_D$ destination bus stops, denoted by $\beta_j$ where $j=1,\cdots,M_D$. In other words, each of the destination bus stop $\beta_j$ has nobody who wants to board from; and nobody wants to alight at any origin bus stop $\alpha_i$. Note that
\begin{align}\label{sum}
\sum_{j=1}^{M_D}{\zeta_{ij}}=1,
\end{align}
since everybody must end up at one of the destination bus stops $\beta_j$.

\item There are $N$ buses serving this loop, all going in the same direction and move at constant speed. The time it takes for each bus to complete the loop (excluding any time spent stopping at a bus stop) is $T$. There is no acceleration/deceleration involved when stopping. Each bus has unlimited capacity.

\item The rate of people boarding/alighting a bus is $l$. The dimensionless parameters $k_i$ are defined as $k_i:=s_i/l$, for $i=1,\cdots,M_O$.
\end{enumerate}

\begin{figure}
\centering
\includegraphics[width=16cm]{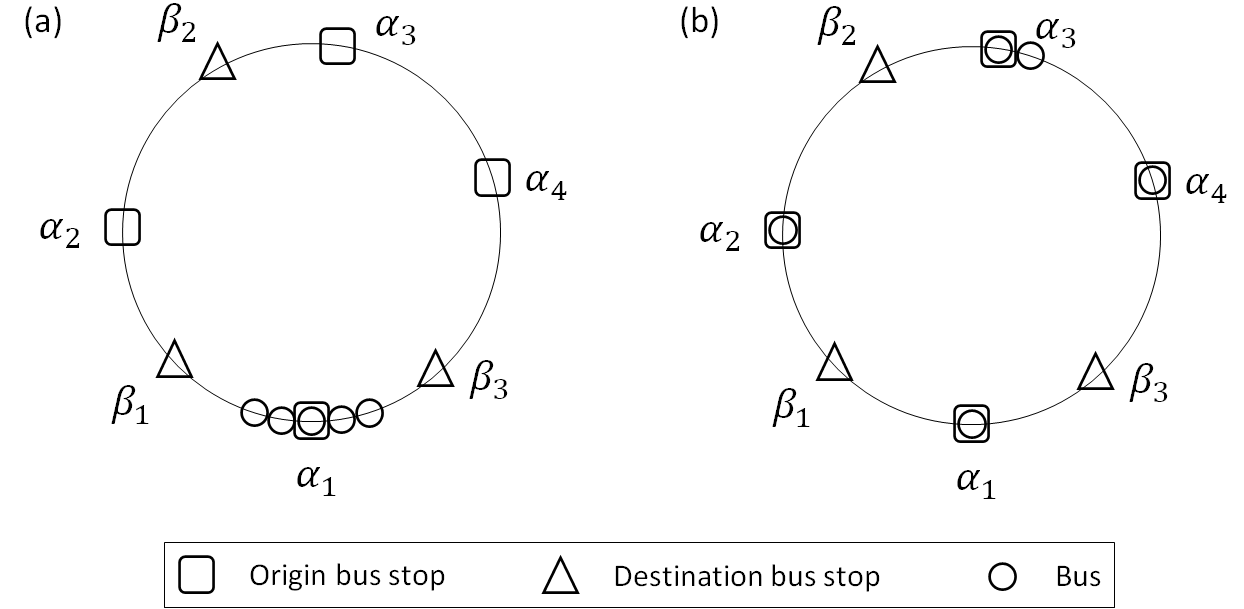}
\caption{(a) $N=5$ \emph{normal} buses serving $M_O=4$ origin bus stops and $M_D=3$ destination bus stops in a loop. All these $M_O+M_D$ bus stops are arbitrarily located along the loop. Since all buses move with the same speed, they all eventually bunch into a single platoon. (b) The same as in (a), but each bus is an \emph{express} bus. An express bus is a bus that only picks up people from one origin bus stop, but always allows alighting at any destination bus stop. Shown here is a setup where the origin bus stop $\alpha_3$ has two express buses serving it, whilst $\alpha_1$, $\alpha_2$ and $\alpha_4$ each only has one express bus serving them.}
\label{fig13}
\end{figure}

Definitions:
\begin{enumerate}
\item For \emph{normal buses}, these $N$ buses would bunch into a single unit (see Fig.\ \ref{fig13}(a) for an example). They form a single platoon with an effective rate of people boarding/alighting of $Nl$.

\item An \emph{express bus} is a bus that serves only one origin bus stop $\alpha_i$ for some $i\in\{1,\cdots,M_O\}$ (see Fig.\ \ref{fig13}(b) for an example). Although it boards people only from one particular $\alpha_i$, it can allow alighting at every $\beta_j$ where $j=1,\cdots,M_D$. More than 1 express bus can serve $\alpha_i$. A platoon of $N_i<N$ express buses would bunch together and serve this specific origin bus stop $\alpha_i$ with an effective boarding/alighting rate of $N_il$. When $N$ is partitioned into these $M_O$ disjoint subsets, we require that $\displaystyle\sum_{i=1}^{M_O}{N_i}=N$.

In other words, express buses are ``express'' only in the sense that they completely skip other origin bus stops. They must still travel the same loop and let people alight at their desired destinations.
\end{enumerate}

The notion of express buses will be extended later in Section \ref{expressdiscussion}, where an express bus can serve more than one bus stop. Without loss of generality, we find it instructive to first build the theory where an express bus only serves one bus stop due to its clarity and simplicity.

\subsection{Normal buses}

Let us now consider a system of normal buses where these $N$ buses bunch into a single unit with effective boarding/alighting rate of $Nl$. Suppose the durations they stop at the origin bus stops are $\tau_i$ where $i=1,\cdots,M_O$, and the durations they stop at the destination bus stops are $\tau_j$ where $j=1,\cdots,M_D$. Hence, the total time taken to complete one loop is
\begin{align}\label{Tbar}
\bar{T}=T+\sum_{i=1}^{M_O}{\tau_i}+\sum_{j=1}^{M_D}{\tau_j}.
\end{align}

With this, note that the total number of people that this single platoon of buses has to pick up at any origin bus stop $\alpha_i$ is $s_i\bar{T}$. These many people are picked up over the duration $\tau_i$ when they stop there, with boarding rate of $Nl$, giving us the following $M_O$ origin equations:
\begin{align}\label{origin}
s_i\bar{T}=Nl\tau_i,
\end{align}
where $i=1,\cdots,M_O$.

At any destination bus stop $\beta_j$, the duration $\tau_j$ this platoon stops there is to let passengers alight. The number of people from $\alpha_i$ who want to alight at $\beta_j$ is $\zeta_{ij}s_i\bar{T}$, so the total number of people who want to alight at $\beta_j$ is the sum over all $i$ from $1$ to $M_O$, giving us the following $M_D$ destination equations:
\begin{align}\label{destination}
\sum_{i=1}^{M_O}{\zeta_{ij}s_i\bar{T}}=Nl\tau_{j},
\end{align}
where $j=1,\cdots,M_D$.

Observe that if we sum the origin equations in Eq.\ (\ref{origin}) over all $i$, the left-hand side turns out to be the same as the left-hand side of summing the destination equations in Eq.\ (\ref{destination}) over all $j$ (where we use Eq.\ (\ref{sum}) to simplify the sum over $j$). Therefore, we arrive at this useful relationship between the total time spent stopping at all origin bus stops and the total time spent stopping at all destination bus stops:
\begin{align}\label{useful}
\sum_{i=1}^{M_O}{\tau_i}=\sum_{j=1}^{M_D}{\tau_j}.
\end{align}
Eq.\ (\ref{useful}) is a highly useful relation that simplifies the subsequent calculations. For instance, together with Eq. (\ref{Tbar}) for $\bar{T}$, the origin equations Eq. (\ref{origin}) become
\begin{align}\label{originNew}
k_i\left(T+2\sum_{i=1}^{M_O}{\tau_i}\right)=N\tau_i,
\end{align}
and the destination equations Eq.\ (\ref{destination}) become
\begin{align}\label{destinationNew}
\left(\sum_{i=1}^{M_O}{\zeta_{ij}k_i}\right)\left(T+2\sum_{j=1}^{M_D}{\tau_j}\right)=N\tau_j.
\end{align}
Here, we have used $k_i:=s_i/l$ where $i=1,\cdots,M_O$. By using the relationship in Eq.\ (\ref{useful}), we have decoupled Eqs.\ (\ref{origin})-(\ref{destination}) into equations exclusively dependent on origins $\tau_i$ in Eq.\ (\ref{originNew}) and equations exclusively dependent on destinations $\tau_j$ in Eq.\ (\ref{destinationNew}).

The solution to the origin equations Eq.\ (\ref{originNew}) is:
\begin{align}
\tau_i=\frac{k_iT}{N-2K},
\end{align}
for $i=1,\cdots,M_O$, where $\displaystyle K:=\sum_{i=1}^{M_O}{k_i}$. This imposes a constraint on how strong the $k_i$ can be, namely that the total number of buses $N$ must be greater than twice the sum of all these $k_i$, otherwise $\tau_i$ becomes negative. Physically, it means that there must be enough buses to serve an extreme demand for service. Otherwise, the system fails with commuters accumulating hopelessly at the bus stops.

The solution to the destination equations Eq.\ (\ref{destinationNew}) is:
\begin{align}
\tau_j=\frac{T}{N-2K}\sum_{i=1}^{M_O}{\zeta_{ij}k_i},
\end{align}
for $j=1,\cdots,M_D$. This again implies that constraint $N>2K$.

With the solution to how long the single platoon of $N$ buses stop at each origin and destination bus stop, we can calculate $W_i$, the average waiting time of commuters waiting for the bus(es) to arrive at bus stop $\alpha_i$. Since people are assumed to arrive uniformly at $\alpha_i$, then $W_i$ is just half the sum of the luckiest person (who has zero waiting time, since the arrival is just before the bus(es) leave(s)), and the unluckiest person when the bus(es) just leave(s) before returning one loop later (which is $\bar{T}-\tau_i$). Hence,
\begin{align}
W_i&=\frac{1}{2}\left((0)+(\bar{T}-\tau_i)\right)\\
&=\frac{1}{2}\left(T-\tau_i+2\sum_{l=1}^{M_O}{\tau_l}\right)\\
&=\frac{1}{2}T\left(\frac{N-k_i}{N-2K}\right).\label{Wi}
\end{align}
Note that the useful relation in Eq.\ (\ref{useful}) enables the elimination of the sum of $\tau_j$ over $j$ in place of the sum of $\tau_i$ over $i$. This turns out to eliminate all traces of $\zeta_{ij}$, i.e. we do not actually need to care where people want to go since all that matters in the average waiting time is the \emph{sum of all incurred durations to deliver them to some places}.

Thus, the overall average waiting time for the system is the average of $W_i$ at each $\alpha_i$ weighted by the number of people arriving there, or equivalently its $k_i$:
\begin{align}
W&=\frac{1}{K}\sum_{i=1}^{M_O}{k_iW_i}\\
&=\frac{T}{2K(N-2K)}\left(KN-\sum_{i=1}^{M_O}{k_i^2}\right).
\end{align}

In the symmetric case where $k_i=k$ for all $i=1,\cdots,M_O$, we have
\begin{align}
W&=\frac{1}{M_O}\sum_{i=1}^{M_O}{\frac{1}{2}T\left(\frac{N-k}{N-2M_Ok}\right)}\\
&=\frac{1}{2}T\left(\frac{N-k}{N-2M_Ok}\right).\label{normalsymmetric}
\end{align}

\subsection{Express buses}

Suppose $N\geq M_O$, and $N_i$ buses serve $\alpha_i$ where $\displaystyle\sum_{i=1}^{M_O}{N_i}=N$, sending commuters from $\alpha_i$ to every destination bus stop. Each disjoint subset of $N_i$ express buses forms a platoon of bunched buses, which is equivalent to the system of normal buses with only one origin bus stop $M_O=1$, i.e. $k_l=0$ if $l\neq i$ and $k_l=k_i$ if $l=i$ (recall Fig.\ \ref{fig13}(b)). Therefore, the average waiting time at $\alpha_i$ for this platoon of $N_i$ buses to arrive can be directly obtained from Eq.\ (\ref{Wi}) to yield:
\begin{align}
W_i=\frac{1}{2}T\left(\frac{N_i-k_i}{N_i-2k_i}\right).
\end{align}
The overall average waiting time over every origin bus stop $\alpha_i$ which is served by its respective platoon of express buses $N_i$, for all $i=1,\cdots,M_O$ is thus:
\begin{align}
W=\frac{1}{K}\sum_{i=1}^{M_O}{k_iW_i}.
\end{align}

In the case where $k_i=k$ for all $i=1,\cdots,M_O$,
\begin{align}\label{expresssamek}
W=\frac{1}{M_O}\sum_{i=1}^{M_O}{\frac{1}{2}T\left(\frac{N_i-k}{N_i-2k}\right)}.
\end{align}
Furthermore if $N_i=N/M_O$ for all $i=1,\cdots,M_O$, we have
\begin{align}\label{expressfullysymmetric}
W=\frac{1}{2}T\left(\frac{N-M_Ok}{N-2M_Ok}\right).
\end{align}

\subsection{Discussion on analytical results for non-interacting express buses}\label{expressdiscussion}

\begin{enumerate}
\item The overall average waiting time $W$ of commuters for a bus to arrive at the bus stop generally depends on the time it takes for a bus to complete the loop $T$, the rates of people arrival at the bus stops per rate of loading/unloading $k_i$, the number of buses serving each origin bus stop $N_i$, and the number of origin bus stops $M_O$. It does \emph{not} depend on the number of destination bus stops $M_D$, nor the probability distribution of the origin-destination of the commuters $\zeta_{ij}$. Furthermore, the locations of the bus stops are arbitrary. The origin bus stops could alternate with destination bus stops, or we could have a stretch of origin bus stops followed by destination bus stops, etc.

The intuition for why $M_D$ and $\zeta_{ij}$ do not show up in any formula for $W_i$ is that what matters is how many people to pick up ($k_i, M_O$) by how many available buses ($N_i$) and how fast they travel ($T$). For this number of people, it does not matter where they want to go. The total time for all of them to eventually alight is the same --- regardless of where they actually alight. This is manifested by Eq.\ (\ref{useful}).

\item In the \emph{fully symmetric case}, where each origin bus stop has the same rate of people arrival per rate of loading/unloading $k$ and for the express buses setup all bus stops are served by the same number of express buses $N/M_O$, then Eqs.\ (\ref{normalsymmetric}) and (\ref{expressfullysymmetric}) imply that having express buses where each platoon of $N_i=N/M_O$ buses serves one distinct origin bus stop is better (i.e. lower $W$) than having a single normal platoon of $N$ buses bunching together and serving all bus stops.

\item If $N$ is not a multiple of $M_O$, then some origin bus stops would have additional express bus(es). As an explicit example, consider the case where there are $M_O=4$ origin bus stops and $N=5$ buses. Let the first four express buses serve one of each origin bus stops. If all people arrival rates are the same, then from Eq.\ (\ref{expresssamek}), it does not matter which of the four origin bus stops that the fifth express bus serves (or which of the four origin bus stops that has two express buses serving it, instead of just one --- Fig.\ \ref{fig13}(b)).

\item We can generalise the notion of express buses to serve not just one particular origin bus stop $\alpha_i$, but a fixed subset of origin bus stops. In this case, we no longer need the condition that $N\geq M_O$. If we partition the number of origin bus stops into $P$ disjoint subsets $\Omega_i$ and also partition the $N$ buses into $P$ disjoint subsets $N_i$ where $i=1,\cdots,P$ (so we assume here that $N\geq P$ and $M_O\geq P$), then we have a situation where there are effectively $P$ subsets of origin bus stops each served exclusively by its own dedicated platoon of $N_i$ express buses. For each subset $\Omega_i$, these $N_i$ buses are like normal buses serving each origin bus stop in $\Omega_i$. Hence from Eq.\ (\ref{Wi}), the average waiting time at bus stop $\gamma$ in this subset $\Omega_i$ is
\begin{align}\label{subset}
W_\gamma=\frac{1}{2}T\left(\frac{N_i-k_\gamma}{N_i-2K}\right),
\end{align}
where here $K$ is the sum of the people arrival rates per rate of loading/unloading for each origin bus stop being served in this subset $\Omega_i$. The overall average waiting time for the entire system is again, the weighted average of these $W_\gamma$. We use the index $\gamma$ here to denote a bus stop within the subset $\Omega_i$, where $i$ labels the partitioning into $P$ disjoint subsets.

With this, we have a general theory of any $N$ express buses serving a loop of $M_O+M_D$ bus stops.

\begin{figure}
\centering
\includegraphics[width=16cm]{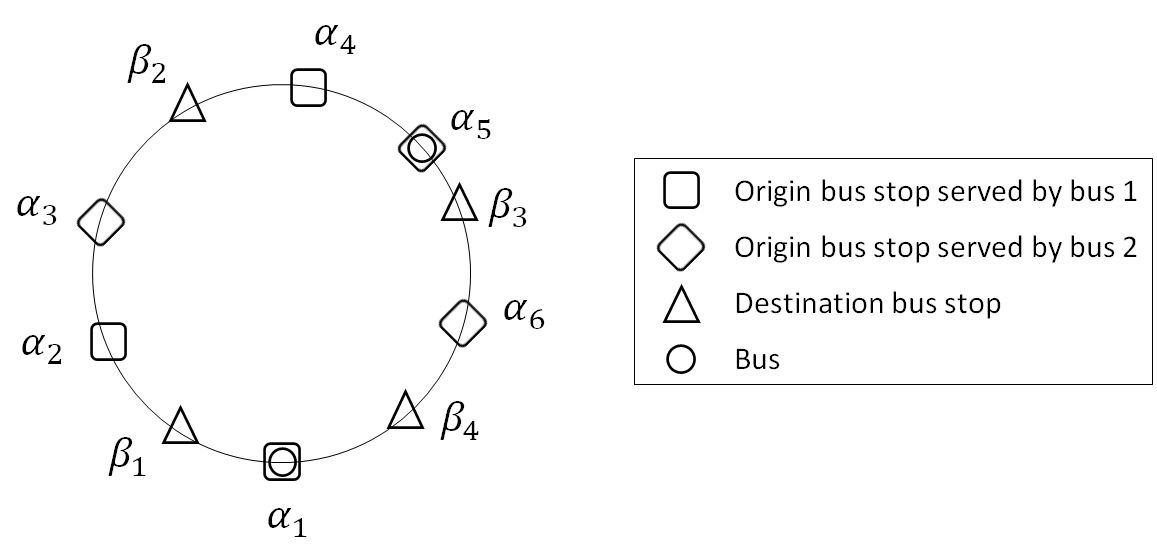}
\caption{$N=2$ express buses serving $M_O=6$ origin bus stops and $M_D=4$ destination bus stops in a loop. All these $M_O+M_D$ bus stops are arbitrarily located along the loop. This is a generalisation where different express buses serve disjoint subsets of origin bus stops (which could comprise more than 1 origin bus stop). On top of that, each disjoint subset of origin bus stops can be served by more than one express bus. Shown in this figure is a \emph{symmetric} partitioning of origin bus stops, where each express bus serves $m=M_O/N=6/2=3$ origin bus stops. Furthermore, if all origin bus stops have the same people arrival rate, then this system is \emph{fully symmetric} and the average waiting time $W$ is given by Eq.\ (\ref{Wmorefullysymmetric}).}
\label{fig14}
\end{figure}

\item As a special case of the generalisation in the previous point, suppose $N\leq M_O$ and $M_O$ is a multiple of $N$, i.e. $mN=M_O$ where $m$ is a positive integer (see Fig.\ \ref{fig14} for an example). Let each bus serve $m$ specific origin bus stops, where different buses do not share any common origin bus stop. Furthermore, let each $k_i=k$, so that we have a fully symmetric setup. Then using Eq.\ (\ref{subset}), the average waiting time for each origin bus stop (which are all identical) is
\begin{align}
W_\gamma&=\frac{1}{2}T\left(\frac{1-k}{1-2mk}\right)\\
&=\frac{1}{2}T\left(\frac{N-Nk}{N-2M_Ok}\right),
\end{align}
with the overall average waiting time being just $W=W_\gamma$ for this fully symmetric setup,
\begin{align}\label{Wmorefullysymmetric}
W&=\frac{1}{2}T\left(\frac{N-Nk}{N-2M_Ok}\right).
\end{align}
By comparing with the system of normal buses given by Eq.\ (\ref{normalsymmetric}), we see that express buses reduce the overall average waiting time. The reduction is greater with more buses $N$ serving the loop. This fully symmetric $N\leq M_O$ setup corresponds to that for $N\geq M_O$ where in Eq.\ (\ref{expressfullysymmetric}), the reduction in overall average waiting time is enhanced with more origin bus stops $M_O$.


\item In general, a bus stop has people who want to board from, and people who want to alight to. A bus stop is therefore \emph{both an origin and a destination}. If we assume that alighting occurs before boarding (i.e. these processes are sequential), then this bus stop first behaves as a destination, and then behaves as an origin. In other words, given $M$ general bus stops, this is equivalent to $M$ origin bus stops with $M$ destination bus stops. Since the locations of the bus stops are arbitrary, a general bus stop comprises an origin bus stop as well as a destination bus stop located ``infinitesimally close by''.

With this, we have a general theory of $N$ express buses serving a loop of $M$ bus stops.
\end{enumerate}

\section{Exact \texorpdfstring{$AB$}{AB} system}\label{appenB}

\begin{figure}
\centering
\includegraphics[width=13cm]{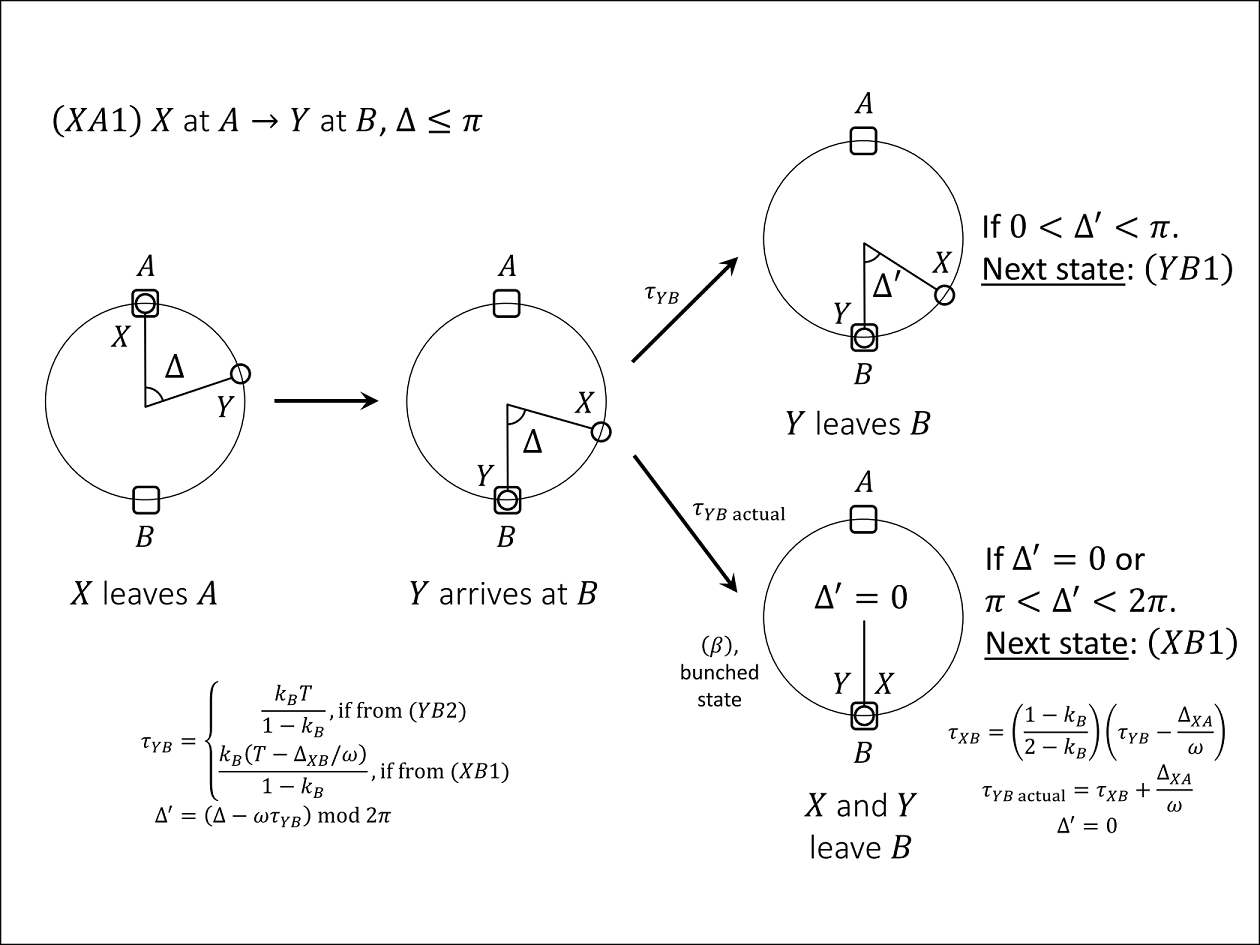}
\caption{After $X$ leaves $A$ with $\Delta_{XA}\leq\pi$, the next possible states are $XB1$ or $YB1$.}\label{fig15}
\end{figure}

\begin{figure}
\includegraphics[width=13cm]{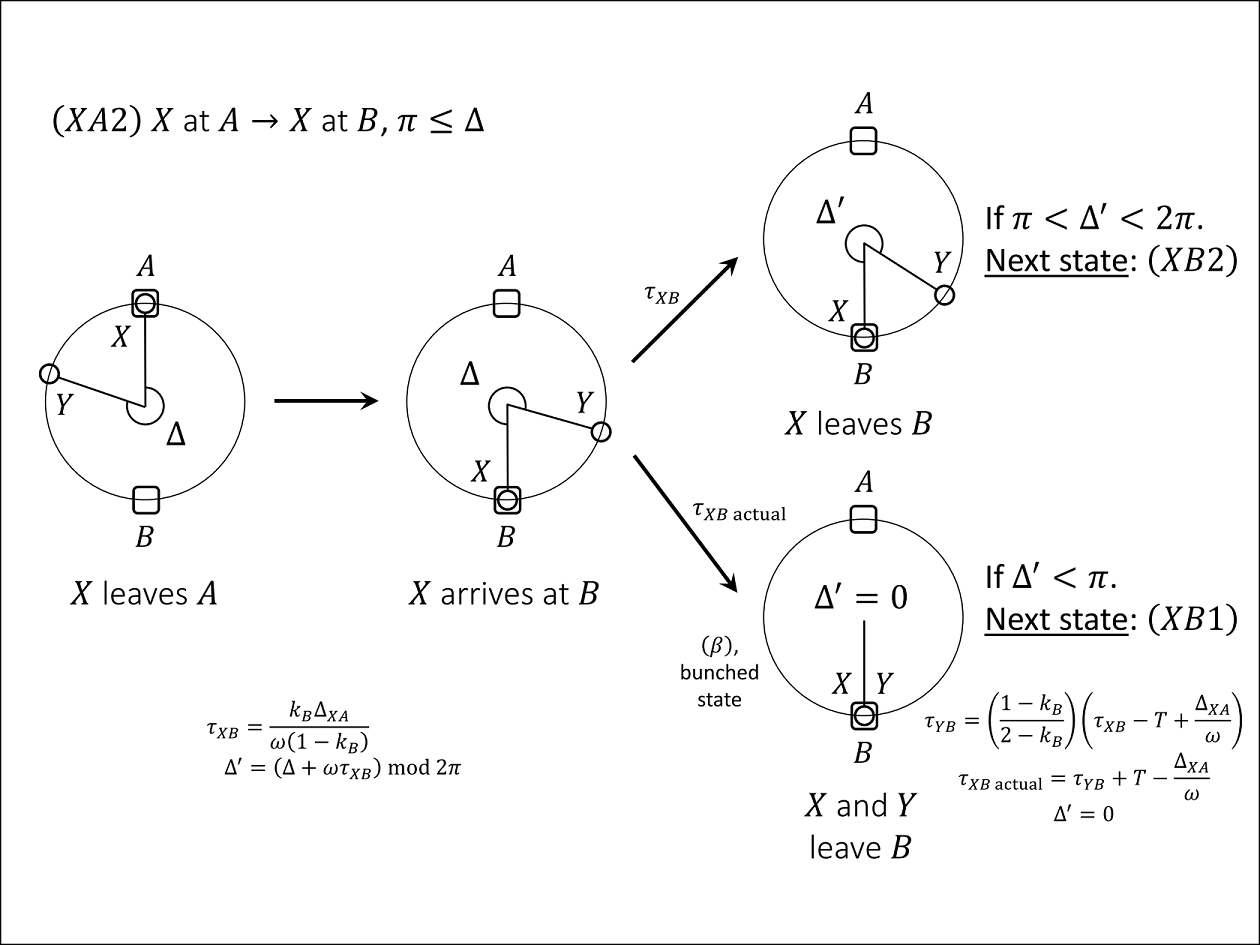}
\caption{After $X$ leaves $A$ with $\Delta_{XA}\geq\pi$, the next possible states are $XB1$ or $XB2$.}\label{fig16}
\end{figure}

In deriving the transition rules between one state to its next state, we enumerate what its possible next states are. For example in Fig.\ \ref{fig15}, $X$ has just left $A$ with the phase difference $\Delta$ (or in this case $\Delta_{XA}$) being less than or equal to $\pi$. This therefore defines the state $XA1$. If $\Delta\geq\pi$, then the state of the bus system is $XA2$, whose next state is described by Fig.\ \ref{fig16}.

It is important to distinguish between $\Delta\leq\pi$ and $\Delta\geq\pi$ because the next possible states are different, as shown in Figs.\ \ref{fig15}-\ref{fig16}. This distinction arises due to the two bus stops being separated by $\pi$ such that when $X$ leaves $A$, then the next event is $Y$ arriving at $B$ if $\Delta\leq\pi$, otherwise it is $X$ arriving at $B$ if $\Delta\geq\pi$. As $X$ would stop at two bus stops and $Y$ would stop at one bus stop, there are three combinations of a bus leaving a bus stop, viz. $XA,XB,YB$. Since each combination has two distinct states corresponding to $\Delta\leq\pi$ or $\Delta\geq\pi$, there is a total of six states.

\begin{figure}
\includegraphics[width=13cm]{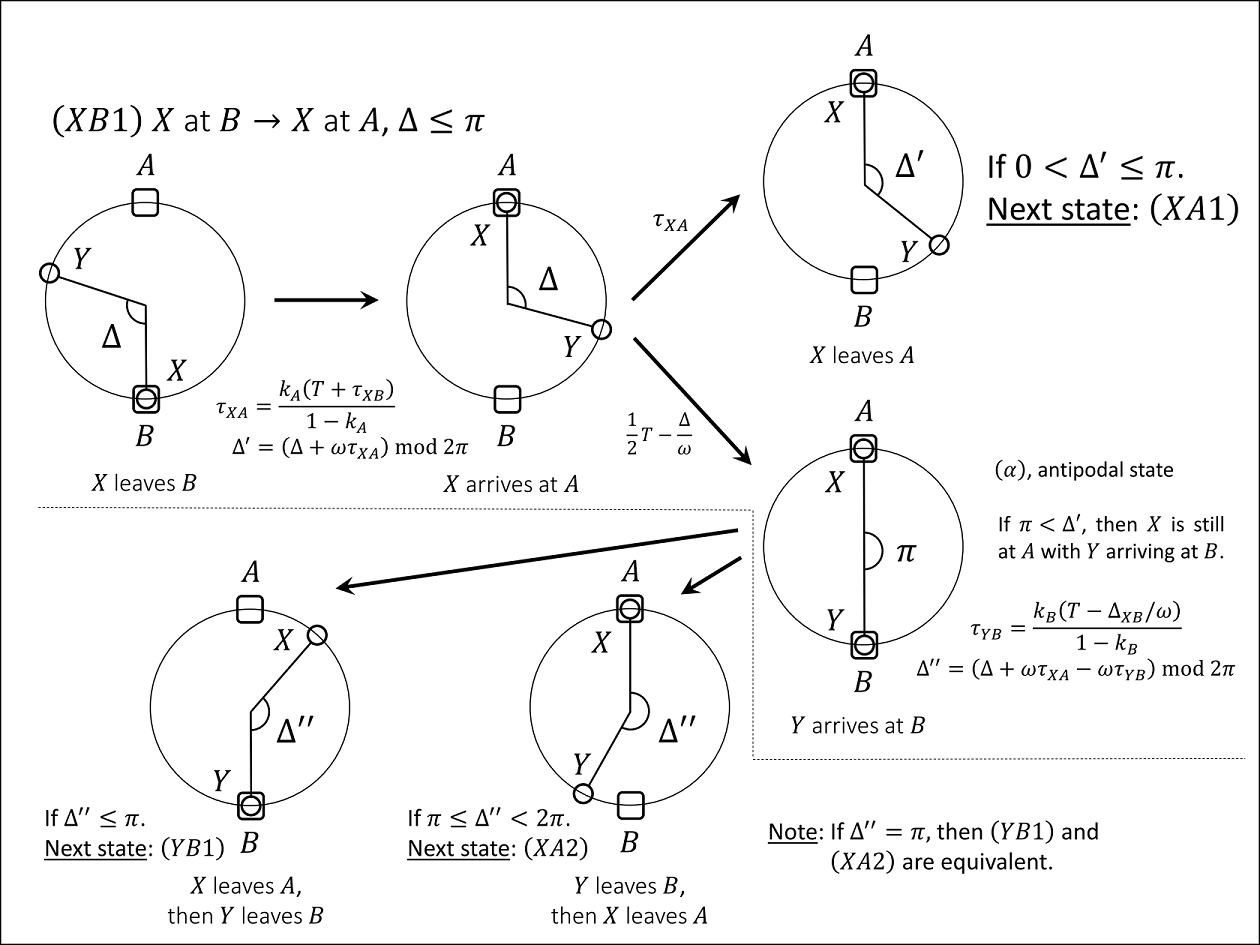}
\caption{After $X$ leaves $B$ with $\Delta_{XB}\leq\pi$, the next possible states are $XA1$, $XA2$ or $YB1$.}\label{fig17}
\end{figure}

\begin{figure}
\includegraphics[width=13cm]{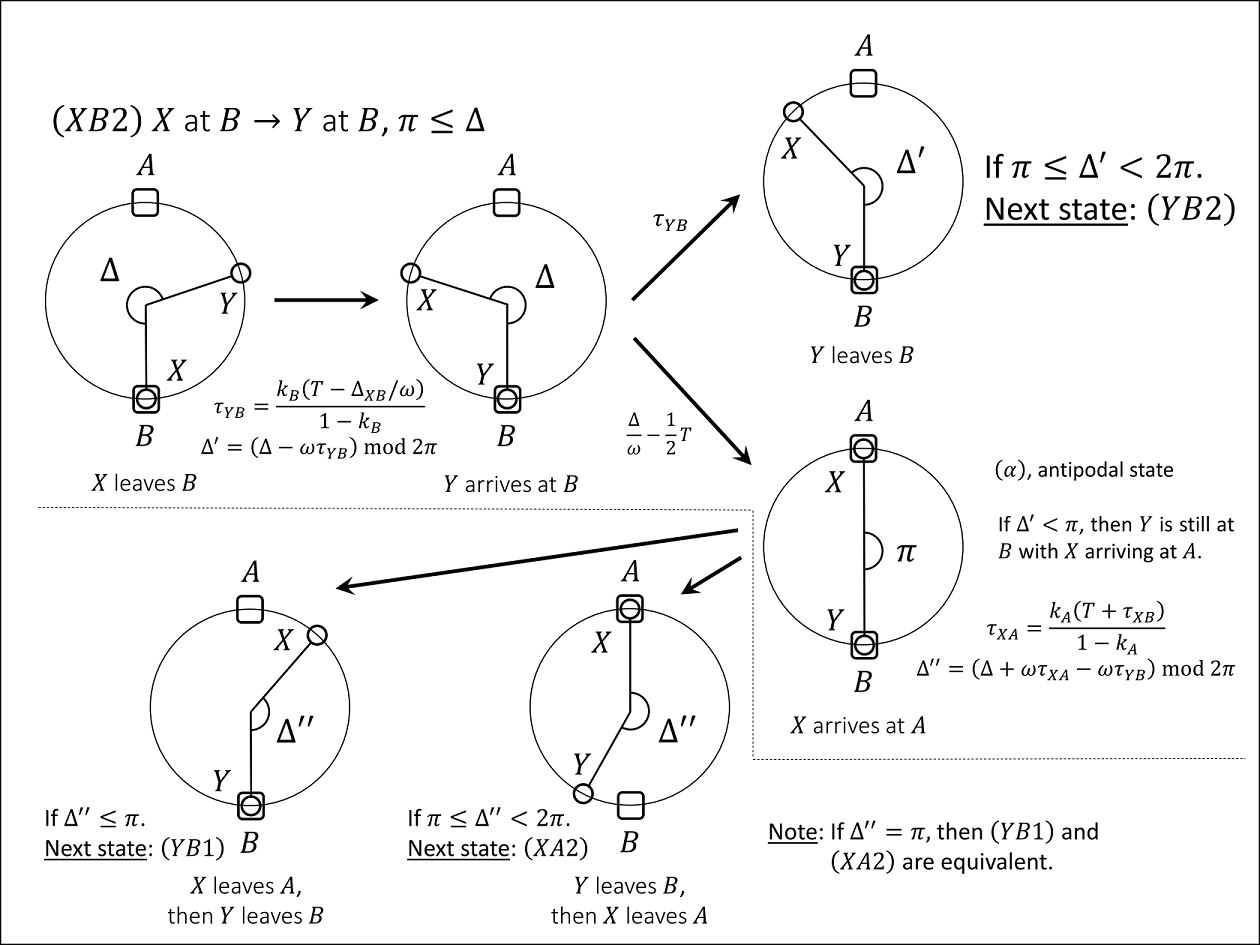}
\caption{After $X$ leaves $B$ with $\Delta_{XB}\geq\pi$, the next possible states are $XA2$, $YB1$ or $YB2$.}\label{fig18}
\end{figure}

\begin{figure}
\includegraphics[width=13cm]{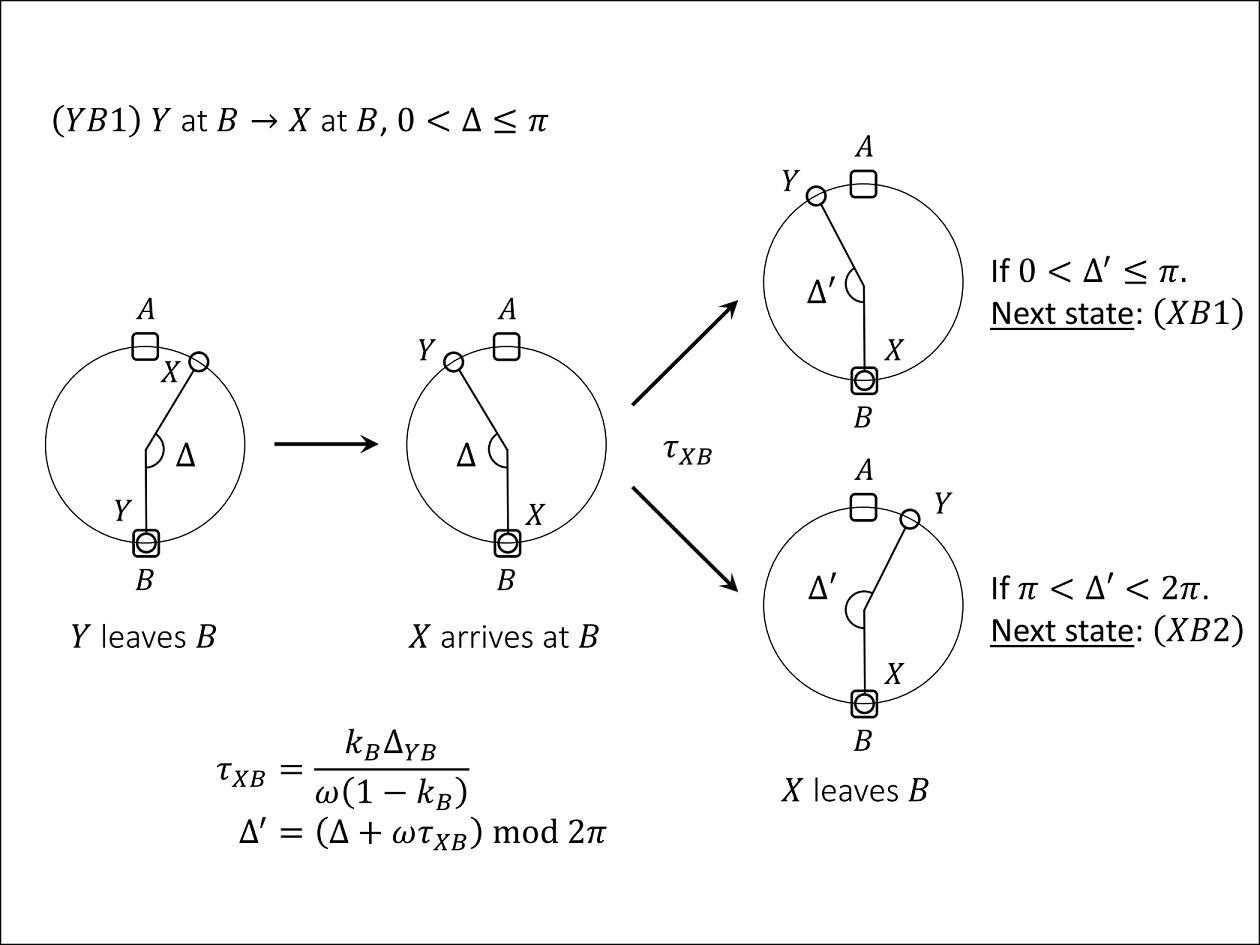}
\caption{After $Y$ leaves $B$ with $0<\Delta_{YB}\leq\pi$, the next possible states are $XB1$ or $XB2$.}\label{fig19}
\end{figure}

\begin{figure}
\includegraphics[width=13cm]{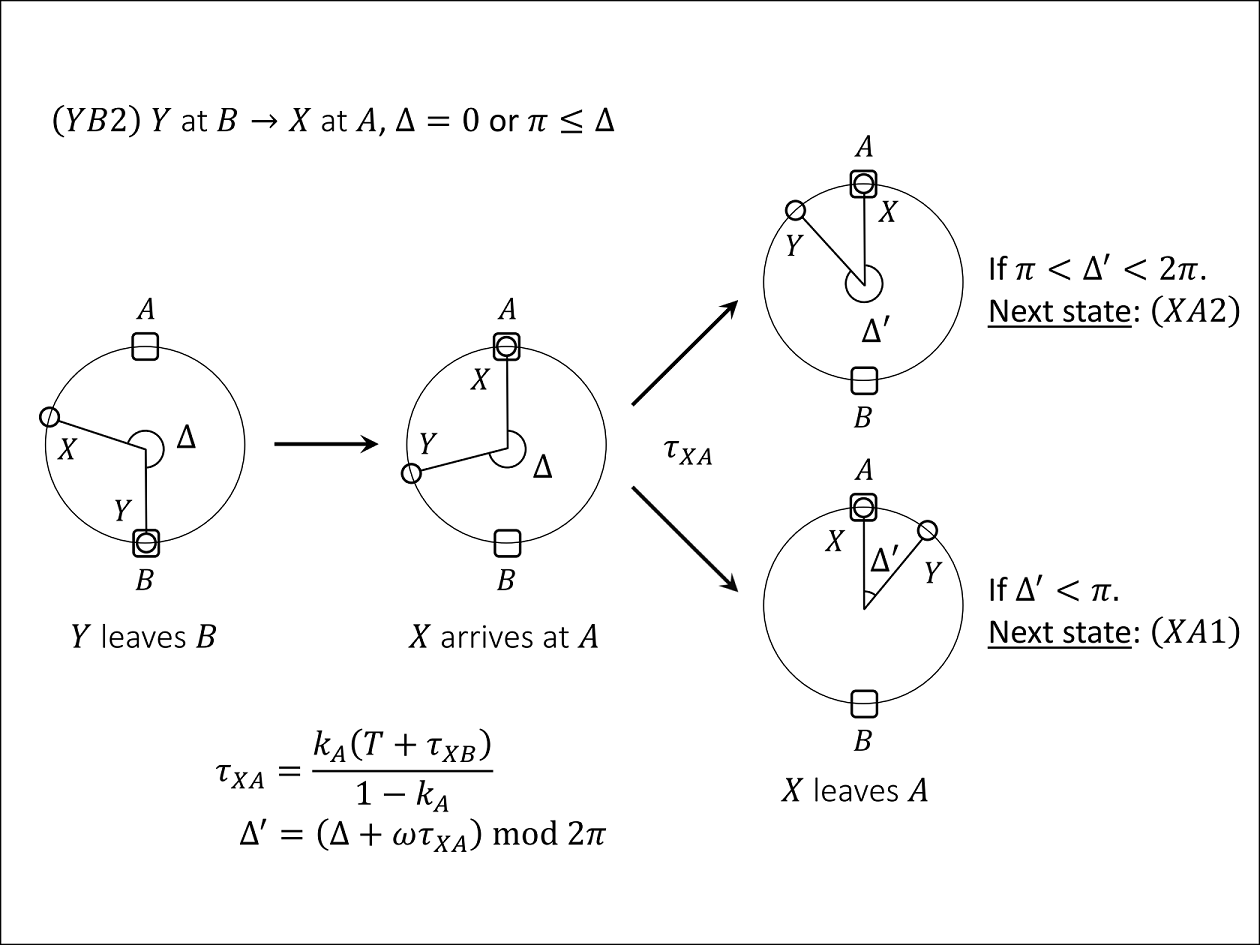}
\caption{After $Y$ leaves $B$ with $\Delta_{YB}=0$ or $\Delta_{YB}\geq\pi$, the next possible states are $XA1$ or $XA2$.}\label{fig20}
\end{figure}

In the next event after a bus leaves a bus stop, a bus will be arriving at a bus stop and thus we need to calculate the time it spends stopping there. This is a straightforward calculation, depending on when the last bus left this bus stop which would have reset the number of people to zero. Figs.\ \ref{fig15}-\ref{fig20} show the results for $\tau_i$, based on when the last bus left that bus stop to determine the number of people to pick up and hence $\tau_i$. Sometimes, the expression for $\tau_i$ at a bus stop depends on which previous states it came from since it could have been $X$ or $Y$ which last left that bus stop (for example, $\tau_{YB}$ in $XA1$ in Fig.\ \ref{fig15}).

Once this bus has finished picking up everybody, it leaves. This gives a new phase difference $\Delta'$ which is obtained from the previous phase difference $\Delta$ plus (if it is $X$ who is stopping) or minus (if it is $Y$ who is stopping) $\omega$ times $\tau_i$. The value of $\Delta'$ lies within $[0,2\pi)$ since it is modulo $2\pi$. We summarise this algorithm as follows:

\begin{enumerate}
\item Given some state (e.g. $XA1$). The next event where a bus arrives at a bus stop is definite. (E.g. if the state is $XA1$, then definitely what happens next is $Y$ arrives at $B$ as shown in Fig.\ \ref{fig15}.)
\item Calculate how long this bus spends stopping at the bus stop, $\tau_i$. To do so, we need to know the number of people to pick up. This requires the knowledge of when a bus last left this bus stop, which depends on the particular history that leads to this state. We exhaustively enumerate all possibilities. Sometimes there is more than one possible expression for $\tau_i$ (e.g. $\tau_{YB}$ in $XA1$ as shown in Fig.\ \ref{fig15}), depending on the historical path along the transition graph shown in Fig.\ \ref{fig1}.
\item Whilst this bus is stopping at the bus stop, the other bus keeps moving on the road. If $\tau_i$ is not too long, then this bus leaves the bus stop before the other bus arrives at a bus stop. Therefore, this defines the next state and we are done.

However, if $\tau_i$ is too long such that the other bus arrives at some bus stop, then the next state is different:

($\beta$) If the other bus arrives at the same bus stop, then they are bunched (e.g. from $XA1$ or $XA2$ in Figs.\ \ref{fig15}-\ref{fig16}). The overall $\tau_i$ is recalculated to account for the fact that bunched buses share loading. The new phase difference $\Delta'$ becomes zero, and we have a next state.

($\alpha$) If the other bus arrives at a different bus stop, then we need to calculate how long this bus spends stopping at this bus stop (e.g. from $XB1$ or $XB2$ in Figs.\ \ref{fig17}-\ref{fig18}). Again, this calculation requires knowing the historical path to find out when a bus last left this bus stop. With this, we can determine which of these two buses first leaves its respective bus stop and thus define the next state.
\end{enumerate}
Note that for states $YB1$ and $YB2$ (Figs.\ \ref{fig19}-\ref{fig20}), since $Y$ does not pick up people from $A$, it traverses this bus stop without stopping, leading to two possible next states.

In the case where the two buses bunch at $B$ which may happen from $XA1$ or $XA2$, they share the loading of people. The calculations of $\tau_i$ take into account that the first bus picks up people at a loading rate $l$ up to the point where the second bus arrives. Then, they collectively pick up people at a rate of $2l$, and leave together with phase difference $\Delta'=0$. As an explicit example, we show how this is calculated for $XA1$ in Fig.\ \ref{fig15}. After $Y$ arrives at $B$, if $X$ does not also arrive at $B$ before $Y$ leaves, then $Y$ would have stopped at $B$ over a duration of $\tau_{YB}$. This implies that the total number of people $Y$ would have picked up is $l\tau_{YB}$, as it picks up people at a rate of $l$ people per second. Hence when $Y$ just arrives at $B$, there are only $l\tau_{YB}-s_B\tau_{YB}$ people there. Now if after a duration of $\Delta/\omega$, $Y$ is still at $B$ because there are still people there, this number of people is $(l-s_B)\tau_{YB}+(s_B-l)\Delta/\omega$. At this point, $X$ just arrives at $B$ and shares loading with $Y$ so that people are collectively boarded at a rate of $2l$ people per second. From then on, the two buses spend a further duration of $\tau_{XB}$ to clear the load, i.e.
\begin{gather}
(l-s_B)\tau_{YB}+(s_B-l)\frac{\Delta}{\omega}+(s_B-2l)\tau_{XB}=0\\
\tau_{XB}=\left(\frac{1-k_B}{2-k_B}\right)\left(\tau_{YB}-\frac{\Delta}{\omega}\right),
\end{gather}
where $k_B:=s_B/l$. Recall that $\tau_{YB}$ is the duration that $Y$ spends at $B$ to pick up everybody by itself if $X$ does not bunch with it. Since $X$ bunches with it and shares the load, the actual time that $Y$ stops would be first $\Delta/\omega$ where it picks up by itself, and then a further $\tau_{XB}$ when $X$ helps it out, i.e.
\begin{align}
\tau_{YB\textrm{ actual}}=\tau_{XB}+\frac{\Delta}{\omega}.
\end{align}
These results are summarised in Fig.\ \ref{fig15}.

Our consideration of the future states, given a present state, allows for one bus to be at a bus stop whilst the other bus possibly traverses one bus stop. We do not consider the situation where one bus is at a bus stop for too long such that the other bus possibly traverses two bus stops, because the number of possible future states would dramatically increase (and even blow up, when there is a third bus stop $C$ in the $A+B\rightarrow C$ system). This leads to an upper bound to the values of $k_A$, given some $k_B$. When $k_B=0.01$, we find that this consideration works for $k_A$ up till $0.3325$, which is more than enough to account for realistic demands for buses.


The results of these calculations for $\tau_i$ and new $\Delta'$ together with the possible next states are summarised in Figs.\ \ref{fig15}-\ref{fig20}. One can refer to these figures to evaluate the system and evolve it forward in time, deterministically, given some initial state. A simple computer programme can be written with conditional statements to determine the next state, given a current state. We have done this and presented the results in Fig.\ \ref{fig2} for $k_B=0.01$, $k_A\in[0,0.3325]$.


\section{Exact \texorpdfstring{$A+B\rightarrow C$}{A + B -> C} system}\label{appenC}

\begin{figure}
\centering
\includegraphics[width=13cm]{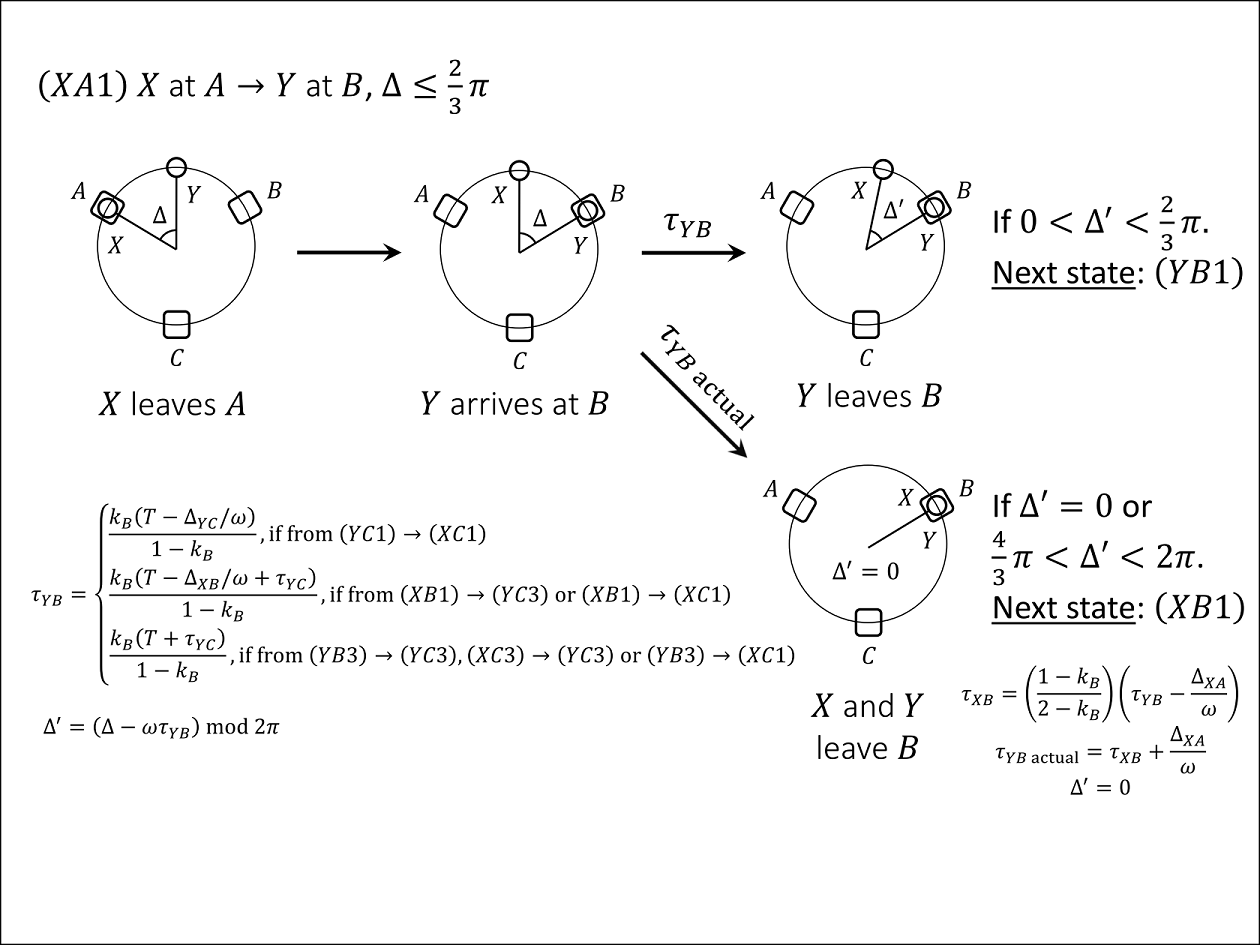}
\caption{After $X$ leaves $A$ with $\Delta_{XA}\leq2\pi/3$, the next possible states are $XB1$ or $YB1$.}\label{fig21}
\end{figure}
\begin{figure}
\includegraphics[width=13cm]{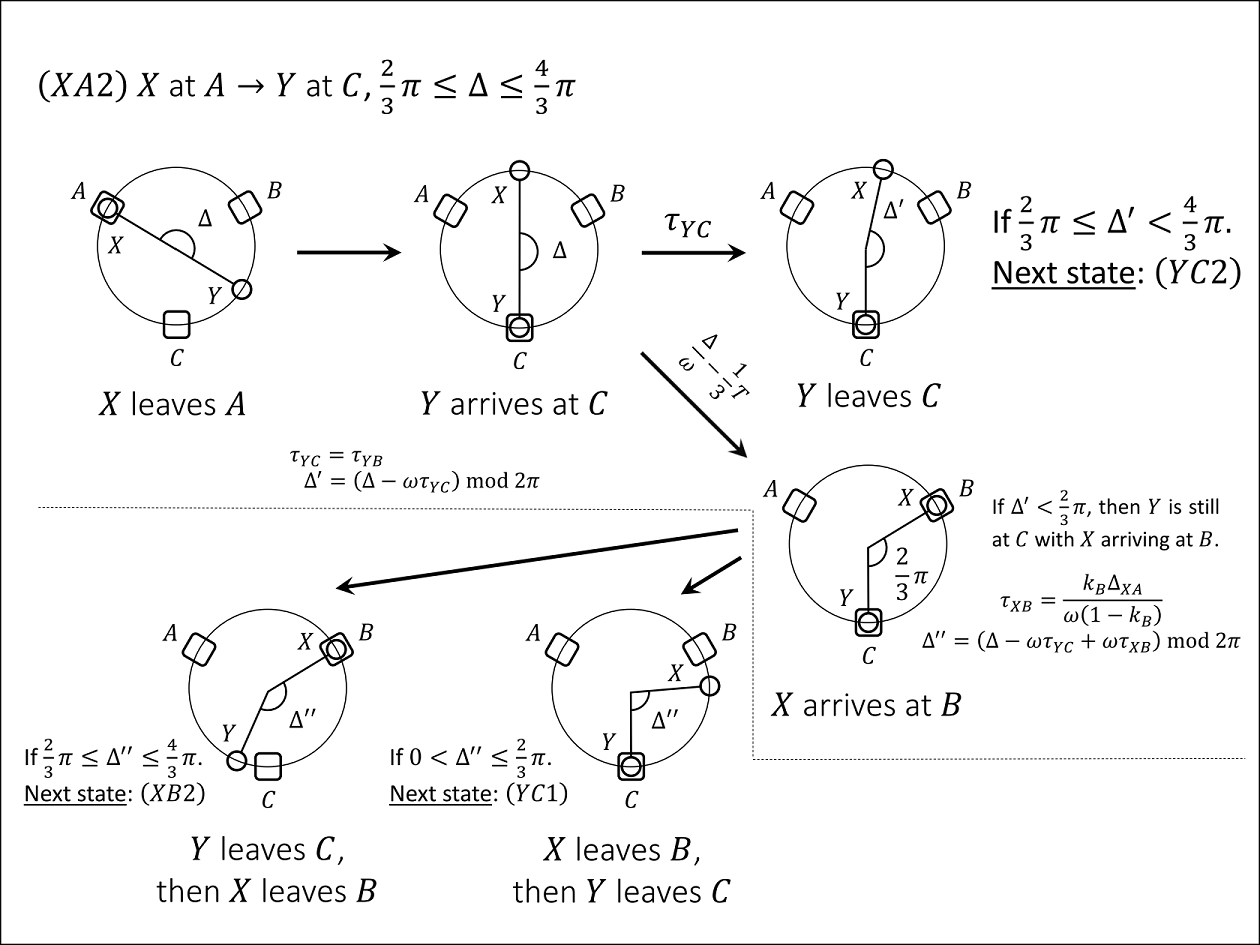}
\caption{After $X$ leaves $A$ with $2\pi/3\leq\Delta_{XA}\leq4\pi/3$, the next possible states are $XB2$, $YC1$ or $YC2$.}\label{fig22}
\end{figure}
\begin{figure}
\centering
\includegraphics[width=13cm]{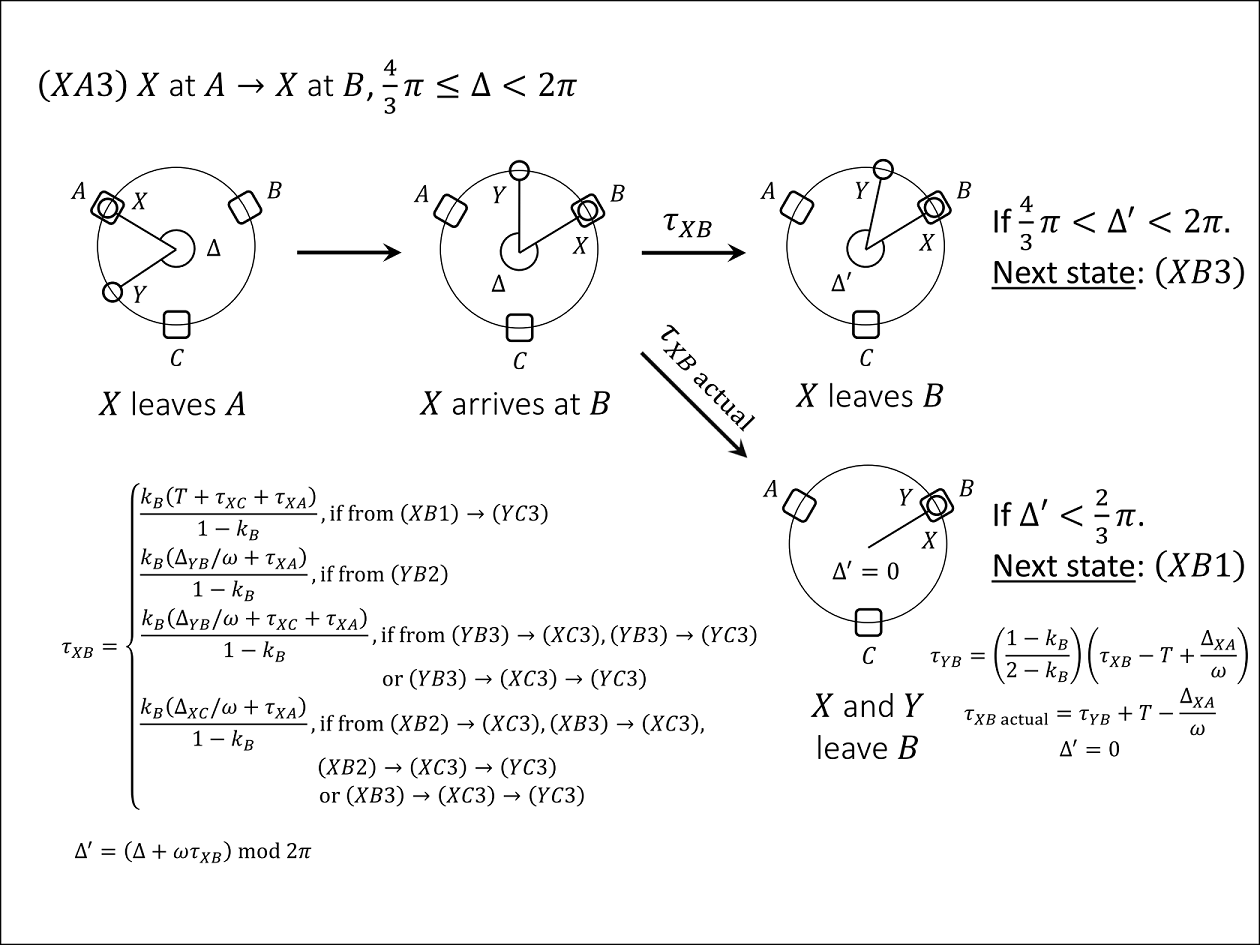}
\caption{After $X$ leaves $A$ with $4\pi/3\leq\Delta_{XA}$, the next possible states are $XB1$ or $XB3$.}\label{fig23}
\vskip 0.5cm
\includegraphics[width=13cm]{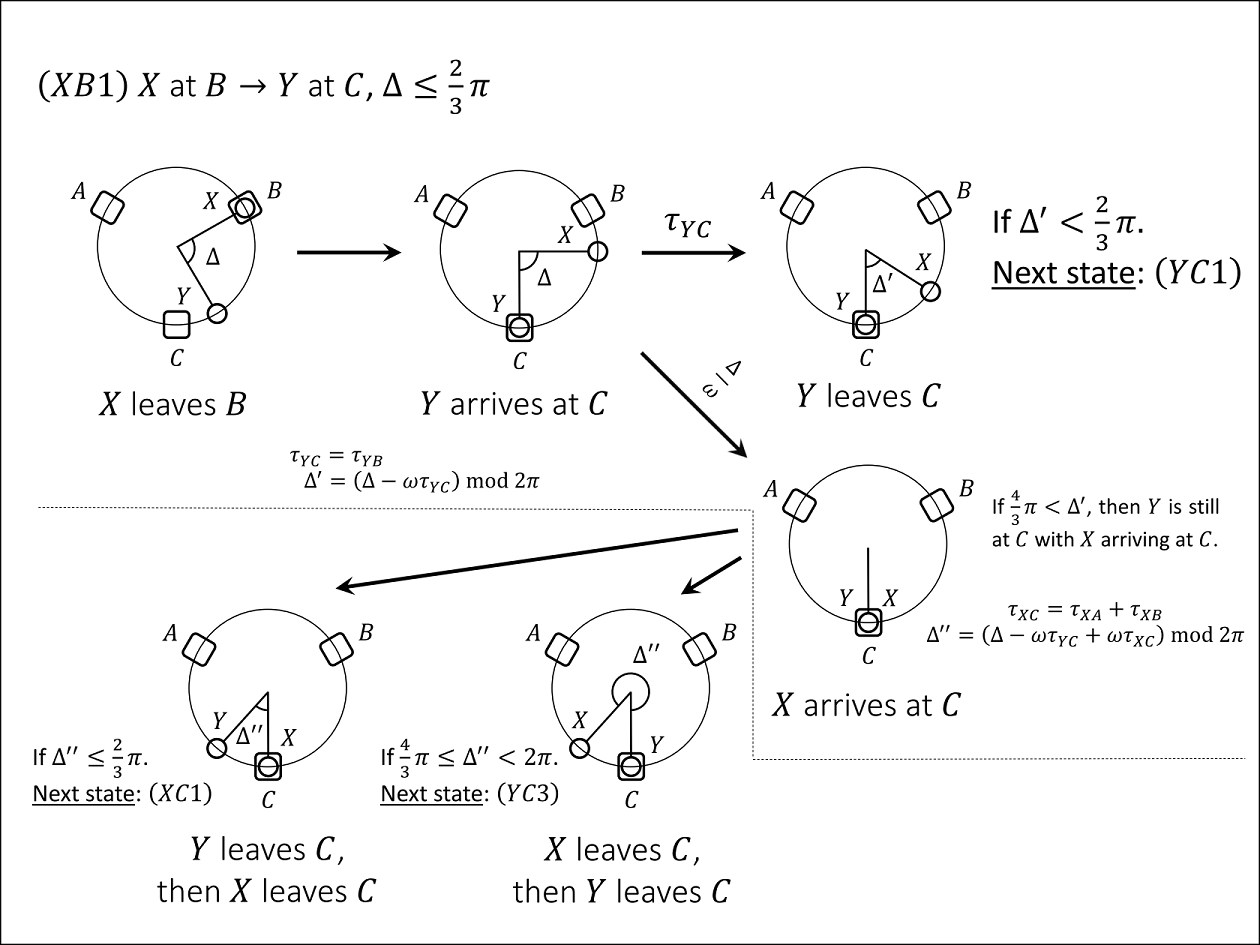}
\caption{After $X$ leaves $B$ with $\Delta_{XB}\leq2\pi/3$, the next possible states are $XC1$, $YC1$ or $YC3$.}\label{fig24}
\end{figure}
\begin{figure}
\centering
\includegraphics[width=13cm]{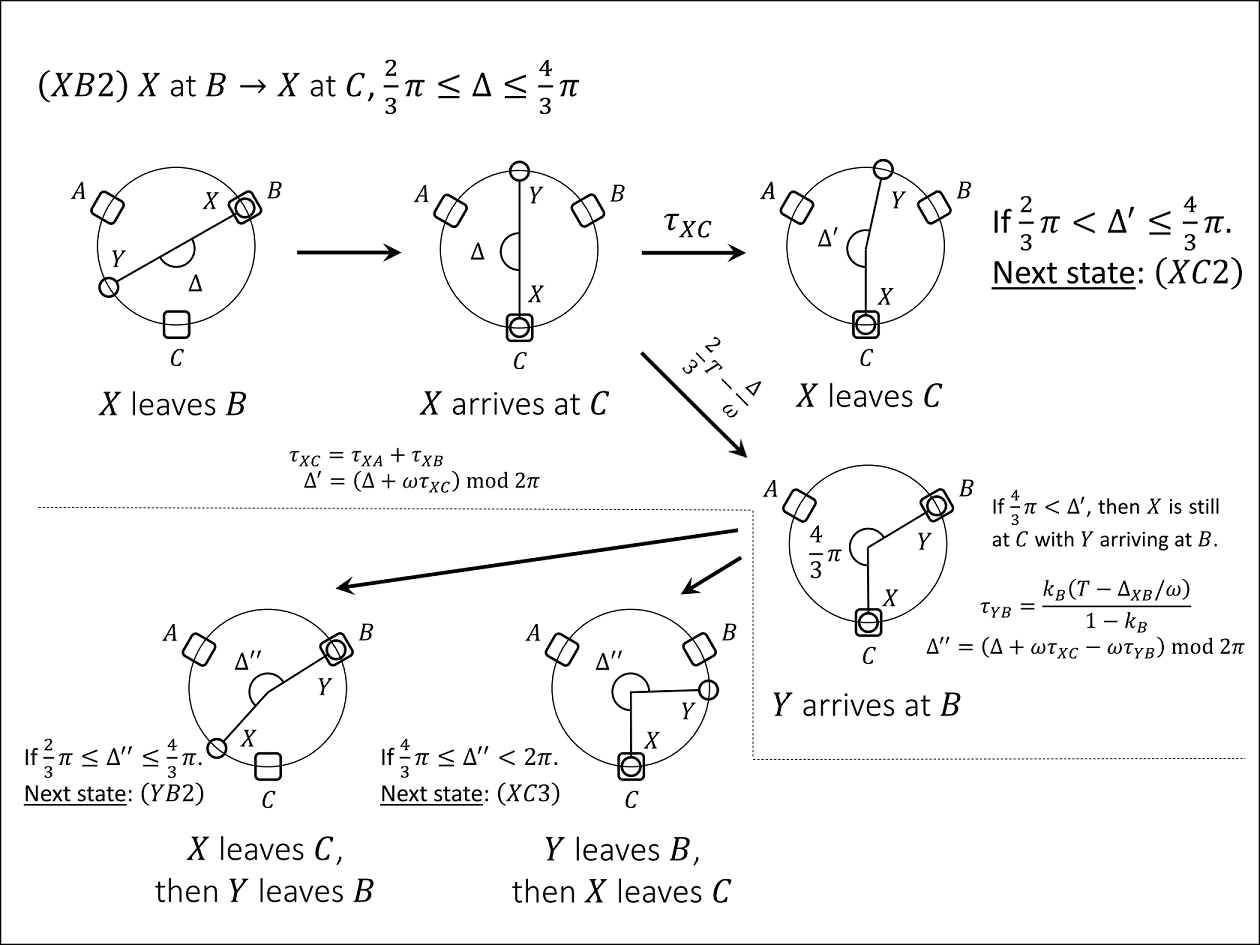}
\caption{After $X$ leaves $B$ with $2\pi/3\leq\Delta_{XB}\leq4\pi/3$, the next possible states are $XC2$, $XC3$ or $YB2$.}\label{fig25}
\end{figure}
\begin{figure}
\includegraphics[width=13cm]{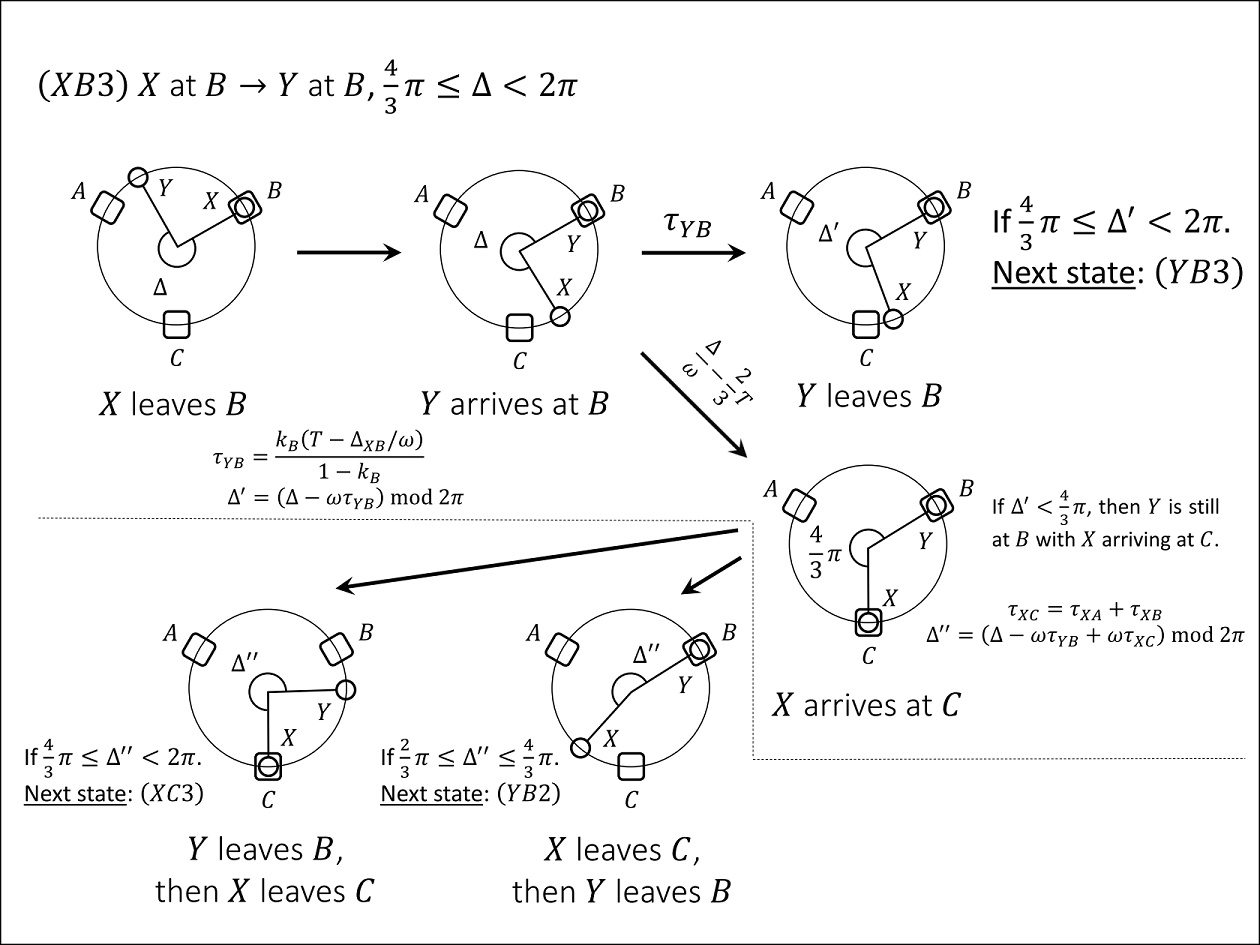}
\caption{After $X$ leaves $B$ with $4\pi/3\leq\Delta_{XB}$, the next possible states are $XC3$, $YB2$ or $YB3$.}\label{fig26}
\end{figure}
\begin{figure}
\centering
\includegraphics[width=13cm]{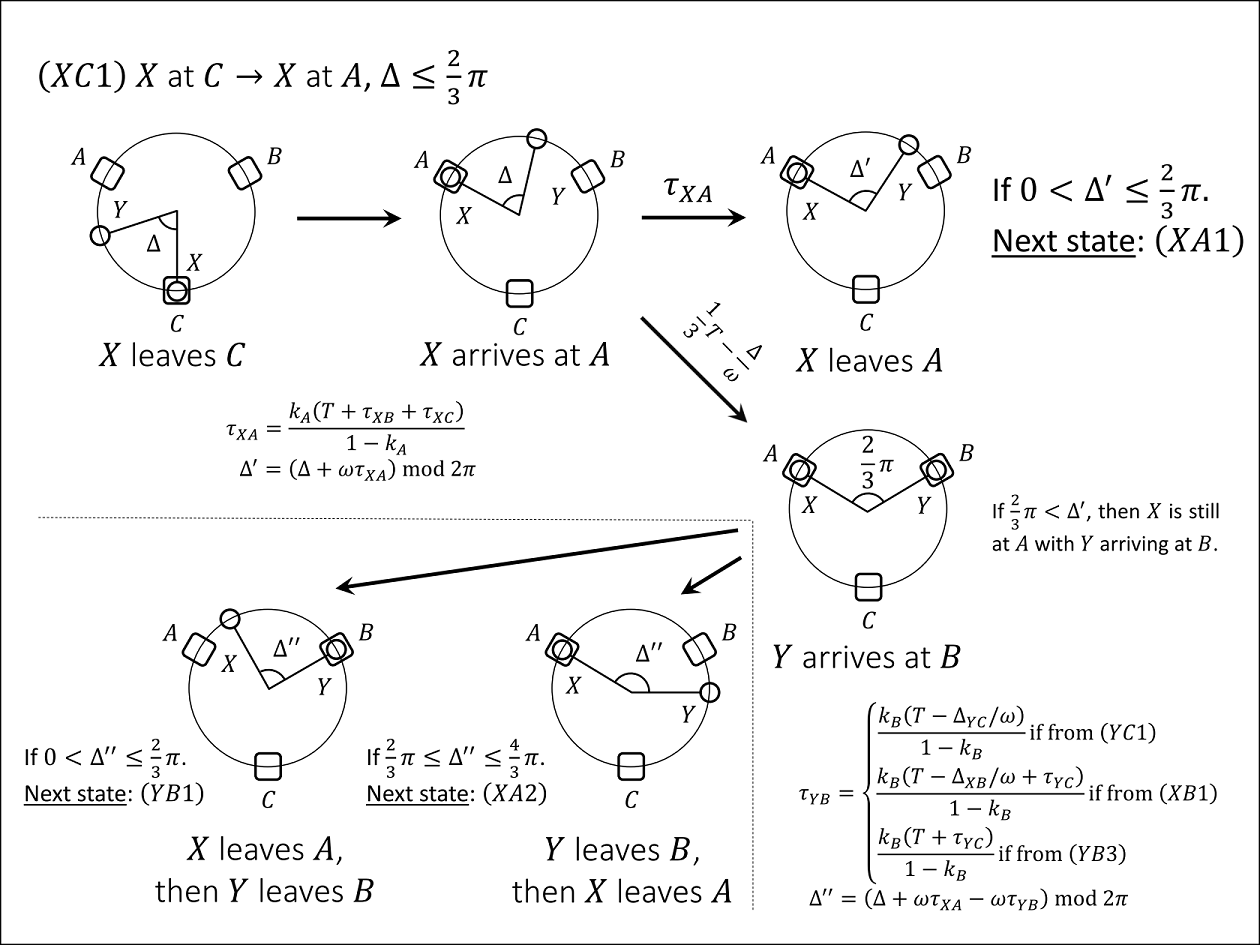}
\caption{After $X$ leaves $C$ with $\Delta_{XC}\leq2\pi/3$, the next possible states are $XA1$, $XA2$ or $YB1$.}\label{fig27}
\end{figure}
\begin{figure}
\includegraphics[width=13cm]{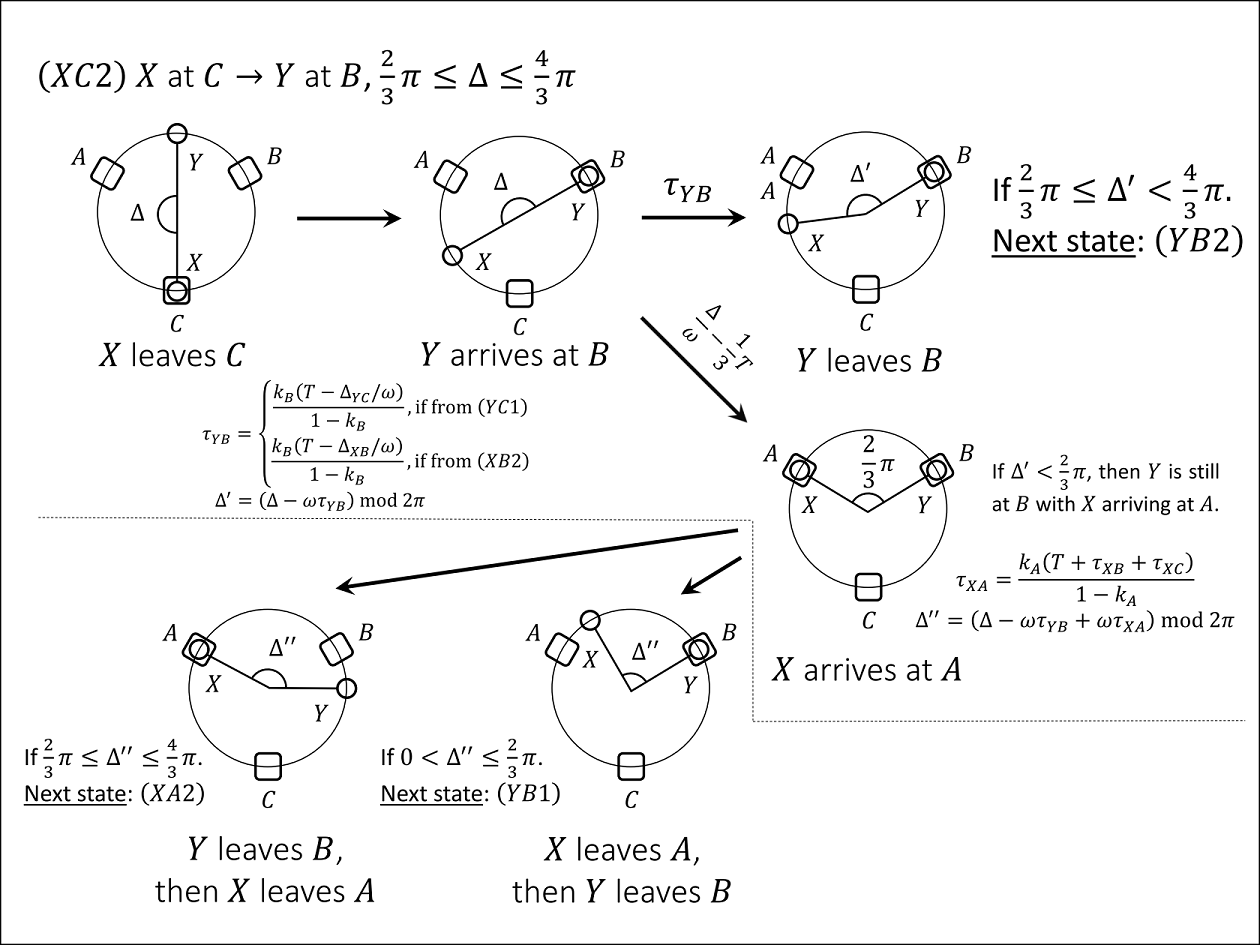}
\caption{After $X$ leaves $C$ with $2\pi/3\leq\Delta_{XC}\leq4\pi/3$, the next possible states are $XA2$, $YB1$ or $YB2$.}\label{fig28}
\end{figure}
\begin{figure}
\centering
\includegraphics[width=13cm]{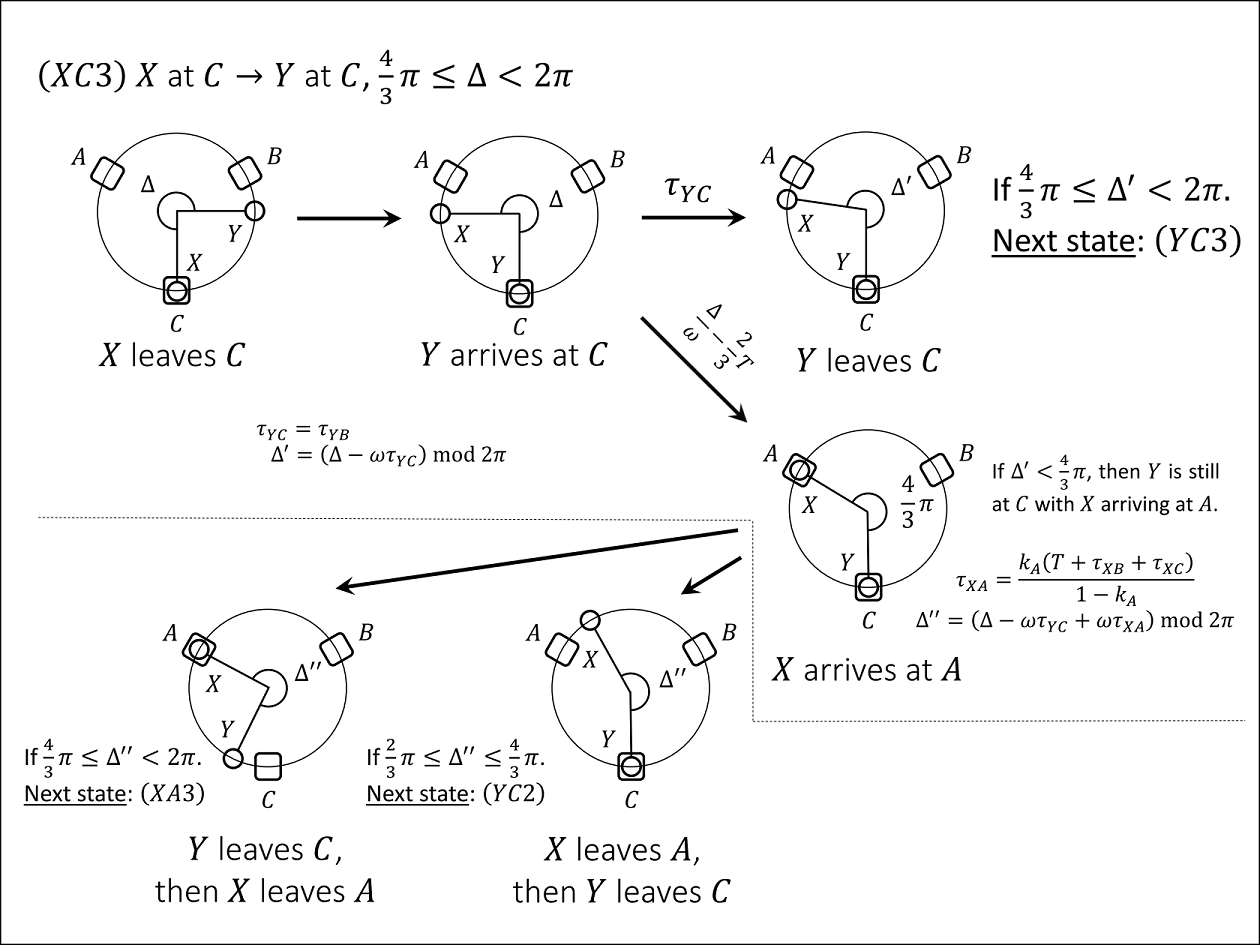}
\caption{After $X$ leaves $C$ with $4\pi/3\leq\Delta_{XC}$, the next possible states are $XA3$, $YC2$ or $YC3$.}\label{fig29}
\end{figure}
\begin{figure}
\includegraphics[width=13cm]{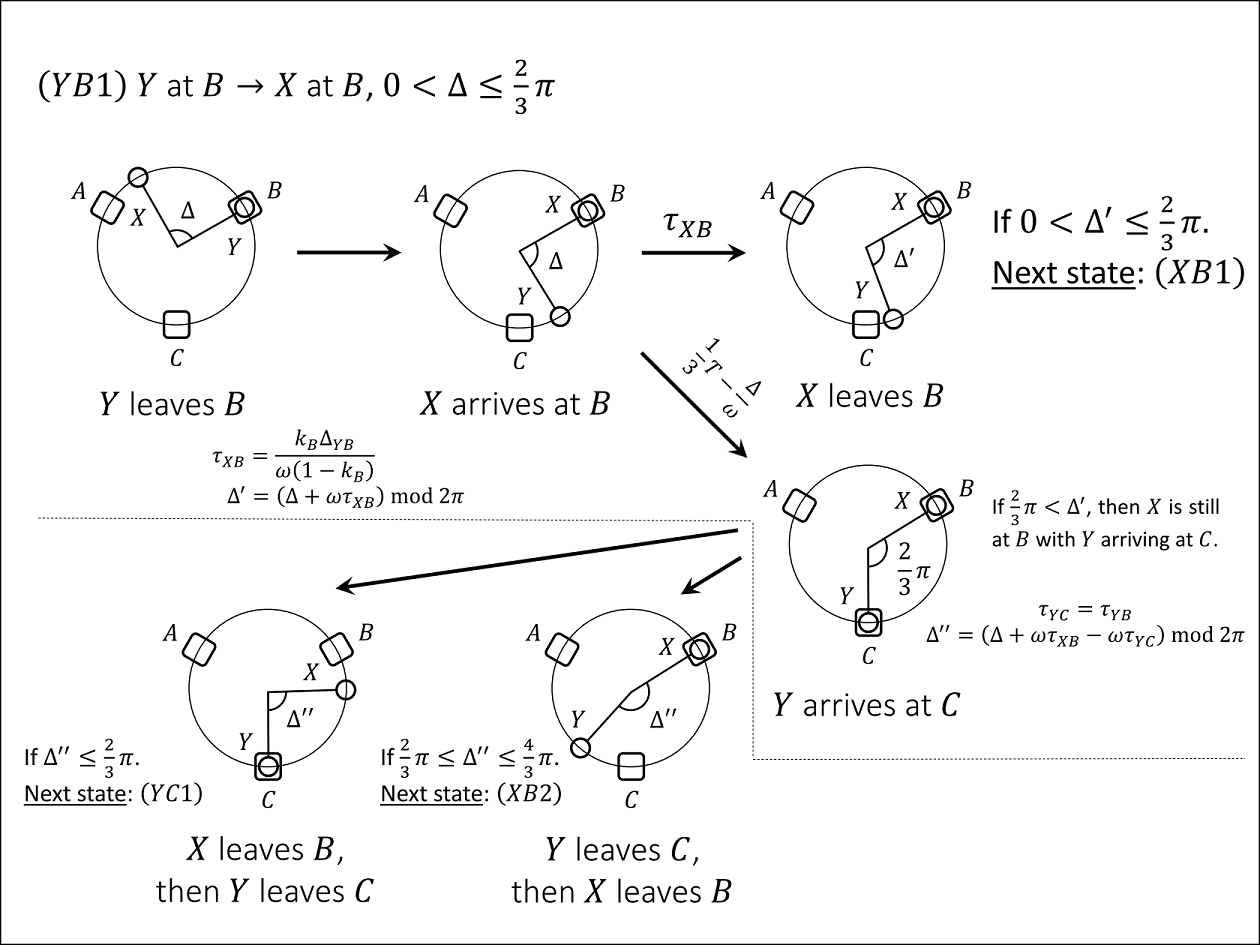}
\caption{After $Y$ leaves $B$ with $0<\Delta_{YB}\leq2\pi/3$, the next possible states are $XB1$, $XB2$ or $YC1$.}\label{fig30}
\end{figure}
\begin{figure}
\centering
\includegraphics[width=13cm]{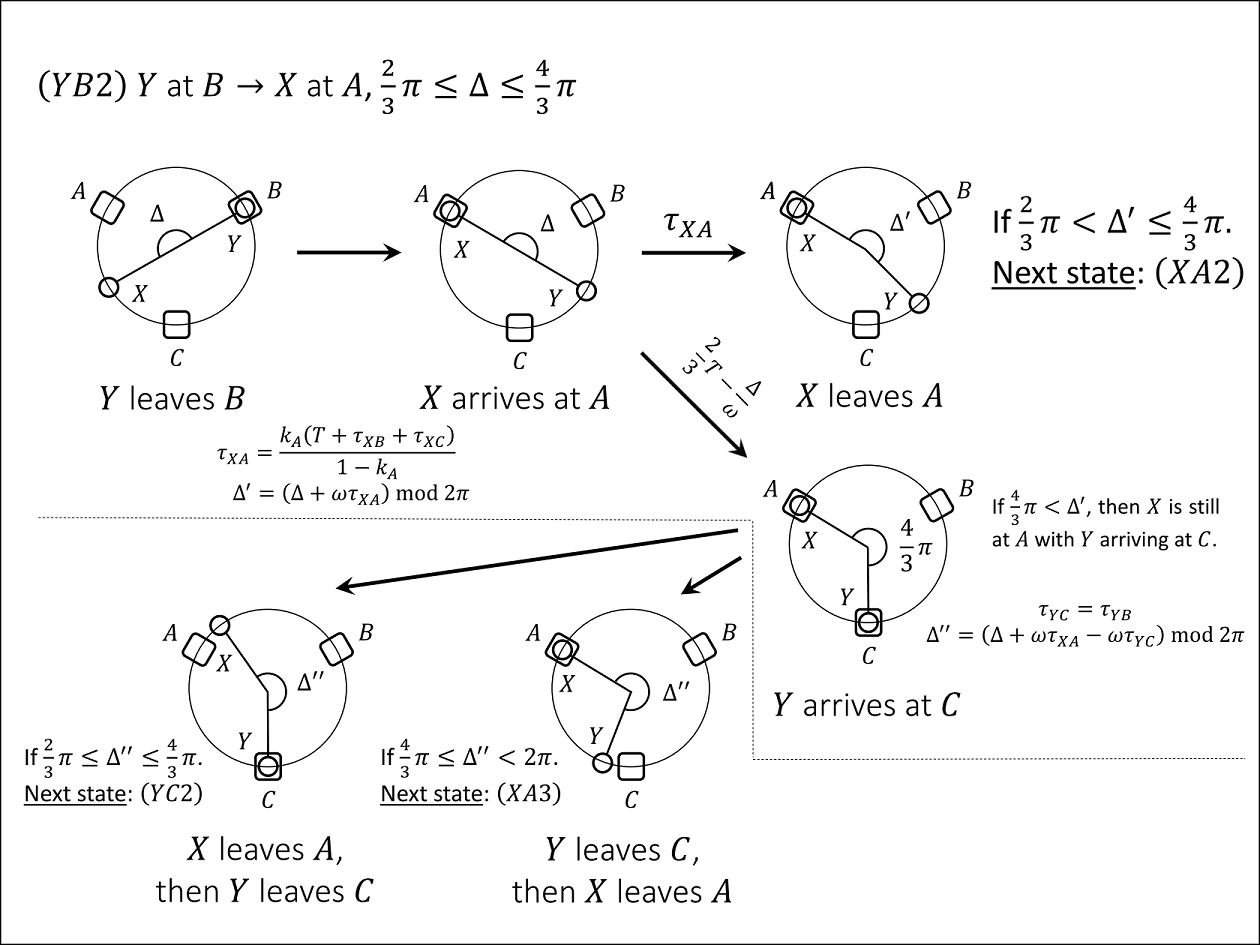}
\caption{After $Y$ leaves $B$ with $2\pi/3\leq\Delta_{YB}\leq4\pi/3$, the next possible states are $XA2$, $XA3$ or $YC2$.}\label{fig31}
\end{figure}
\begin{figure}
\includegraphics[width=13cm]{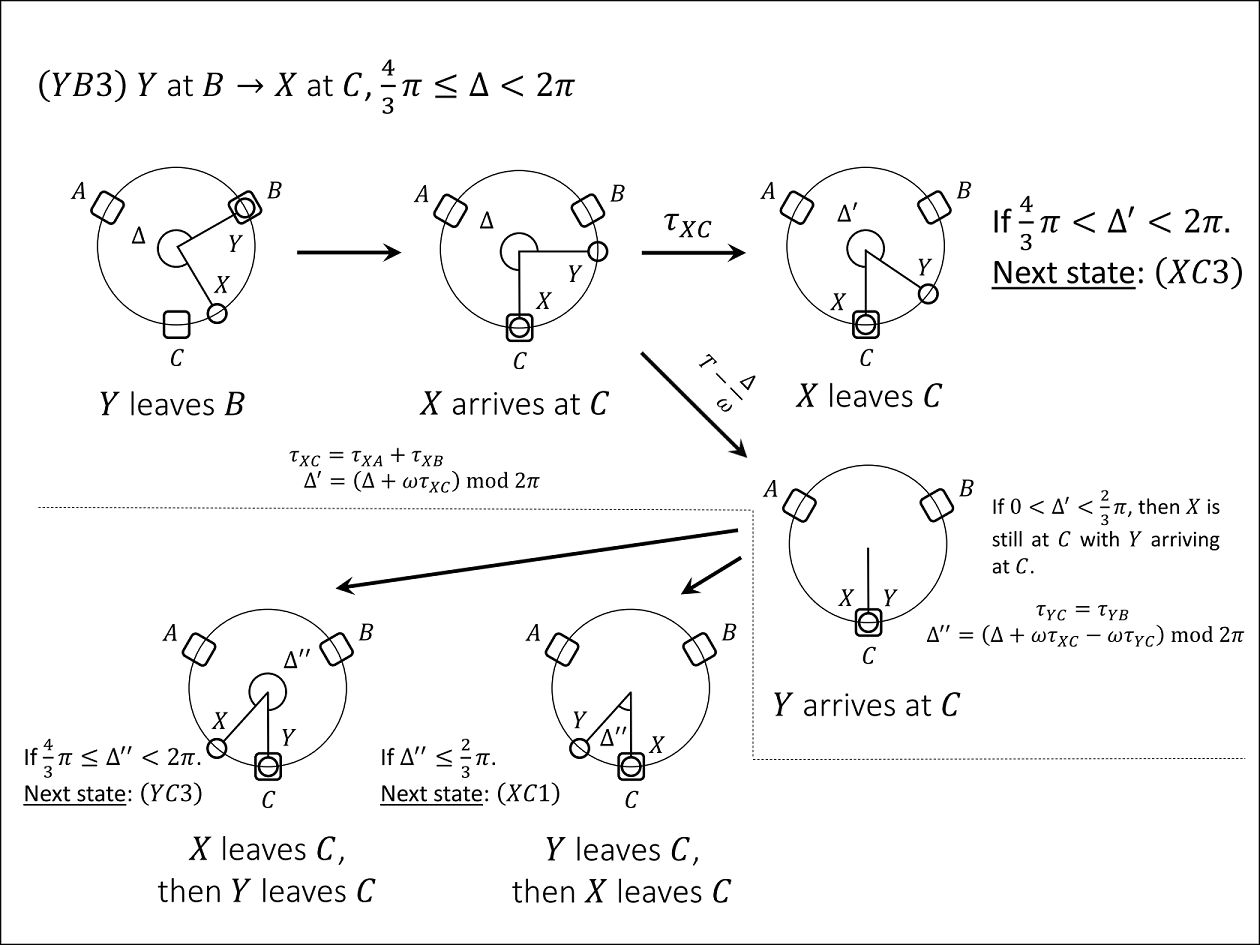}
\caption{After $Y$ leaves $B$ with $4\pi/3\leq\Delta_{YB}$, the next possible states are $XC1$, $XC3$ or $YC3$.}\label{fig32}
\end{figure}
\begin{figure}
\centering
\includegraphics[width=13cm]{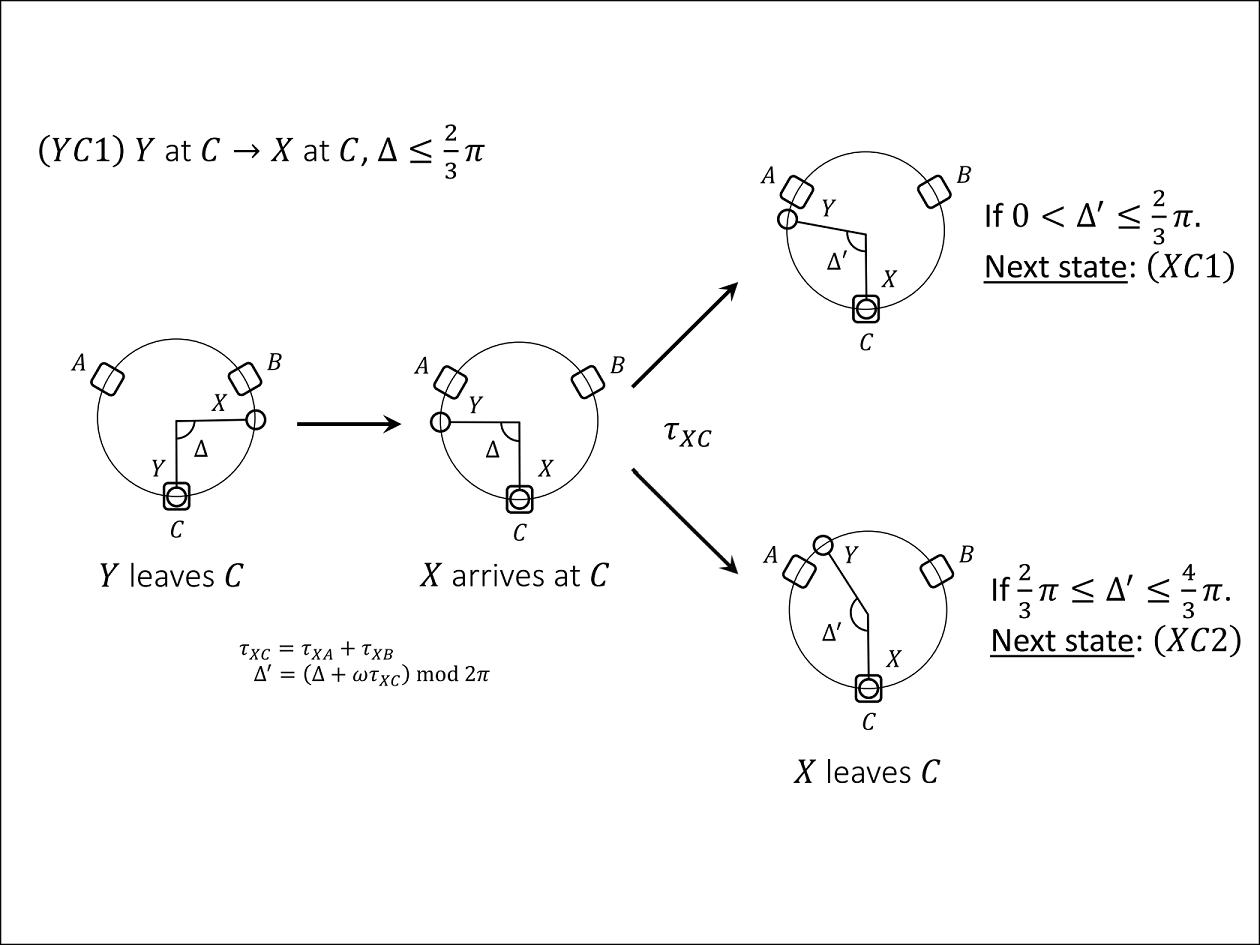}
\caption{After $Y$ leaves $C$ with $\Delta_{YC}\leq2\pi/3$, the next possible states are $XC1$ or $XC2$.}\label{fig33}
\end{figure}
\begin{figure}
\includegraphics[width=13cm]{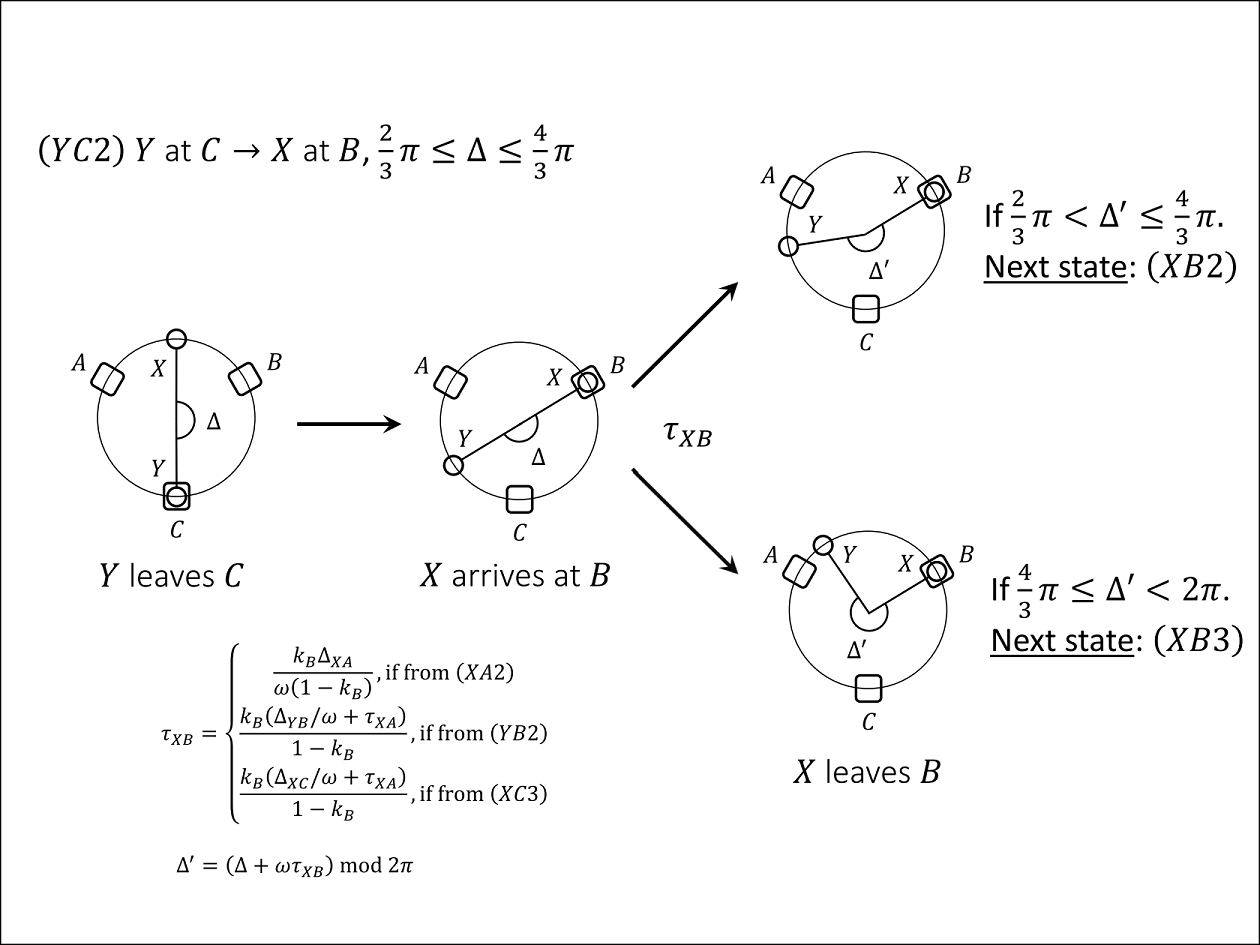}
\caption{After $Y$ leaves $C$ with $2\pi/3\leq\Delta_{YC}\leq4\pi/3$, the next possible states are $XB2$ or $XB3$.}\label{fig34}
\end{figure}
\begin{figure}
\centering
\includegraphics[width=13cm]{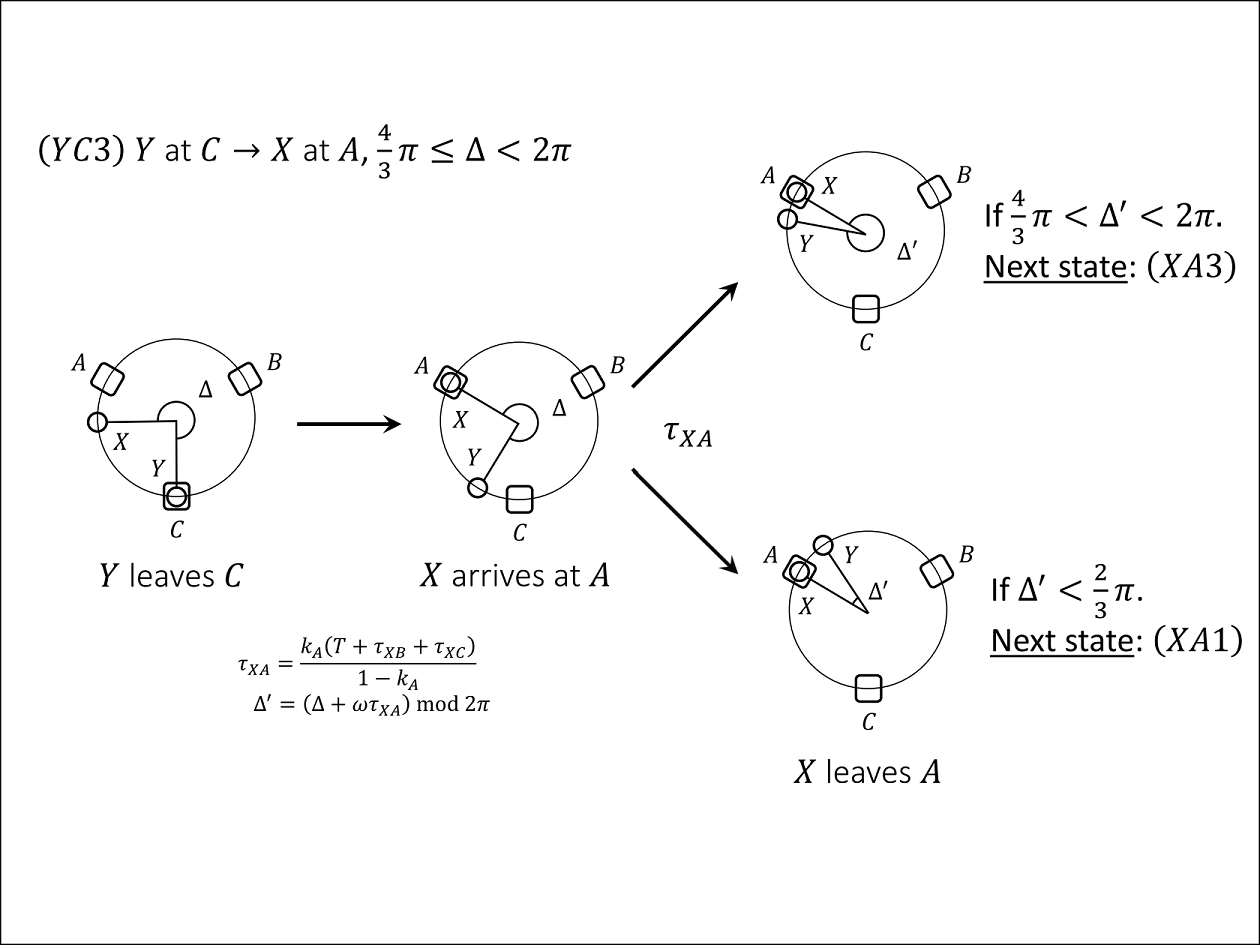}
\caption{After $Y$ leaves $C$ with $4\pi/3\leq\Delta_{YC}$, the next possible states are $XA1$ or $XA3$.}\label{fig35}
\end{figure}

The $A+B\rightarrow C$ system extends the brute-force enumeration of future states given a present state from the $AB$ system, to now include a third bus stop $C$ where people would alight. This implies that $\tau_{XC}=\tau_{XA}+\tau_{XB}$ and $\tau_{YC}=\tau_{YB}$. As in the $AB$ system, we consider the situation where when one bus is at a bus stop, the other bus possibly traverses at most one other bus stop, where the three bus stops are separated by $2\pi/3$ along the circle. For $k_B=0.01$, the upper limit to $k_A$ before the other bus possibly traverses two bus stops is $k_A=0.1945$. This is more than enough to account for realistic demands for buses.

With three bus stops here, when a bus leaves a bus stop, the other bus can be in three different situations depending generally on whether the phase difference is $\Delta\leq2\pi/3,2\pi/3\leq\Delta\leq4\pi/3$ or $4\pi/3\leq\Delta$. The specific rules are summarised in Figs.\ \ref{fig21}-\ref{fig35}. As $X$ would stop at three bus stops and $Y$ would stop at two bus stops, there are five combinations of a bus leaving a bus stop, viz. $XA,XB,XC,YB,YC$. Since each combination has three distinct states corresponding to $\Delta$, there is a total of fifteen states.  The results on our enumeration for this exact $A+B\rightarrow C$ system is presented in Fig.\ \ref{fig3} for $k_B=0.01$, $k_A\in[0,0.1945]$.

\section{Periodic orbits in the \texorpdfstring{$AB$}{AB} system}\label{appenD}

\subsection{Period-\texorpdfstring{$2$}{2} orbits for \texorpdfstring{$k_A<k_B$}{kA < kB}}

The $AB$ system cycles around a period-$2$ orbit for its phase difference $\Delta$ if $k_A<k_B$. Recall that $X$ serves both $A$ and $B$ with $Y$ serving only $B$. The role of $A$ is to hold back $X$ such that if $X$ and $Y$ are bunched at $B$, then they can unbunch with $Y$ just proceeding away when $X$ stops at $A$. However, with $k_A<k_B$, this effect of holding back $X$ to unbunch is insufficient. From the evolution of the states according to Fig.\ \ref{fig1}, we find that the system cycles around the following states:
\begin{align}\label{period2}
\beta\rightarrow XB1\rightarrow XA1\rightarrow\beta.
\end{align}

\begin{figure}
\centering
\includegraphics[width=16cm]{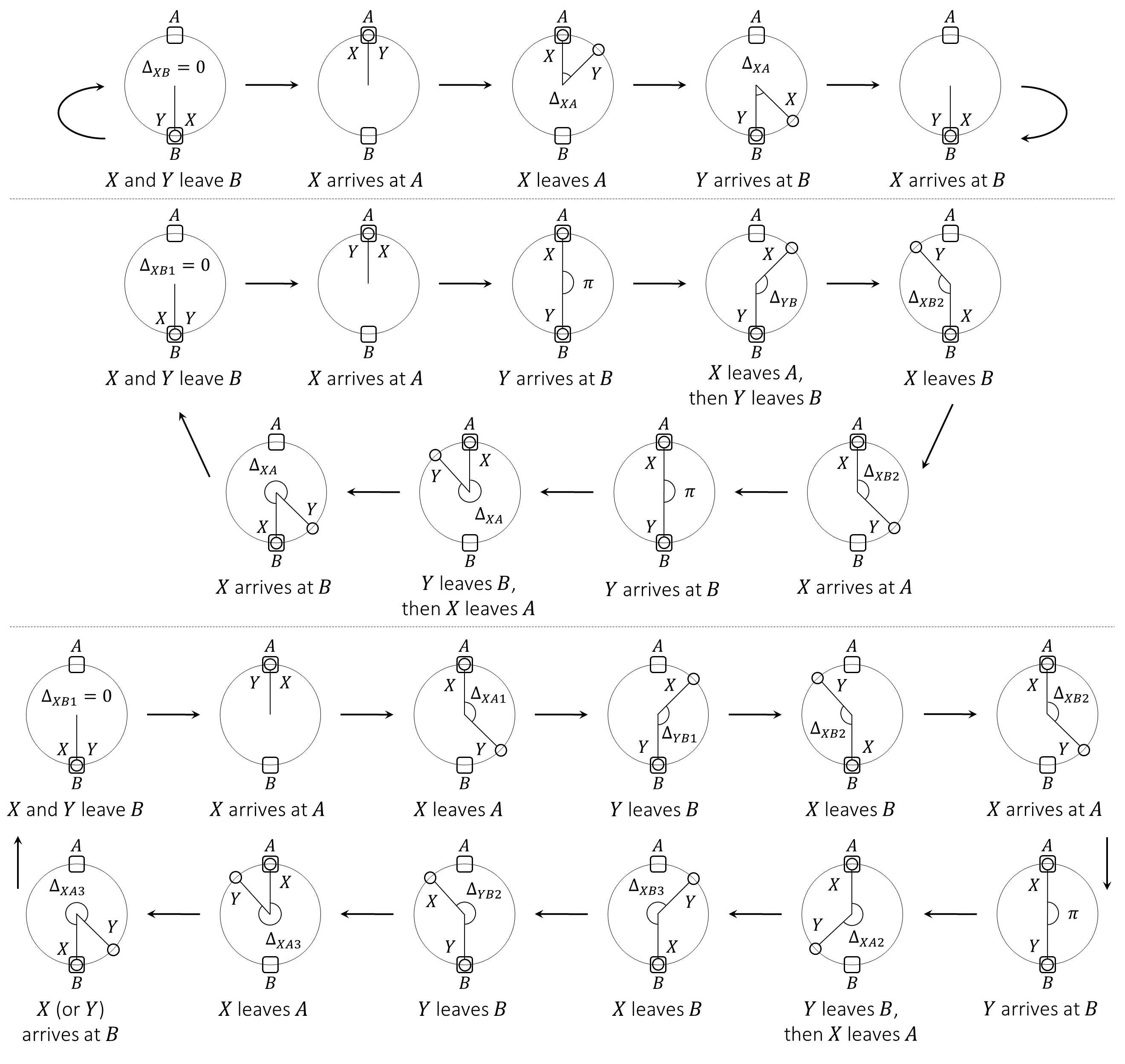}
\caption{The phase difference $\Delta$ of the $AB$ system cycles through periodic orbits. Top: Period-$2$ orbits for $k_A<k_B$. Middle: A period-$4$ orbit when $k_A=0.3325$, $k_B=0.01$. Bottom: A window of period-$8$ orbits when $k_A\in[0.2495,0.2530]$, $k_B=0.01$.}
\label{fig36}
\end{figure}

The top of Fig.\ \ref{fig36} shows this cycle where $\Delta$ undergoes a period-$2$ orbit. We can calculate this from the state transition rules given in Appendix \ref{appenB}. When $X$ and $Y$ leave $B$ being bunched together, the phase difference is $\Delta_{XB}=0$. Next, $X$ arrives at $A$ with $Y$ carrying on with its journey. The time that $X$ stops at $A$ is
\begin{align}
\tau_{XA}=\frac{k_A(T+\tau_{XB})}{1-k_A}.
\end{align}
When $X$ leaves $A$, the phase difference $\Delta_{XA}$ is given by
\begin{align}
\Delta_{XA}=0+\omega\tau_{XA},
\end{align}
since $Y$ opens up this phase difference over a time $\tau_{XA}$ by moving at its constant angular velocity of $\omega=2\pi/T$. Subsequently, $Y$ arrives at $B$, with $X$ eventually bunching with $Y$ at $B$ since $k_B$ is sufficiently large to keep $Y$ there until $X$ arrives. Buses share loading when they are bunched. The times that $X$ and $Y$ spend stopping at $B$ are respectively
\begin{align}
\tau_{XB}&=\left(\frac{1-k_B}{2-k_B}\right)\left(\frac{k_BT}{1-k_B}-\frac{\Delta_{XA}}{\omega}\right)\\
\tau_{YB}&=\left(\frac{1-k_B}{2-k_B}\right)\left(\frac{k_BT}{1-k_B}-\frac{\Delta_{XA}}{\omega}\right)+\frac{\Delta_{XA}}{\omega}.
\end{align}
The phase difference is back to $\Delta_{XB}=0$ when they leave $B$, with the cycle repeating. We can solve this to obtain the following in terms of the parameters $k_A$ and $k_B$:
\begin{align}
\Delta_{XB}&=0\\
\Delta_{XA}&=\frac{4\pi k_A}{2-k_A-k_B}\\
\tau_{XA}&=\frac{2k_AT}{2-k_A-k_B}\label{XAperiod2}\\
\tau_{XB}&=\left(\frac{k_B-k_A}{2-k_A-k_B}\right)T\label{XBperiod2}\\
\tau_{YB}&=\left(\frac{k_A+k_B}{2-k_A-k_B}\right)T\label{YBperiod2}.
\end{align}
When these solution curves are drawn onto Fig.\ \ref{fig2} for $k_A<k_B$, they would precisely fit through all the points plotted from the evolution of the system via the transition rules.

\subsection{Windows of periodic orbits for \texorpdfstring{$k_A>k_B$}{kA > kB}}

For $k_A>k_B$, the values of $\Delta$ generally fill up space over $[0,2\pi)$ or some finite-size intervals. Intriguingly, in the midst of filling up space, there exist some windows of periodic orbits for different values of $k_A$. One such periodic orbit occurs right at the upper limit of $k_A=0.3325$ with $k_B=0.01$, before these exact evolution rules given in Appendix \ref{appenB} break down. At this value of $k_A$, the system cycles through the following states:
\begin{align}\label{period4}
\beta\rightarrow XB1\rightarrow\alpha\rightarrow YB1\rightarrow XB1\rightarrow\alpha\rightarrow XA2\rightarrow\beta.
\end{align}
We can again explicitly calculate this periodic orbit using the evolution rules as in the period-$2$ case where $k_A<k_B$. These details are given below. We will state the results here: $\Delta$ cycles around a period-$4$ orbit, as described in the middle of Fig.\ \ref{fig36}, taking values $0, 3.093, 3.125, 6.238$; $\tau_{XA}$ cycles around a period-$2$ orbit with values $0.5006, 0.5024$; $\tau_{XB}$ cycles around a period-$2$ orbit with values $0.0050,0.0086$; $\tau_{YB}$ cycles around a period-$3$ orbit with values $0.0101,0.0051,0.0014$.

Perhaps the largest window of periodic orbits occurs for $k_A\in[0.2490, 0.2530]$. Here, $\Delta$ cycles through eight values (hence a period-$8$ orbit), and the system evolves through the following states:
\begin{align}
\beta\rightarrow XB1\rightarrow XA1\rightarrow YB1\rightarrow XB1\rightarrow\alpha\rightarrow XA2\rightarrow XB2\rightarrow YB2\rightarrow XA2\rightarrow\beta,\nonumber\\\textrm{ for }k_A\in[0.2490, 0.2505]\label{period8-1}\\
\beta\rightarrow XB1\rightarrow XA1\rightarrow YB1\rightarrow XB1\rightarrow\alpha\rightarrow XA2\rightarrow XB2\rightarrow YB2\rightarrow XA1\rightarrow\beta,\nonumber\\\textrm{ for }k_A\in[0.2510, 0.2530].\label{period8-2}
\end{align}
Somewhere between $k_A\in[0.2505, 0.2510]$, the system switches from the former path to the latter path. The existence of such periodic windows embedded within a region where the quantities fill up the space is due to the fact that when $k_A$ gets too large, then the system cannot remain in such an evolution of states because $\Delta$ would change from being less than $\pi$ to greater than $\pi$ or vice versa. Consequently, the system evolves via other routes in Fig.\ \ref{fig1}. For many values of $k_A$, the system does not cycle through some finite set of states with $\Delta$ filling up space. But for some $k_A$, the system finds itself in such periodic orbits.

Explicit calculations for this period-$8$ orbit of $\Delta$, as well as the periodic orbits of $\tau_{XA},\tau_{XB},\tau_{YB}$ are given below.

\subsubsection{A period-\texorpdfstring{$4$}{4} orbit right at the upper limit of \texorpdfstring{$k_A$}{kA}}

We shall calculate the periodic orbits for $k_A=0.3325$, $k_B=0.01$ as well as those for $k_A\in[0.2490,0.2530]$, $k_B=0.01$ of the $AB$ system. The former is the highest $k_A$ described by these state evolution rules and is perhaps the shortest periodic orbit for $k_A>k_B=0.01$ (length $4$ for $\Delta$), whilst the latter is the largest \emph{window} of periodic orbits for $k_A>k_B=0.01$.

For the former, the evolution of the states is given by Eq.\ (\ref{period4}) and depicted in the middle of Fig.\ \ref{fig36}. According to the evolution rules in Appendix \ref{appenB}, we get the following equations:
\begin{align}
\Delta_{XB1}&=0\\
\tau_{XA1}&=\frac{k_A(T+\tau_{XB2})}{1-k_A}\\
\tau_{YB1}&=\frac{k_BT}{1-k_B}\\
\Delta_{YB}&=\omega(\tau_{XA1}-\tau_{YB1})\\
\tau_{XB1}&=\frac{k_B\Delta_{YB}}{\omega(1-k_B)}\\
\Delta_{XB2}&=\Delta_{YB}+\omega\tau_{XB1}\\
\tau_{XA2}&=\frac{k_A(T+\tau_{XB1})}{1-k_A}\\
\tau_{YB2}&=\frac{k_B(T-\Delta_{XB2}/\omega)}{1-k_B}\\
\Delta_{XA}&=\Delta_{XB2}+\omega(\tau_{XA2}-\tau_{YB2})\\
\tau_{XB2}&=\frac{T}{2-k_B}-\left(\frac{1-k_B}{2-k_B}\right)\frac{\Delta_{XA}}{\omega}\\
\tau_{YB3}&=\frac{\Delta_{XA}}{\omega(2-k_B)}-\left(\frac{1-k_B}{2-k_B}\right)T.
\end{align}
We can carry out some substitutions to eliminate all $\tau_i$, and end up with two linear equations with two variables $\Delta_{XA}$ and $\Delta_{YB}$:
\begin{align}
\Delta_{XA}&=P+Q\Delta_{YB}\\
\Delta_{YB}&=R-S\Delta_{XA},
\end{align}
where
\begin{align}
P&=2\pi\left(\frac{k_A}{1-k_A}-\frac{k_B}{1-k_B}\right)\\
Q&=\frac{1}{1-k_B}\left(\frac{k_Ak_B}{1-k_A}+\frac{1}{1-k_B}\right)\\
R&=2\pi\left(\frac{k_A(3-k_B)}{(1-k_A)(2-k_B)}-\frac{k_B}{1-k_B}\right)\\
S&=\frac{k_A(1-k_B)}{(1-k_A)(2-k_B)}.
\end{align}
Then, we have the following period-$4$ values of $\Delta$:
\begin{align}
\Delta_{XB1}&=0\\
\Delta_{XA}&=\frac{P+QR}{1+QS}\\
\Delta_{YB}&=\frac{R-PS}{1+QS}\\
\Delta_{XB2}&=\frac{\Delta_{YB}}{1-k_B}.
\end{align}
The two periodic values of $\tau_{XA}$, two periodic values of $\tau_{XB}$ and three periodic values of $\tau_{YB}$ can then be calculated one by one.

\subsubsection{The largest window of periodic orbits for \texorpdfstring{$k_A>k_B$}{kA>kB}}

For the largest window of periodic orbits for $k_A>k_B=0.01$, the evolution of the states is given by Eqs. (\ref{period8-1})-(\ref{period8-2}), depending on whether the second last state is $XA2$ or $XA1$, and depicted at the bottom of Fig.\ \ref{fig36}. We show the case given by Eq.\ (\ref{period8-1}). According to the evolution rules in Appendix \ref{appenB}, we get the following equations:
\begin{align}
\Delta_{XB1}&=0\\
\tau_{XA1}&=\frac{k_A(T+\tau_{XB3})}{1-k_A}\\
\Delta_{XA1}&=\omega\tau_{XA1}\\
\tau_{YB1}&=\frac{k_BT}{1-k_B}\\
\Delta_{YB1}&=\Delta_{XA1}-\omega\tau_{YB1}\\
\tau_{XB1}&=\frac{k_B\Delta_{YB1}}{\omega(1-k_B)}\\
\Delta_{XB2}&=\Delta_{YB1}+\omega\tau_{XB1}\\
\tau_{XA2}&=\frac{k_A(T+\tau_{XB1})}{1-k_A}\\
\tau_{YB2}&=\frac{k_B(T-\Delta_{XB2}/\omega)}{1-k_B}\\
\Delta_{XA2}&=\Delta_{XB2}+\omega(\tau_{XA2}-\tau_{YB2})\\
\tau_{XB2}&=\frac{k_B\Delta_{XA2}}{\omega(1-k_B)}\\
\Delta_{XB3}&=\Delta_{XA2}+\omega\tau_{XB2}\\
\tau_{YB3}&=\frac{k_B(T-\Delta_{XB3}/\omega)}{1-k_B}\\
\Delta_{YB2}&=\Delta_{XB3}-\omega\tau_{YB3}\\
\tau_{XA3}&=\frac{k_A(T+\tau_{XB2})}{1-k_A}\\
\Delta_{XA3}&=\Delta_{YB2}+\omega\tau_{XA3}\\
\tau_{YB4}&=\frac{\Delta_{XA3}}{\omega(2-k_B)}-\left(\frac{1-k_B}{2-k_B}\right)T\\
\tau_{XB3}&=\frac{T}{2-k_B}-\left(\frac{1-k_B}{2-k_B}\right)\frac{\Delta_{XA3}}{\omega}.
\end{align}
Eliminating all $\tau_i$, we get the following eight $\Delta$'s:
\begin{align}
\Delta_{XB1}&=0\\
\Delta_{XA1}&=\frac{2\pi k_A(3-k_B)}{(1-k_A)(2-k_B)}-\frac{k_A(1-k_B)}{(1-k_A)(2-k_B)}\Delta_{XA3}\\
\Delta_{YB1}&=\Delta_{XA1}-\frac{2\pi k_B}{1-k_B}\\
\Delta_{XB2}&=\frac{\Delta_{YB1}}{1-k_B}\\
\Delta_{XA2}&=2\pi\left(\frac{k_A}{1-k_A}-\frac{k_B}{1-k_B}\right)+\frac{\Delta_{YB1}}{1-k_B}\left(\frac{k_Ak_B}{1-k_A}+\frac{1}{1-k_B}\right)\\
\Delta_{XB3}&=\frac{\Delta_{XA2}}{1-k_B}\\
\Delta_{YB2}&=\frac{\Delta_{XB3}}{1-k_B}-\frac{2\pi k_B}{1-k_B}\\
\Delta_{XA3}&=2\pi\left(\frac{k_A}{1-k_A}-\frac{k_B}{1-k_B}\right)+\frac{\Delta_{XA2}}{1-k_B}\left(\frac{k_Ak_B}{1-k_A}+\frac{1}{1-k_B}\right).
\end{align}
These $\Delta$'s can be manipulated to end up with two linear equations with two variables $\Delta_{XA3}$ and $\Delta_{XA1}$ satisfying:
\begin{align}
\Delta_{XA3}&=P+Q\Delta_{XA1}\\
\Delta_{XA1}&=R-S\Delta_{XA3},
\end{align}
where
\begin{align}
P&=2\pi\Bigg(\frac{k_A}{1-k_A}-\frac{k_B}{1-k_B}+\frac{1}{1-k_B}\left(\frac{k_Ak_B}{1-k_A}+\frac{1}{1-k_B}\right)\times\nonumber\\&\phantom{=2\pi\Bigg(}\left(\frac{k_A}{1-k_A}-\frac{k_B}{1-k_B}-\frac{k_Ak_B^2}{(1-k_A)(1-k_B)^2}-\frac{k_B}{(1-k_B)^3}\right)\Bigg)\\
Q&=\frac{1}{(1-k_B)^2}\left(\frac{k_Ak_B}{1-k_A}+\frac{1}{1-k_B}\right)^2\\
R&=\frac{2\pi k_A(3-k_B)}{(1-k_A)(2-k_B)}\\
S&=\frac{k_A(1-k_B)}{(1-k_A)(2-k_B)}.
\end{align}
Then,
\begin{align}
\Delta_{XA3}&=\frac{P+QR}{1+QS}\\
\Delta_{XA1}&=\frac{R-PS}{1+QS},
\end{align}
and all the quantities can be calculated one by one.

\section{A \texorpdfstring{$6$}{6}-d analytical approximation to the \texorpdfstring{$AB$}{AB} system}\label{appenE}

In general, the order of events is not fixed, especially when overtaking between $X$ and $Y$ occurs when $X$ is stuck at $A$ whilst $Y$ just goes past. Appendices \ref{appenB} and \ref{appenC} describe the systems exactly by brute-force enumeration of the possible future states. The evolving order of events makes writing a set of closed form equations formidable as it necessarily requires tracking back endless historical data to determine the current number of people accumulated at the bus stops. To circumnavigate this hurdle in an effort to at least arrive at an approximate analytical model, we would consider a version where the order of events are assumed to be fixed. In particular, we proceed to derive the evolution of the $AB$ system in the following order as we now describe.

After $X$ just leaves $A$ and $Y$ just leaves $B$ simultaneously from our initial condition, the following event is $X$ stopping at $B$ to pick up people. The number of people accumulated at $B$ is in general given by $s_B(\Delta_{YB}/\omega+\tau_{XA}+\tau_{XB})$. The quantity $\Delta_{YB}$ denotes the phase difference between bus $Y$ as measured from bus $X$ after $Y$ leaves $B$, etc. We use the phase difference $\Delta_{YB}$ from the event where $Y$ last finished serving $B$ and left, to calculate the number of people accumulated at $B$ because that was the previous time when the number of people at $B$ went to zero. It is from the moment that $Y$ last left $B$ that people started accumulating at $B$ at the rate of $s_B$. An assumption is made to calculate this, viz. $X$ has to stop at $A$ after $Y$ last served $B$, which incurs the additional time of $\tau_{XA}$. This should not be included if $\Delta_{YB}<\pi$ since $X$ does not need to traverse $A$ prior to arriving at $B$. However, writing multiple equations conditional upon $\Delta_{YB}$ (like the exact enumeration in Appendices \ref{appenB} and \ref{appenC}) would make them complicated when trying to analytically calculate the Liapunov exponents. Thus, we would slightly overestimate the number of people to be picked up at $B$ by always including $\tau_{XA}$ with the benefit of having one general equation for any $\Delta_{YB}$. Anyway, with this number of people to pick up over a time interval of $\tau_{XB}$ at a loading rate of $l$, we have:
\begin{gather}
s_B\left(\frac{\Delta_{YB}}{\omega}+\tau_{XA}+\tau_{XB}\right)=l\tau_{XB}\\
\tau_{XB}=\frac{k_B(\Delta_{YB}/\omega+\tau_{XA})}{1-k_B},\label{6d1}
\end{gather}
where $k_B:=s_B/l$. After $X$ finishes and leaves $B$, the phase difference $\Delta_{XB}$ now becomes:
\begin{gather}
\Delta_{XB}=(\Delta_{XA}+\omega\tau_{XB})\textrm{ mod }2\pi.\label{6d2}
\end{gather}
Here, $\Delta_{XA}$ was the phase difference of $Y$ with respect to $X$ in the event before $X$ stopping at $B$, which we show below would be the event where $X$ stopped at $A$.

After the event $X$ stopping at $B$, the next event is $Y$ stopping at $B$. The number of people for $Y$ to pick up is $s_B(T-\Delta_{XB}/\omega+\tau_{YB})$. This is because the moment when $X$ leaves $B$, their phase difference is $\Delta_{XB}$ which is measured from $Y$ with respect to $X$. Hence $Y$ has to traverse a phase difference of $2\pi-\Delta_{XB}$ at an angular velocity of $\omega$, which takes a time of $T-\Delta_{XB}/\omega$. Once $Y$ arrives at $B$, it spends a dwell time of $\tau_{YB}$ to pick up people at a rate of $l$. Thus,
\begin{gather}
s_B\left(T-\frac{\Delta_{XB}}{\omega}+\tau_{YB}\right)=l\tau_{YB}\\
\tau_{YB}=\frac{k_B(T-\Delta_{XB}/\omega)}{1-k_B}.\label{6d3}
\end{gather}
After $Y$ finishes and leaves $B$, the phase difference $\Delta_{YB}$ now becomes:
\begin{gather}
\Delta_{YB}=(\Delta_{XB}-\omega\tau_{YB})\textrm{ mod }2\pi.\label{6d4}
\end{gather}

Now after the event $Y$ stopping at $B$, the final event is $X$ stopping at $A$. The number of people for $X$ to pick up is always $s_A(T+\tau_{XA}+\tau_{XB})$, independent of the phase difference since only $X$ ever picks up people from $A$ and the time it takes is always $T$ to complete the loop plus the time it takes to traverse $B$, before returning to $A$ and spend some time stopping there. With people picked up over $\tau_{XA}$ at a rate of $l$,
\begin{gather}
s_A(T+\tau_{XA}+\tau_{XB})=l\tau_{XA}\\
\tau_{XA}=\frac{k_A(T+\tau_{XB})}{1-k_A},\label{6d5}
\end{gather}
where $k_A:=s_A/l$. After $X$ finishes and leaves $A$, the phase difference $\Delta_{XA}$ now becomes:
\begin{gather}
\Delta_{XA}=(\Delta_{YB}+\omega\tau_{XA})\textrm{ mod }2\pi.\label{6d6}
\end{gather}

After this event, the subsequent event is $X$ stopping at $B$ where the cycle of these three distinct events repeat. Hence as mentioned earlier, $X$ stopping at $A$ is the event before $X$ stopping at $B$. Let us now define the following six variables $x_1,\cdots,x_6$ as:
\begin{align}
x_1&=\tau_{XB}\\
x_2&=\frac{\Delta_{XB}}{\omega}\\
x_3&=\tau_{YB}\\
x_4&=\frac{\Delta_{YB}}{\omega}\\
x_5&=\tau_{XA}\\
x_6&=\frac{\Delta_{XA}}{\omega}.
\end{align}
In terms of these variables, the dynamics for this bus system is given by the following 6-d map where Eqs. (\ref{6d1}), (\ref{6d2}), (\ref{6d3}), (\ref{6d4}), (\ref{6d5}), (\ref{6d6}) are rewritten as:
\begin{align}
x_1&=\frac{k_B(x_4+x_5)}{1-k_B}\\
x_2&=(x_6+x_1)\textrm{ mod }T\\
x_3&=\frac{k_B(T-x_2)}{1-k_B}\\
x_4&=(x_2-x_3)\textrm{ mod }T\\
x_5&=\frac{k_A(T+x_1)}{1-k_A}\\
x_6&=(x_4+x_5)\textrm{ mod }T.
\end{align}
When iterating through this map, the variables must be updated in the order of $x_1,\cdots,x_6$, and then back to $x_1$ where the events cycle again.

Note that this map has a constant Jacobian (matrix of first order partial derivatives), which would allow for a straightforward and direct calculation of its six Liapunov exponents. The same is true for the $A+B\rightarrow C$ system, and we will carry out further analysis to show that the interacting semi-express bus system is chaotic. Let us proceed with obtaining the corresponding map where people now alight at $C$ in Appendix \ref{appenF}.

\section{A \texorpdfstring{$10$}{10}-d analytical approximation to the \texorpdfstring{$A+B\rightarrow C$}{A+B to C} system}\label{appenF}

For the bus system where people now must alight at $C$, let us put the bus stops $A,B,C$ to be equidistant on the unit circle so that travel time from $A$ to $B$, $B$ to $C$, $C$ to $A$ are all $T/3$. The same arguments apply as in the boarding-only case in the preceding Appendix. There, the result is a $6$-d map because there are three $\tau_i$, viz. $X$ at $A$, $X$ at $B$, $Y$ at $B$, which define three events. For each event, there is also an associated phase difference, giving the total of six variables $x_1,\cdots,x_6$. Here,   there are five events in the following order:
\begin{enumerate}
\item $X$ at $B$.
\item $Y$ at $C$.
\item $X$ at $C$.
\item $Y$ at $B$.
\item $X$ at $A$.
\end{enumerate}
Associated with each event is a $\tau_i$ as well as a corresponding $\Delta_i$, giving a total of ten variables $x_1,\cdots,x_{10}$. We now enunciate the resulting 10-d map:
\begin{align}
x_1&=\frac{k_B(x_8+x_9+x_5)}{1-k_B}\label{aux1}\\
x_2&=(x_{10}+x_1)\textrm{ mod }T\\
x_3&=x_7\\
x_4&=(x_2-x_3)\textrm{ mod }T\\
x_5&=x_9+x_1\\
x_6&=(x_4+x_5)\textrm{ mod }T\\
x_7&=\frac{k_B(T-x_2 + x_3)}{1-k_B}\label{aux2}\\
x_8&=(x_6-x_7)\textrm{ mod }T\\
x_9&=\frac{k_A(T+x_1+x_5)}{1-k_A}\label{auxA}\\
x_{10}&=(x_8+x_9)\textrm{ mod }T,\label{endmap}
\end{align}
where
\begin{align}
x_1&=\tau_{XB}\\
x_2&=\frac{\Delta_{XB}}{\omega}\\
x_3&=\tau_{YC}\\
x_4&=\frac{\Delta_{YC}}{\omega}\\
x_5&=\tau_{XC}\\
x_6&=\frac{\Delta_{XC}}{\omega}\\
x_7&=\tau_{YB}\\
x_8&=\frac{\Delta_{YB}}{\omega}\\
x_9&=\tau_{XA}\\
x_{10}&=\frac{\Delta_{XA}}{\omega}.
\end{align}

Once again, the variables must be iterated in the order of $x_1,\cdots,x_{10}$, and then back to $x_1$ where the events cycle again. This 10-d map also has a constant Jacobian, $J$. To calculate the Liapunov exponents \cite{Yorke96}, all we need to do is calculate $J$ times its transpose, calculate the resulting matrix's eigenvalues (which are guaranteed to be non-negative), take their square roots, and finally take their natural logarithms. The result on the largest Liapunov exponent as a function of the two parameters of the system $k_A$ and $k_B$ are shown in Fig. \ref{fig6}. The largest Liapunov exponent is always positive, indicating sensitivity to initial conditions. None of the 10 Liapunov exponents ever has value zero, which rules out quasi-periodicity.

\begin{figure}
\centering
\includegraphics[width=16cm]{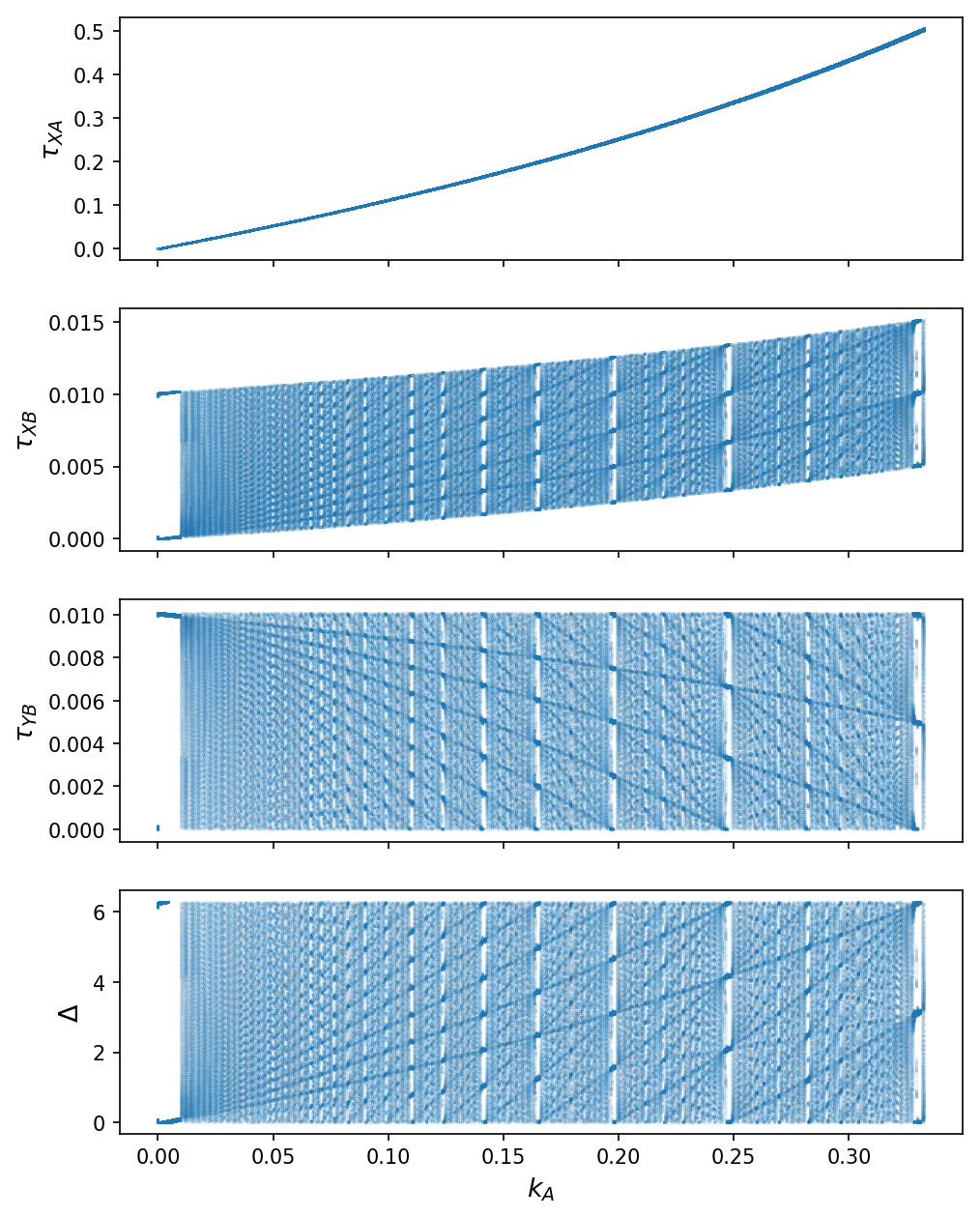}
\caption{Values taken by $\tau_{XA}$, $\tau_{XB}$, $\tau_{YB}$ and $\Delta$ for various values of $k_A$ from $0$ to $0.3325$, in an increment of $0.0005$. The value for $k_B$ is kept at $0.01$. This is an analytical approximation of the $AB$ system, where the map is evaluated based on the usual rules.}
\label{fig37}
\end{figure}

\begin{figure}
\centering
\includegraphics[width=16cm]{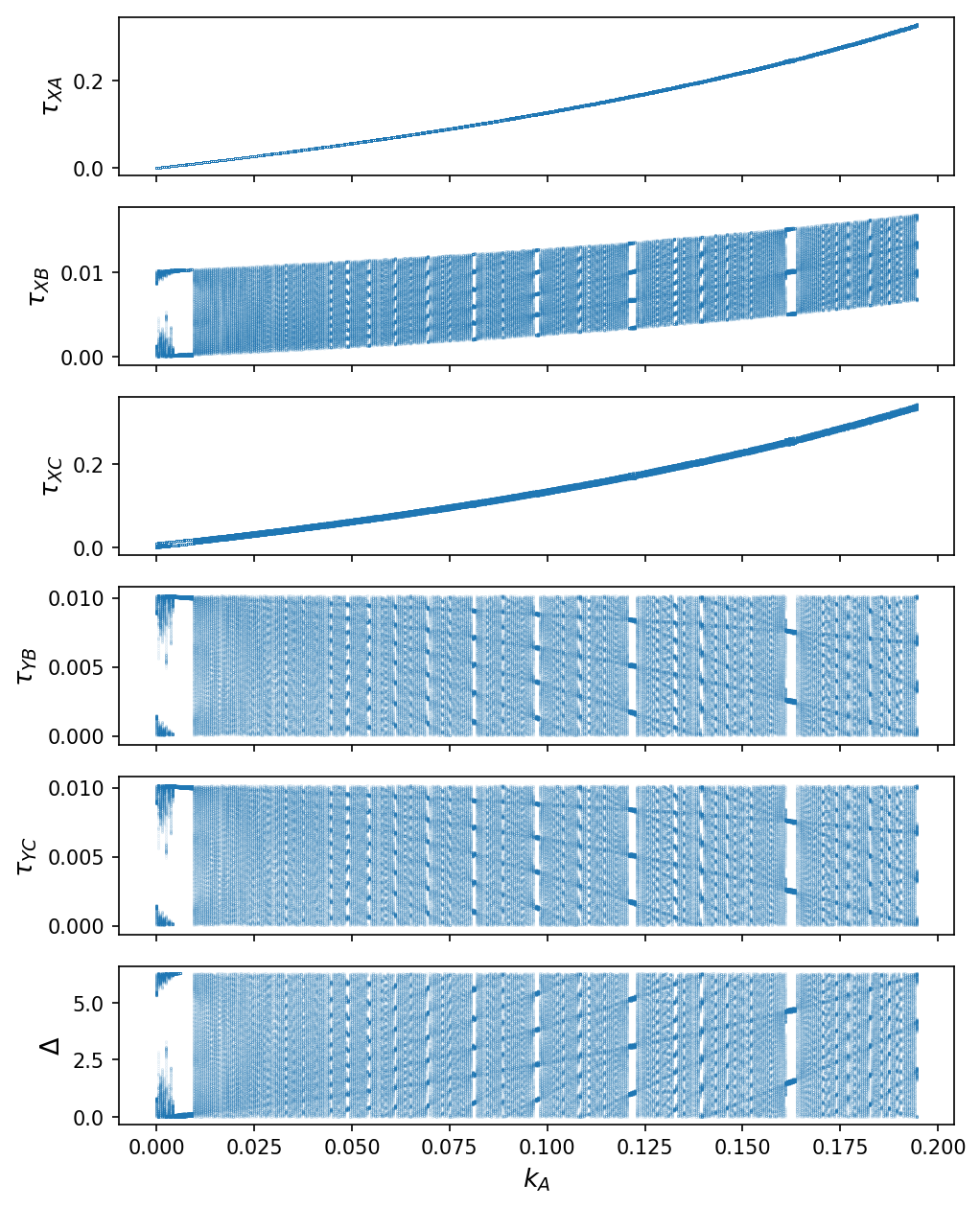}
\caption{Values taken by $\tau_{XA}$, $\tau_{XB}$, $\tau_{XC}$, $\tau_{YB}$, $\tau_{YC}$ and $\Delta$ for various values of $k_A$ from $0$ to $0.1945$, in an increment of $0.0005$. The value for $k_B$ is kept at $0.01$. This is an analytical approximation of the $A+B\rightarrow C$ system, where the map is evaluated based on the usual rules.}
\label{fig38}
\end{figure}

As all variables $x_1,\cdots,x_{10}$ are bounded, we find that the bus system is essentially always in chaos. Note that in evaluating the Jacobian of this $10$-d map, it treats the map as taking values from the previous iteration $x_{1\textrm{ to }10\textrm{ at iteration }t - 1}$ in order to get $x_{1\textrm{ to }10\textrm{ at iteration }t}$. For the bus system however, the evaluation of $x_{2\textrm{ at iteration }t}$ takes the value of $x_{1\textrm{ at iteration }t}$ instead of $x_{1\textrm{ at iteration }t - 1}$, for example, according to the logic upon which this map is derived. Nevertheless, the corresponding figures for Figs.\ \ref{fig4}-\ref{fig5} are essentially the same (see Figs.\ \ref{fig37}-\ref{fig38}). This implies that one may approximate the bus system where the evaluation of $x_{2\textrm{ at iteration }t}$ takes the value of $x_{1\textrm{ at iteration }t}$ by the usual rules of a map where the evaluation of $x_{2\textrm{ at iteration }t}$ takes the value of $x_{1\textrm{ at iteration }t - 1}$ and calculate the Jacobian of the map as usual.

\bibliography{Citation}
\begin{acknowledgments}
This work was supported by the Joint WASP/NTU Programme (Project No. M4082189) and the DSAIR@NTU Grant (Project No. M4082418).
\end{acknowledgments}

\section*{Data availability statement}
Data sharing is not applicable to this article as no new data were created or analysed in this study.




\end{document}